\documentclass[twocolumn]{aastex63}
\newcommand{\psj}{Planet. Sci. J.}
\submitjournal{AJ}

\shortauthors{Vokrouhlick{\' y} et~al.}
\graphicspath{{./}{figures/}}
\usepackage{tikz}
\usetikzlibrary{arrows,shapes,positioning,shadows,trees}

\tikzset{
  basic/.style  = {draw, text width=4cm, drop shadow, font=\sffamily, rectangle},
  root/.style   = {basic, rounded corners=2pt, thin, align=center, fill=blue!10},
  level 2/.style = {basic, rounded corners=6pt, thin,align=center, fill=pink!40, text width=11em},
  level 3/.style = {basic, thin, align=center, fill=white!60, text width=8em}
}
\usepackage[utf8]{inputenc}
\usepackage{dirtytalk}
\begin{document}

\title{Orbital and absolute magnitude distribution of Hilda population}

\correspondingauthor{David Vokrouhlick\'y}
\email{vokrouhl@cesnet.cz}

\author[0000-0002-6034-5452]{David Vokrouhlick\'y}
\affiliation{Astronomical Institute, Charles University, V Hole\v{s}ovi\v{c}k\'ach 2,
             CZ 18000, Prague 8, Czech Republic}
\author[0000-0002-4547-4301]{David Nesvorn{\' y}}
\affiliation{Department of Space Studies, Southwest Research Institute, 1050 Walnut St., Suite 300,
             Boulder, CO 80302, United States}
\author[0000-0003-2763-1411]{Miroslav Bro{\v z}}
\affiliation{Astronomical Institute, Charles University, V Hole\v{s}ovi\v{c}k\'ach 2,
             CZ 18000, Prague 8, Czech Republic}
\author[0000-0002-1804-7814]{William F. Bottke}
\affiliation{Department of Space Studies, Southwest Research Institute, 1050 Walnut St., Suite 300,
             Boulder, CO 80302, United States}
\author[0000-0001-6730-7857]{Rogerio Deienno}
\affiliation{Department of Space Studies, Southwest Research Institute, 1050 Walnut St., Suite 300,
             Boulder, CO 80302, United States}
\author{Carson D. Fuls}
\affiliation{Lunar and Planetary Laboratory, The University of Arizona, 1629 E. University
             Boulevard, Tucson, AZ 85721-0092, USA}             
\author{Frank C. Shelly}
\affiliation{Lunar and Planetary Laboratory, The University of Arizona, 1629 E. University
             Boulevard, Tucson, AZ 85721-0092, USA}
%\author{...}

\begin{abstract}
The Hilda population of asteroids is located in a large orbital zone of long-term stability associated with the Jupiter J3/2 mean motion resonance. They are a sister population of the Jupiter Trojans, since both of them are likely made up of objects captured from the primordial Kuiper belt early in the solar system history. Comparisons between the orbital and physical properties of the Hilda and Trojan populations thus represent a test of outer planet formation models. Here we use a decade of observations from the Catalina Sky Survey (G96 site) to determine the bias-corrected orbital and magnitude distributions of Hildas. We also identify collisional families and the background population by computing a new catalog of synthetic proper elements for Hildas. We model the cumulative magnitude distribution of the background population using a local power-law representation with slope $\gamma(H)$, where $H$ is the absolute magnitude. For the largest Hildas, we find $\gamma\simeq 0.5$ with large uncertainty due to the limited population. Beyond $H\simeq 11$, we find that $\gamma$ transitions to a mean value ${\bar \gamma}=0.32\pm 0.04$ with a slight dependence on $H$ (significantly smaller than Jupiter Trojans with ${\bar \gamma}=0.43\pm 0.02$). We find that members of identified collisional families represent more than $60$\% of the total population (both bias counts). The bias-corrected populations contain about the same number of Hildas within the families and the background for $H\leq 16$, but this number may increase to $60$\% families when their location in the orbital space is further improved in the future. 
\end{abstract}


% ............................................................................
\section{Introduction} \label{intro}
The importance of small bodies in the solar system is twofold: (i) they are interesting as the building blocks or fragments of larger bodies (planets or dwarf planets, in particular) and (ii) if put in the right context, their orbits and size frequency distributions can tell us about the evolution of the largest bodies. This is especially true if the population of small bodies is directly connected to the larger worlds. A prime example of this are populations of asteroids located in the mean-motion resonances with planets. Restricting ourselves to the case of Jupiter, we have two such populations at hand: (i) the Jupiter Trojans, which are locked in the J1/1 mean motion resonance (librating about the L$_4$ and L$_5$ stationary points), and (ii) the Hilda population, which resides in the J3/2 mean motion resonance. In this paper we focus on the latter.

A large amount of information about Hilda population has been collected
to date (see Appendix~\ref{hildas} for a brief overview, especially topics relevant
to our findings in this paper). Nevertheless, we still lack a bias-corrected
global picture of its orbital architecture and magnitude distribution. Here we attempt to do
so, following a methodology used in a sister effort designed to analyze the Jupiter Trojan
$L_4$ and $L_5$ populations \citet{vok2024}. Using the same formalism is advantageous because
our ultimate science goal is to put the Hilda and Trojan populations within the context of
a single model explaining their origin. 

As in \citet{vok2024}, we use observations between 2013-2022 from the Catalina Sky Survey (CSS) at Mt.~Lemmon (G96) (Sec.~\ref{datag96}). We constructed a parametric model that allowed us to describe the orbital architecture and magnitude distribution of the Hilda population from these observations (Sec.~\ref{model}).  For each point (or rather a small bin) in the parameter space, we determined the survey detection probability (Sec.~\ref{det_prob}). We used a maximum likelihood technique to optimize the model parameters to determine the bias-corrected Hilda population (Sec.~\ref{opti}). 

Our main findings are presented in Sec.~\ref{res}, with our outlook for further work summarized in Sec.~\ref{concl}. Similarly to the study of Jupiter Trojans \citep{vok2024}, we found that our new catalog of synthetic proper elements for the Hilda population was an important byproduct of our work (see Appendix~\ref{propel}). A second byproduct is the identification of asteroid families in the Hilda population, with two new families discovered. These data are made available through our webpage \url{https://sirrah.troja.mff.cuni.cz/~mira/tmp/hildas/} and on \url{https://doi.org/10.5281/zenodo.14959239} repository.
% FIG 1 %%%%%%%%%%%%%%%%%%%%%%%%%%%%%%%%%%%%%%%%%%%%%%%%%%%%%%%%%%%%%%%%%%%%%%%%%%%%%%%%%%%%%%%%%%%%%%%
\begin{figure*}[t!]
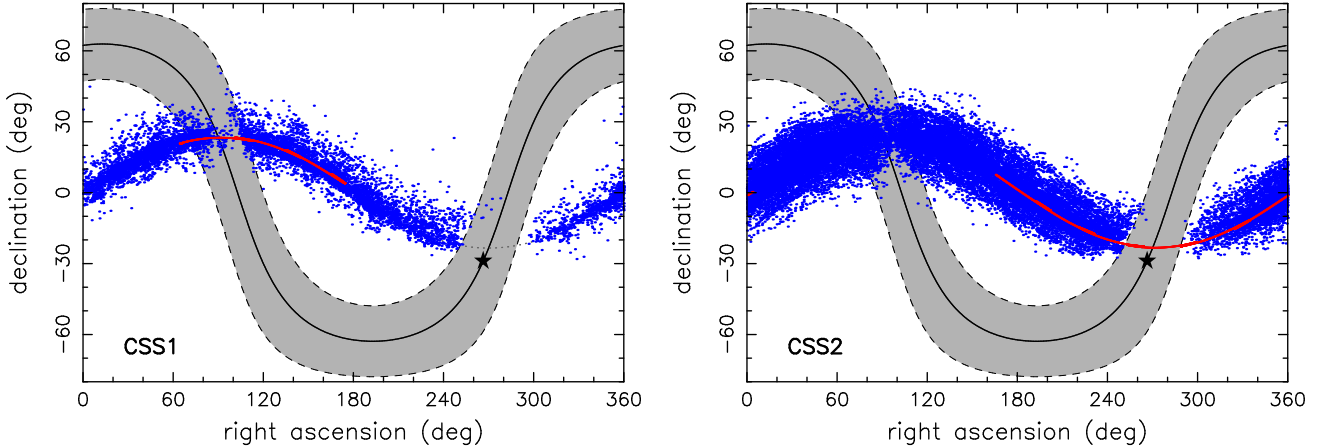

% \plottwo{g96_1_alde_L4.eps}{g96_1_alde_L5.eps}
 \begin{center}
 \begin{tabular}{cc}
  \includegraphics[width=0.47\textwidth]{f1a.eps} &
  \includegraphics[width=0.47\textwidth]{f1b.eps} \\ 
 \end{tabular}
 \end{center}  
 \caption{Detections of Hilda asteroids by the CSS operations in between Jan~2, 2013 and May~14, 2016 (phase~I; left panel), and between May~31, 2016 and June~14, 2022 (phase~II; right panel). There are 8,686 detections of 2,970 individual Hildas during the phase~I, and 28,930 detections of 4,317 individual Hildas during the phase~II. The blue symbols are topocentric right ascension (abscissa) and  declination (ordinate) of the observations. The red line is the Jupiter track in the same period of time near the ecliptic plane (the sense of motion during the displayed period of time is along the increasing right ascension value). The solid black line is the projection of the galactic
 plane surrounded by the $\pm 15^\circ$ latitude strip (gray region). The black star is the direction to the galactic center.}
 \label{fig1}
\end{figure*}
%%%%%%%%%%%%%%%%%%%%%%%%%%%%%%%%%%%%%%%%%%%%%%%%%%%%%%%%%%%%%%%%%%%%%%%%%%%%%%%%%%%%%%%%%%%%%%%%%%%%%%%

% ----------------------------------------------------------------------------
% SEC ???
\section{The observed Hilda population}\label{datag96}
In this section we present the observational data used in our determination of the bias-corrected Hilda population. We proceed in two steps. 

First, we downloaded all known asteroids from the Minor Planet Center (MPC) database and identified those that were members of the Hilda population. However, this is a collection of asteroids detected by many observing stations, all having different detection strategies and sensitivities that are unknown to us. To evaluate the detection probability of the Hildas on different orbits and with different absolute magnitudes, we need to work with data provided by one (or more) stations with an observing strategy and detector performance that we know. Fortunately, this is the case for Catalina Sky Survey (CSS) observations. In the second step, we selected the known Hildas detected by CSS during the interval of operations for which we have good knowledge of the survey's performance. These observations were later used to calibrate the parameters of the Hilda population.
\smallskip
% FIG 1 %%%%%%%%%%%%%%%%%%%%%%%%%%%%%%%%%%%%%%%%%%%%%%%%%%%%%%%%%%%%%%%%%%%%%%%%%%%%%%%%%%%%%%%%%%%%%%%
\begin{figure}[t!]
% \plottwo{g96_nobs_L4.eps}{g96_nobs_L5.eps}
 \begin{center}
 \includegraphics[width=0.47\textwidth]{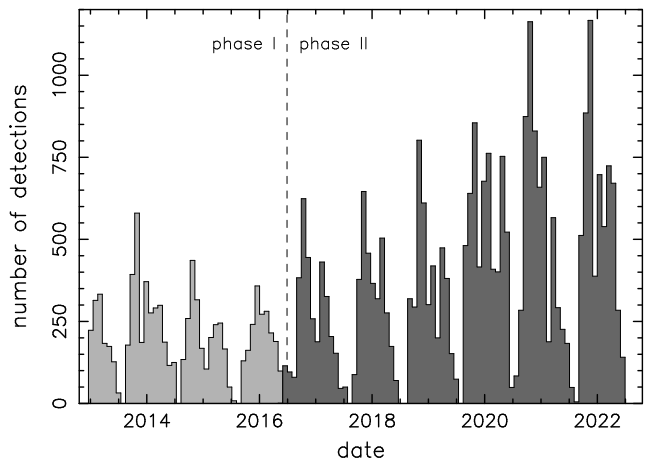} 
 \end{center} 
 \caption{Number of Hilda observations in the CSS operations (phases~I and II separated by the vertical dashed line). Apart from the obvious annual variation (determined both by the seasonal changes and summer maintenance gaps) and more data collected during the phase~II due to the camera upgrade, there is also a longer term variation due to the Jupiter motion and its relative position to the galactic plane.}
 \label{fig2}
\end{figure}
%%%%%%%%%%%%%%%%%%%%%%%%%%%%%%%%%%%%%%%%%%%%%%%%%%%%%%%%%%%%%%%%%%%%%%%%%%%%%%%%%%%%%%%%%%%%%%%%%%%%%%%
% FIG 1 %%%%%%%%%%%%%%%%%%%%%%%%%%%%%%%%%%%%%%%%%%%%%%%%%%%%%%%%%%%%%%%%%%%%%%%%%%%%%%%%%%%%%%%%%%%%%%%
\begin{figure*}[t!]
 \plottwo{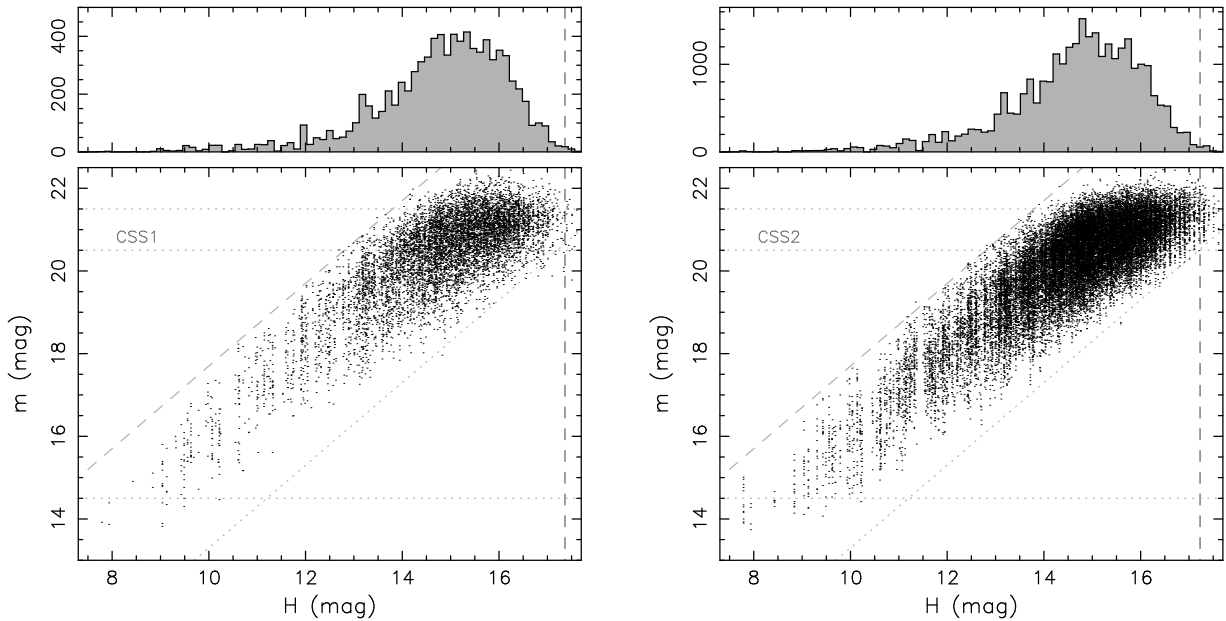}{f3b.eps}
 \caption{Upper panel: Histogram of the absolute magnitude $H$ values for detections of Hildas during the CSS phase~I (left panels) and II (right panels) operations (the ordinate is number of such detections within the bin). The vertical dashed line shows magnitude at the center of a bin in which number of detections dropped below $5$\% of the bin with maximum number of detections. This value is about $\simeq 17.2$ for both phases. Bottom panel: Correlation between the absolute magnitude $H$ at the abscissa and visual magnitude $m$ at the ordinate for Hilda detections during the CSS phase I and II operations. At an ideal opposition, when a high-eccentricity Hilda asteroid would be near perihelion, $m-H\simeq 3.3$ (neglecting the phase function correction) from simple distance arguments (slanted dotted line). In the opposite worst situation, when the same Hilda object would be near aphelion and high phase, $m-H\simeq 7.7$ (slanted dashed line). There are more than four magnitude difference between these cases.  The reason is mainly due to the large eccentricity of some Hilda orbits and the possibility to observe them at phase angles up to $\simeq 20^\circ$ \citep[both values are larger than for Jupiter Trojans, for which the apparent magnitude spread is only about 2 magnitudes, e.g., Fig.~4 in][]{vok2024}. The horizontal dotted lines between visual magnitudes 20.5 and 21.5 indicate typical 50\% photometric efficiency of detections. Some objects brighter than $m\simeq 14.5$ (lower dotted line) may saturate exposures and be undetected due to their slow motion.}
 \label{fig3}
\end{figure*}
%%%%%%%%%%%%%%%%%%%%%%%%%%%%%%%%%%%%%%%%%%%%%%%%%%%%%%%%%%%%%%%%%%%%%%%%%%%%%%%%%%%%%%%%%%%%%%%%%%%%%%%

\noindent{\it Hilda population known to-date.-- }We started by downloading the MPC orbit database ({\tt MPCORB.DAT}) as of Feb~14, 2024 (osculating orbital elements corresponding to the epoch MJD60200.0). Next, we proceed in several steps. We selected all bodies with semimajor axis in the range $3.7$~au and $4.15$~au. We dropped bodies on orbits determined from astrometric data that lasted less than a week that are quite uncertain, and obtained $6,466$ inputs. However, many of these bodies are still unrelated to long-term stable Hildas (residing, for instance, on high eccentricity orbits compatible with the Jupiter family comets or other unstable bodies that cross the orbits of Jupiter or the terrestrial planets). We numerically propagated the preselected orbits for $20$~Myr forward in time using the well-tested integration package {\tt swift}%
\footnote{\citet{swift1994} and \url{http://www.boulder.swri.edu/~hal/swift.html}.}.
We included the Sun and all planets, whose initial conditions were taken from the JPL ephemerides DE420.  We used a short integration step of $3$~days. The massless Hilda candidate particles were eliminated when they reached the heliocentric limits of $1$ au, $10$~au, or they hit Jupiter. This step trimmed $277$ bodies, leaving us with $6,189$ asteroids. We also monitored oscillations of the heliocentric semimajor axis $a$ during the integrations and further eliminated all bodies with a maximum $a$ value smaller than $3.95$~au. Bodies that survived in the integration and had $a$ smaller than this limit belong to a population of low eccentricity and low inclination orbits adjacent to the J3/2 resonance. They benefit from its protection against close encounters with Jupiter on a short- or medium-timescale, but are not truly members of the Hilda population residing in the stable zone of the J3/2 resonance. Another $95$ bodies were thus eliminated. This includes the well-known object (334)~Chicago \citep[see already][]{laves1904}, which has the absolute magnitude%
\footnote{We adopt absolute magnitude values provided by the MPC database.}
$H=7.69$ and was previously the largest object in the sample.  This leaves (153)~Hilda, with $H=7.78$, as the largest body.

This procedure left us with 6,094 individual Hildas. Traditionally, members of the Hilda population are distinct from other asteroids by a long-period and librating behavior of the resonant angle
$\sigma=3\lambda'-2\lambda-\varpi$, where $\lambda$ and $\varpi$ are longitudes in orbit and perihelion of the asteroid, respectively, while $\lambda'$ is the longitude in orbit of Jupiter. We checked the time series of $\sigma$ for the 6,094 bodies and found the expected behavior in the majority of the asteroids. Only $170$ of them in $\sigma$ exhibited slow circulation. Although not exactly ``resonant'', according to our definition, these objects are very close to resonant objects, so that we decided to include them in our work. As discussed in Sec.~\ref{concl}, they might have actually leaked from the resonant zone via Yarkovsky diffusion.
\smallskip

\noindent{\it Hilda asteroids detected by CSS.-- }Similar to our earlier study of Jupiter Trojans \citep{vok2024}, we used observations of the CSS $1.5$m survey telescope located at Mt.~Lemmon%
\footnote{\url{https://catalina.lpl.arizona.edu/}.}
\citep[MPC observatory code G96, e.g.,][]{CSSEPSC2019}. Both studies profit from an earlier effort of \citet{nes2023}, who carefully analysed the G96 detection performance in the period between January~2013 and June~2022. An important upgrade of the CCD camera mounted on the G96 telescope in May~2016 makes a logical division of the survey interval into two phases: (i) observations
before May~14, 2016 (phase I), and (ii) observations after May~31, 2016 (phase II). The second phase is therefore longer, allowing the survey to collect more observations (the available database contains 61,585 
well-characterized frames --sequences of four, typically $30$~s exposure images-- during phase~I and 162,280 well-characterized frames during phase II). Not less important is the fact that the new camera has a $4$~times larger field of view, allowing it to cover a much
larger latitude region about the ecliptic. The only caveat of the second-phase operation has to do with the pixel size of the new camera, which is twice as large as the older camera. This decreases the camera's ability to detect very slow moving objects \citep[plate motion of about $0.07^\circ$~day$^{-1}$ or less, see Fig.~1 of][]{vok2024}.
% FIG 1 %%%%%%%%%%%%%%%%%%%%%%%%%%%%%%%%%%%%%%%%%%%%%%%%%%%%%%%%%%%%%%%%%%%%%%%%%%%%%%%%%%%%%%%%%%%%%%%
\begin{figure}[t!]
 \begin{center}
  \includegraphics[width=0.47\textwidth]{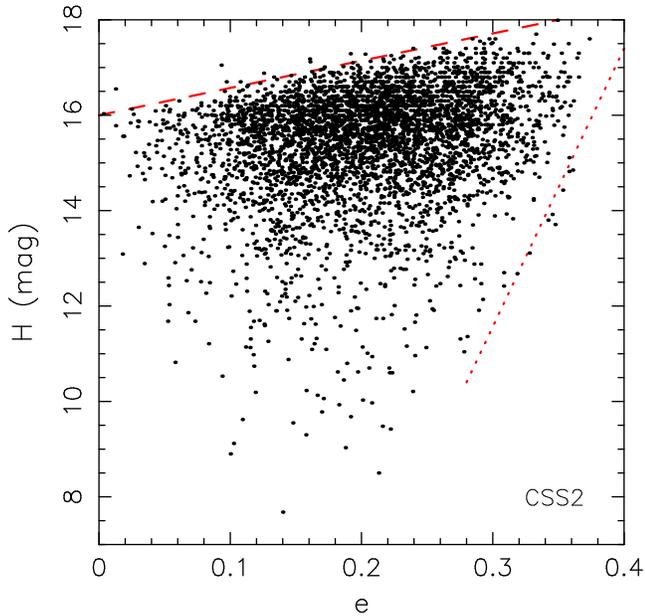}
 \end{center}
 \caption{Osculating orbital eccentricity $e$ (abscissa) vs absolute magnitude $H$ (ordinate) for 4,316 Hildas detected during the CSS~II period (a similar trend is seen also in the CSS~I detections). The slanted lines show two different type of correlations: (i) the dashed line expresses an observational bias allowing to detect fainter Hildas only when they are near perihelion of their orbit, while (ii) the dotted line indicates an intrinsic property of the Hildas having larger eccentricity orbits populated by fainter members. The latter is likely due to the Yarkovsky-driven diffusion of resonant population like Hildas toward larger eccentricities.}
 \label{fig4}
\end{figure}
%%%%%%%%%%%%%%%%%%%%%%%%%%%%%%%%%%%%%%%%%%%%%%%%%%%%%%%%%%%%%%%%%%%%%%%%%%%%%%%%%%%%%%%%%%%%%%%%%%%%%%%

Having previously selected the sample of known Hilda asteroids, we searched which of them were detected by the CSS G96 telescope during phase~I and II operations. We identified 8,686 detections of 2,970 individual Hildas during phase~I, and 28,930 detections of 4,317 individual Hildas during phase~II. Figure~\ref{fig1} shows the equatorial coordinates (right ascension and declination) of the detected objects (blue symbols) together with Jupiter's track along the ecliptic during the corresponding phase (red line). Apart from the avoidance of the galactic plane (especially during the autumn period when the very crowded fields toward the galactic center are on the night sky), there is much less bias in the visibility of the Hilda population compared to the Jupiter Trojans \citep[compare with Figs.~2 and 3 of][]{vok2024}. Due to the characteristic triangular pattern of their occurrence in the co-rotation frame with Jupiter, Hilda detections are spread along all ecliptic longitudes, with only a slight preference for $0^\circ$ or $\pm 120^\circ$ angular distance from Jupiter (when the Hildas are at perihelia). 

A notable feature is a substantial spread of the CCS phase II fields of view around the ecliptic, reaching large latitude values. Since the inclination distribution of Hildas does not extend to extremely high values, we do not expect much of the population to be lost at larger latitudes. The situation is slightly worse during phase I, which had most of the fields of view confined close to the ecliptic, implying that some Hildas on higher-inclination orbits might have been missed.
% FIG 1 %%%%%%%%%%%%%%%%%%%%%%%%%%%%%%%%%%%%%%%%%%%%%%%%%%%%%%%%%%%%%%%%%%%%%%%%%%%%%%%%%%%%%%%%%%%%%%%
\begin{figure*}[t!]
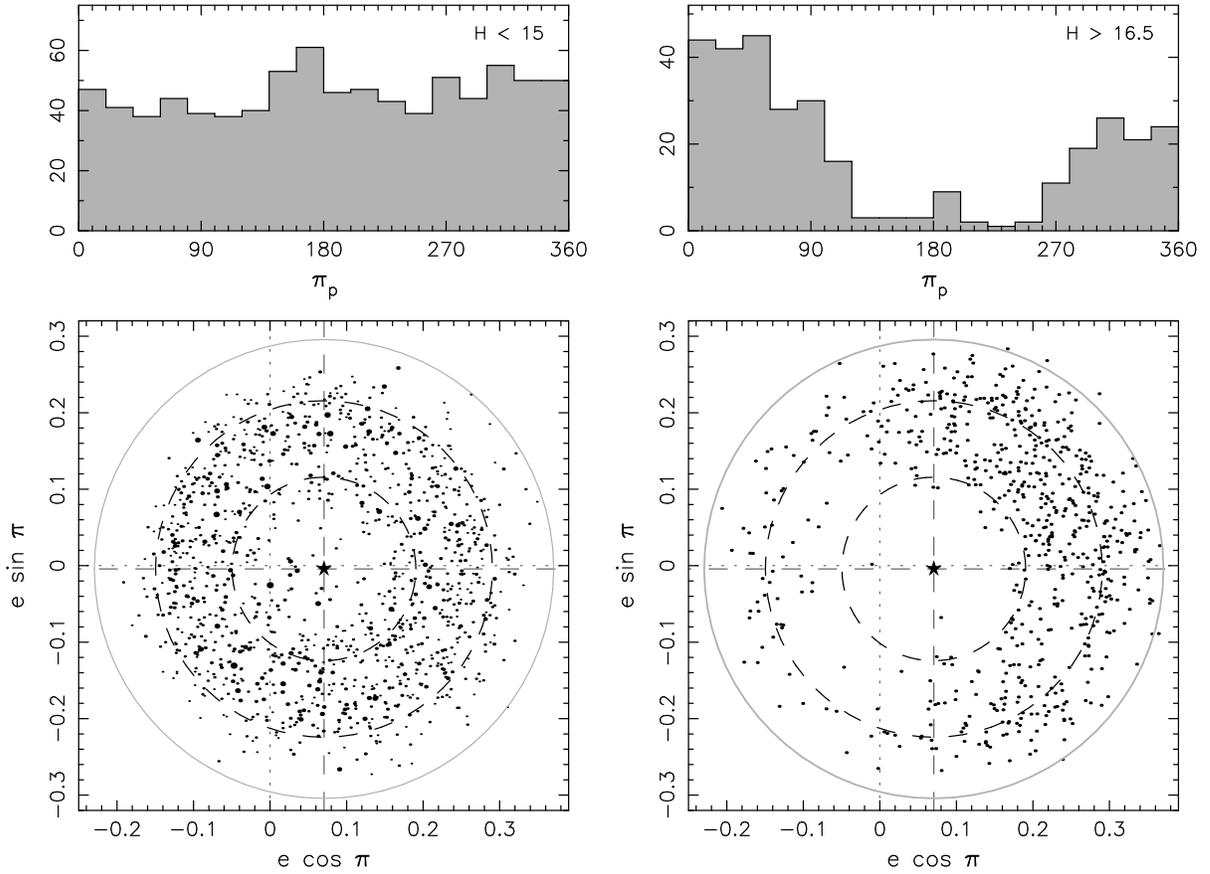

 \begin{center}
 \begin{tabular}{cc}
  \includegraphics[width=0.43\textwidth]{f5a.eps} &
  \includegraphics[width=0.43\textwidth]{f5b.eps} \\ 
 \end{tabular}
 \end{center}  
 \caption{Bottom panels: Detected Hilda asteroids during the phase~II of CSS operations projected onto the plane of Cartesian folding of eccentricity $e$ vs perihelion $\varpi$, namely $e\,\exp(\imath\,\varpi)$: (i) the dotted axes (and the numerical values at the abscissa and on the ordinate) are the osculating heliocentric values of $e$ and $\varpi$ as of MJD 60200.0 epoch, and (ii) the dashed axes (with the origin shifted to the forced planetary term) are the proper parameters $e_{\rm p}$ and $\varpi_{\rm p}$. Left panels for $H\leq 15$ (bright) Hildas (the size of the symbol indicating the size of the asteroid), right panels for $H\geq 16.5$ (faint) Hildas. The enveloping gray circle approximately delimits the observed population. The dashed circles with $e_{\rm p}=0.12$ and $e_{\rm p}=0.22$ specify the annulus, from which data shown at the   upper panels registered. These show distribution of number of detected Hildas in $\varpi_{\rm p}$ using $20^\circ$ bins. Bright Hildas (top and left) are consistent with uniform distribution within that area. Faint Hildas (top and right) have a non-uniform distribution with maximum at $\varpi_{\rm p}\simeq 60^\circ$ and minimum at $\varpi_{\rm p}\simeq 240^\circ$. This is due to the shift of the origin by the forced (Jupiter's) term $e_{\rm forced}\simeq 0.08$ along the $\varpi\simeq 0^\circ$ axis. Small (faint) Hildas can be detected only on large-eccentricity orbits with low-enough perihelion.}
 \label{fig5}
\end{figure*}
%%%%%%%%%%%%%%%%%%%%%%%%%%%%%%%%%%%%%%%%%%%%%%%%%%%%%%%%%%%%%%%%%%%%%%%%%%%%%%%%%%%%%%%%%%%%%%%%%%%%%%%

Figure~\ref{fig2} shows number of Hilda-population asteroid detections through phases~I and II. The annual pattern is punctuated by summer maintenance periods. The slight increase in detections towards the end of the interval has to do with Jupiter's position with respect to the galactic center.

Figure~\ref{fig3} shows a correlation between absolute magnitude $H$ (abscissa) and apparent magnitude $m$ (ordinate) for Hilda detections, which are divided between the two phases of CSS G96 operations. Three horizontal dotted lines are of particular interest: (i) $m\simeq 14.5$ is the saturation limit, such that brighter objects spread so much on the detector that they might not be resolved by the Hilda's slow apparent motion,%
\footnote{The median value of the apparent sky motion of detected Hildas was $\simeq 0.17^\circ$~day$^{-1}$, slightly larger than for Jupiter Trojans \citep[see Fig.~1 of][]{vok2024}. Still some Hildas, may happen to be undetected with motion $\leq 0.05^\circ-0.07^\circ$~day$^{-1}$. We take this bias into account in our work.}
and (ii) $m\simeq 20.5-21.5$ range shows the typical limiting detection magnitude (varying according to nightly conditions) that prevents fainter objects from being detected. The latter effect is obvious, but even the former effect has implications for Hilda detections. Note, for example, that the brightest object in the population --(153)~Hilda with $H\simeq 7.7$-- was detected in both phases, but during the phase~I operations only once. The slanted lines roughly delimit the range of the apparent magnitudes $m$ at which Hildas with the same $H$ were detected. This is principally due to the large eccentricity values of some Hilda orbits and the difference between detecting the body near perihelion and/or aphelion of its orbit. A smaller contribution to the spread comes from detections made at different phase angles.

Another selection effect is documented in Fig.~\ref{fig4}. We show the absolute magnitude $H$ of Hildas detected during the phase~II operations, given as a function of the osculating eccentricity of their heliocentric orbits. The slanted lines, delimiting the bulk of the data, indicate two trends. The upper dashed line tells us that the smallest Hildas are detected only near perihelion for those with the highest-eccentricity orbits. This effect is clearly an observational bias; there is no reason why objects with the same value of $H$ would not also populate lower-eccentricity orbits. The lower boundary of the distribution (dotted line) has a different nature. It tells us that high eccentricity orbits are preferentially populated by smaller Hildas. In all likelihood this is due to the Yarkovsky-driven flow of smaller Hildas towards larger eccentricities near the stability border (see also Sec.~\ref{concl}).

At the same time, it might seem peculiar that large members of the Hilda population ($H\leq 12$, say) do not reach stable orbits with larger eccentricities. If Hildas were captured during the planetary instability phase, the large objects should have been captured onto arbitrary orbits in the stable zone. We return to this issue in Sec.~\ref{concl}.  We will remap the population of large synthetic Hildas into the space of proper orbital elements.  From there, we will reconsider the phase-space zone that is long-term stable.  We believe the larger lost Hildas originally did populate the unstable spaces, thereby explaining this apparent puzzle.

Yet another look at the selective detection of small Hildas whose perihelion values come from highly eccentric orbits is shown in Fig.~\ref{fig5}. The bottom panels show heliocentric osculating values $e\,\exp(\imath\varpi)$ for detected asteroids during the CSS phase~II: (i) left panels for bright Hildas ($H<15$) and (ii) right panels for faint Hildas ($H>16.5$). The star is the forced Jupiter term $\kappa\,e'\, \exp(\imath\varpi')$, where $e'$ and $\varpi'$ are the osculating values of the Jupiter eccentricity and longitude of perihelion at the reference epoch MJD 60200.0 and $\kappa=1.47$ (Appendix~\ref{propar}). The axes are highlighted by dashed horizontal and vertical lines. The origin, shifted to Jupiter's forced term, defines the proper orbital parameters $e_{\rm p}\,\exp(\imath\varpi_{\rm p})$ introduced in Appendix~\ref{map1} (Eq.~\ref{eprop2}). The annulus, shown by the two dashed circles and delimited by $e_{\rm p}=0.12$ and $e_{\rm p}=0.22$ values, is the target zone of our test. 

The upper histograms show the number of Hildas detected within our annulus in $20^\circ$ bins of $\varpi_{\rm p}$. In the case of bright Hildas (left), the distribution is statistically compatible with uniform statistics. Conversely, the same distribution for faint Hildas (right), the distribution has a strong preference for $\varpi_{\rm p}$ values in between $270^\circ$ and $90^\circ$. As $\varpi'\simeq 0^\circ$, these values determine orbits with a large osculating eccentricity $e$ (compare with Appendix~B in \citet{vok2024}, where a similar effect has been discussed for Jupiter Trojans). If we aim to characterize the Hilda population all the way to small sizes, results from this experiment argue in favor of considering both $e_{\rm p}$ and $\varpi_{\rm p}$ as independent variables within the parametric space over which we will fit the observed population.
% FIG 1 %%%%%%%%%%%%%%%%%%%%%%%%%%%%%%%%%%%%%%%%%%%%%%%%%%%%%%%%%%%%%%%%%%%%%%%%%%%%%%%%%%%%%%%%%%%%%%%
\begin{figure*}[t!]
 \begin{center}
  \includegraphics[width=0.9\textwidth]{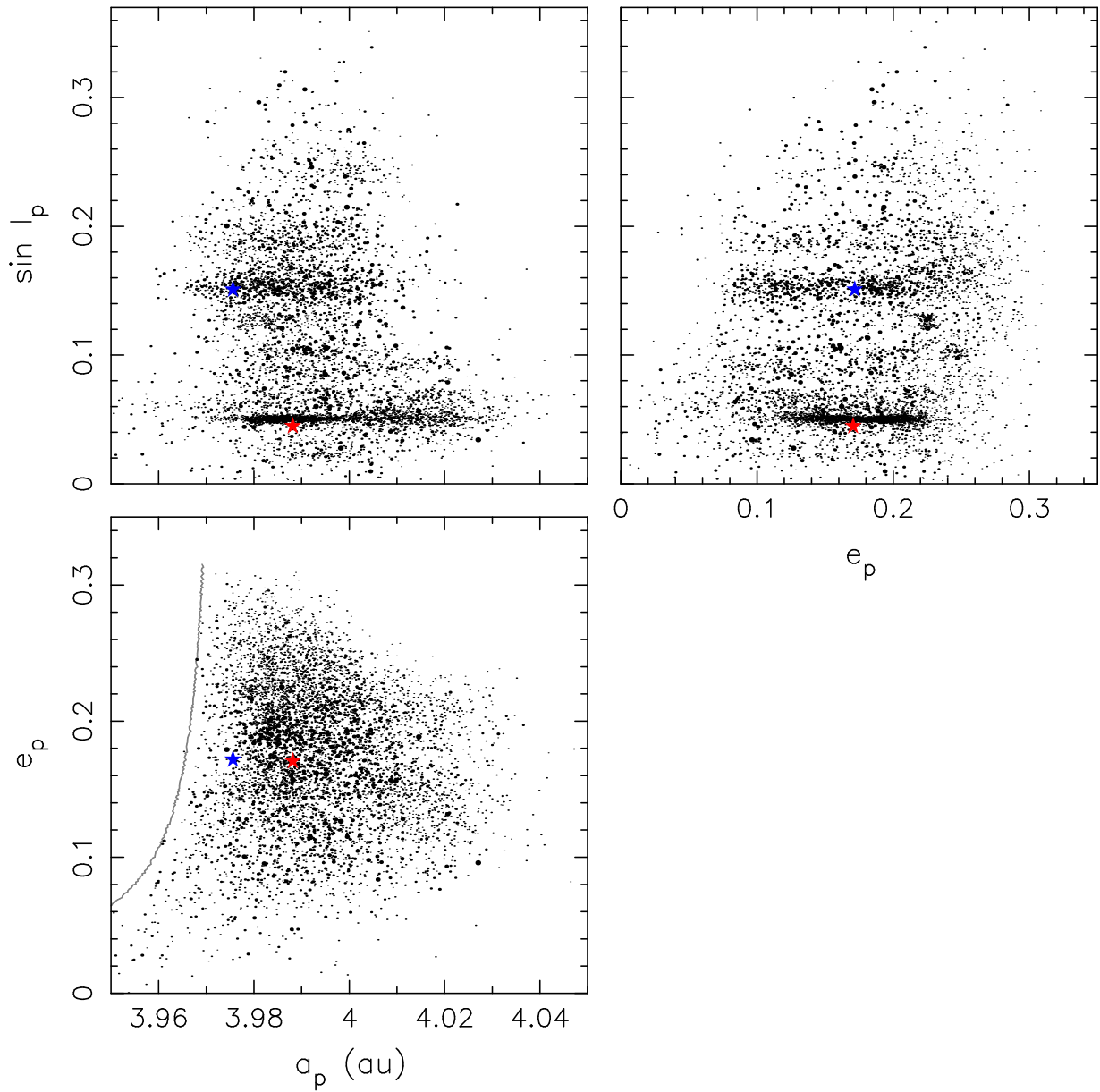} 
 \end{center}
 \caption{Population of 6,094 Hildas known as of February~2024
  projected onto 2D planes of proper parameters: (i) $(a_{\rm p},\sin I_{\rm p})$ top left, (ii) $(e_{\rm p},\sin I_{\rm p})$ top right, and (iii) $(a_{\rm p},e_{\rm p})$ bottom. Size of the symbol correlates with the size of the asteroid. The statistically significant clusters, prominently concentrated in $\sin I_{\rm p}$, are collisional families. The two largest are associated with (153)~Hilda (blue star) and (1911)~Schubart (red star). The gray line in the bottom panel shows location of the stationary point of the J3/2 mean motion resonance with Jupiter
  at $I_{\rm p}=0^\circ$ (see Appendix~\ref{cano}).}
 \label{fig6}
\end{figure*}
%%%%%%%%%%%%%%%%%%%%%%%%%%%%%%%%%%%%%%%%%%%%%%%%%%%%%%%%%%%%%%%%%%%%%%%%%%%%%%%%%%%%%%%%%%%%%%%%%%%%%%%
% FIG 1 %%%%%%%%%%%%%%%%%%%%%%%%%%%%%%%%%%%%%%%%%%%%%%%%%%%%%%%%%%%%%%%%%%%%%%%%%%%%%%%%%%%%%%%%%%%%%%%
\begin{figure*}[t!]
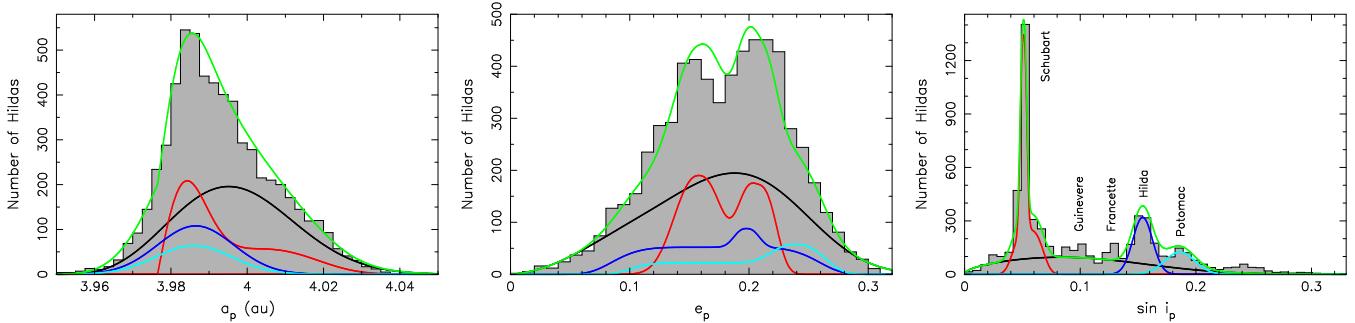

 \begin{center}
 \begin{tabular}{ccc}
  \hspace*{-8pt}
  \includegraphics[width=0.32\textwidth]{f7a.eps} &
  \hspace*{-5.5pt}
  \includegraphics[width=0.32\textwidth]{f7b.eps} &
  \hspace*{-5.5pt}
  \includegraphics[width=0.32\textwidth]{f7c.eps} \\
 \end{tabular}
 \end{center}
 \caption{Hilda population represented by the 1-D distributions of the proper parameters: (i) $a_{\rm p}$ (left), (ii) $e_{\rm p}$ (middle), and (iii) $\sin I_{\rm p}$ (right). Location of the principal families, most distinctly featured by peaks of the inclination distribution, is indicted by labels (rightmost panel). Data shown by the gray histogram with binsize given in Table~\ref{bias} and used in Secs.~\ref{method} and \ref{res} to determine properties of bias-corrected Hilda population. The color-coded solid lines illustrate distribution functions $D_i(p_i)$ for various sub-populations: (i) background population (black line), (ii) Schubart family (red line), (iii) Hilda family (blue line), and (iv) Potomac family (cyan line). The green line is their composite which, after adjustment of free coefficients in $D_i(p_i)$ should match the observed data. Since the purpose of this figure is to illustrate capability of the chosen $D_i(p_i)$ functions to match the data, their free coefficients were preliminarily guessed here. Their values after optimization of the model to the observations will be given in Sec.~\ref{res}.}
 \label{fig7}
\end{figure*}
%%%%%%%%%%%%%%%%%%%%%%%%%%%%%%%%%%%%%%%%%%%%%%%%%%%%%%%%%%%%%%%%%%%%%%%%%%%%%%%%%%%%%%%%%%%%%%%%%%%%%%%

% ----------------------------------------------------------------------------
% SEC ???
\section{Orbital and magnitude distribution of Hilda population}\label{model}
In this section, we present the model $\mathcal{M}$ for the Hilda population, whose free coefficients will be adjusted to fit the available observations. $\mathcal{M}$ is defined on the parametric space $(\mathbf{p};H)$, where (i) $\mathbf{p}$ specifies the orbital architecture of the population and (ii) the absolute magnitude $H$ parameterizes the magnitude distribution. To keep things simple, we assume that all parameters are uncorrelated, and the population is defined for each of them using a suitable 1-D distribution function. In general, for $n_{\rm par}$ of $\mathbf{p}$, we consider number of Hilda asteroids $dN_{\rm model}$ in the bin $(\mathbf{p}+d\mathbf{p};H+dH)$ given by
\begin{equation}
 dN_{\rm model}(\mathbf{p};H)= dN_{\rm model}(\mathbf{p})\,dN(H)\; , \label{modeltot}
\end{equation}
where
\begin{equation}
 dN_{\rm model}(\mathbf{p})=\prod_{i=1}^{n_{\rm par}} \,dN_i\left(p_i\right) \; , \label{model1}
\end{equation}
and
\begin{equation}
 dN_i\left(p_i\right) = D_i\left(p_i\right)\,dp_i \; . \label{model2}
\end{equation}
Here, $D_i\left(p_i\right)$ is the 1-D distribution function for parameter $p_i$ whose form will be chosen below. In the same way, differential magnitude distribution $dN(H)=D_H(H)\, dH$.

The subtlety of $\mathcal{M}$ consists of the fact that it is actually composed of several sub-populations, each obeying the properties outlined by Eqs.~(\ref{modeltot}) to (\ref{model2}). The entire Hilda population is eventually their linear superposition. In particular, we have in mind partitioning to: (i) the background population extending continuously over the net region of stability, and (ii) a set of collisional families localized in a certain orbital region (the complete list of identified families is given in Table~\ref{mb_fams_2023}, and the individual membership is provided through our webpage). The selection of the families used in the model will be specified below.

\subsection{Suitable orbital parameters}\label{orbdist}
The definition of the proper orbital parameters $\mathbf{p}$ for the Hilda population in the J3/2 resonance proves to be a more complicated task than what we did for Jupiter's Trojan population in the J1/1 resonance \citep[Sec.~3 in][]{vok2024}. For this reason, we address this problem at some length in Sec.~\ref{propar} of the Appendix, and we assume here that the reader is familiar with the method. The resulting set
$\mathbf{p}=\{p_i\}=(a_{\rm p},e_{\rm p},\sin I_{\rm p},\varpi_{\rm p})$ has $n_{\rm par}=4$, and consists of the semimajor axis $a_{\rm p}$, eccentricity $e_{\rm p}$, sine of inclinations $\sin I_{\rm p}$ and the proper longitude of perihelion $\varpi_{\rm p}$. The first three in the set are the traditional basis of proper parameters and have a close counterpart in the more accurate synthetic proper elements defined in Appendix~\ref{propel}. They are fundamental to define the orbital archirecture in the population. The last in the set, the longitude of perihelion $\varpi_{\rm p}$, plays a different role in our model. Data for bright Hildas shown on the left panel of Fig.~\ref{fig5} suggest that the intrinsic distribution does not depend on $\varpi_{\rm p}$. Therefore $\varpi_{\rm p}$ does not hold information about the intrinsic orbit distribution of Hildas. However, it is needed in the model to determine the bias-corrected population by including corresponding $\varpi_{\rm p}$-dependent detection probability of faint Hildas (right panel of Fig.~\ref{fig5}).

Figure~\ref{fig6} shows the population of 6,094 selected Hilda asteroids (Sec.~\ref{datag96}) projected onto 2D planes defined by combinations of $(a_{\rm p},e_{\rm p},\sin I_{\rm p})$. Lower-dimensional 1-D projections of the population, which characterize the distribution of the individual proper parameters, is then shown in Fig.~\ref{fig7}. The notable role of collisional families in the population can be observed mainly in the $\sin I_{\rm p}$ distribution. 

Consider, for example, that the population at the location of the Schubart family (the peak near $\sin I_{\rm p}\simeq 0.05$, right panel of Fig.~\ref{fig7}) contains nearly 1,400 members in a single bin in our representation (with only limited contamination by $\lesssim 100$ background asteroids). As a result, population-wise, the Schubart family may represent more than $21$\% of the net. Adding members from other clusters, almost $\simeq 60$\% of the observed Hilda population may belong to collisional families (see Table~\ref{mb_fams_2023}). Because their individual magnitude distributions $D_H(H)$ may differ from one another, and in particular could be substantially steeper than that of the background population \citep[as already noted, e.g., by][]{bv2008,wb2017}, the population located within families must be analyzed separately from the diffuse background.
\smallskip

\noindent{\it Background population.-- }In order to proceed further, we should make a choice of the proper-parameter distribution functions in (\ref{model2}). Here we seek inspiration from \citet{vok2024}, where quasi-Maxwellian distributions (or their combinations) were found suitable to represent the Jupiter Trojan population. Additionally, we have some concerns about possible problems in over-parametrization of the task.  For this reason, we treat the orbital parameter distribution functions $D_i(p_i)$ for the background population and for the individual collisional families in different ways. In the former case, we consider the coefficients of $D_i(p_i)$ to be solved-for parameters in the observational data fitting. Conversely, $D_i(p_i)$ for the families will be taken as fixed templates estimated in an educated way from the observed (biased) population without further adjusting their coefficients.

In the case of $\{p_i\}=(a_{\rm p},e_{\rm p},\sin I_{\rm p})$ proper parameters of the background population we assume 
\begin{equation}
 D_i(p_i) = (p_i-c_i)^{\alpha}\,\exp\left[-\frac{1}{2}\left(\frac{p_i-\overline{c}_i}{\sigma_i}\right)^{\beta}\right]\; , \label{distp}
\end{equation}
with the free coefficients $(\alpha,\beta,\sigma_i)$ and fixed constants $(c_i,\overline{c}_i)$. For the latter, we chose: (i) $c_1=\overline{c}_1=3.95$~au for $a_{\rm p}$, (ii) $c_2=0$ and $\overline{c}_2=0.12$ for $e_{\rm p}$, and (iii) $c_3=\overline{c}_3=0$ for $\sin I_{\rm p}$.

As for the proper longitude of perihelion $\varpi_{\rm p}$, we argued above that none orbital preference in the intrinsic Hilda population is expected. Therefore we consider $D_4(\varpi_{\rm p})$ to be a uniform function.
\smallskip
% FIG 1 %%%%%%%%%%%%%%%%%%%%%%%%%%%%%%%%%%%%%%%%%%%%%%%%%%%%%%%%%%%%%%%%%%%%%%%%%%%%%%%%%%%%%%%%%%%%%%%
\begin{figure*}[t!]
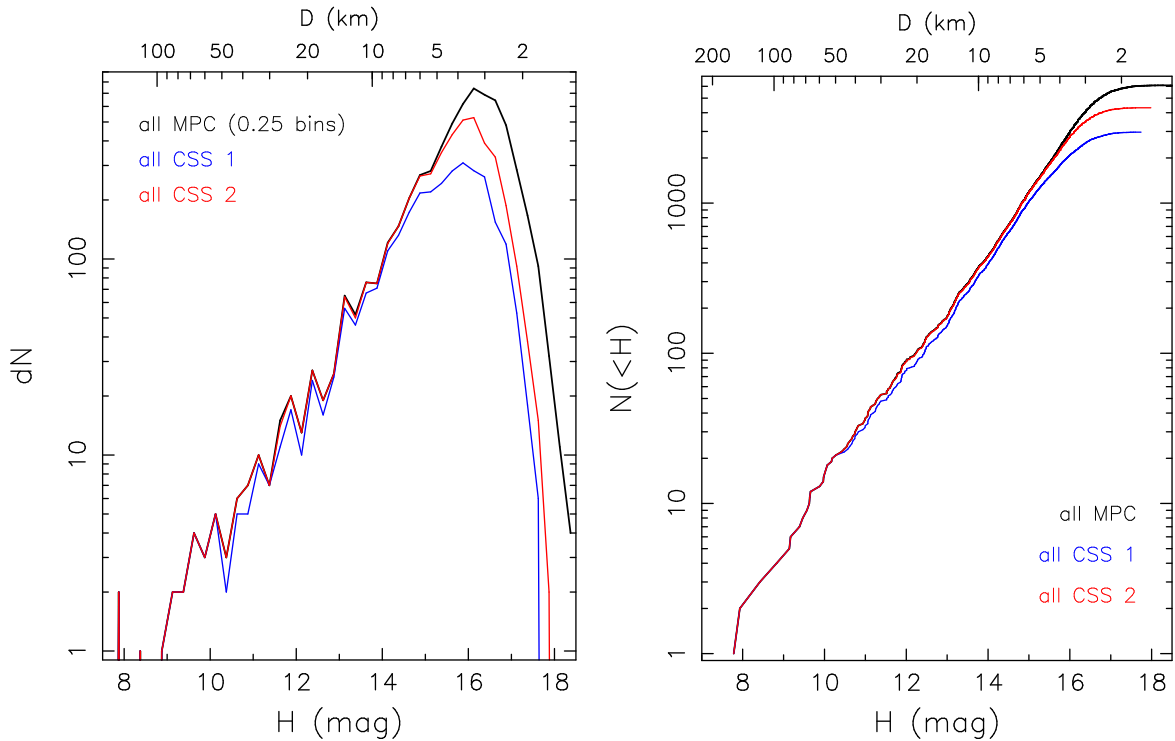

 \begin{center}
 \begin{tabular}{cc}
  \includegraphics[width=0.42\textwidth]{f8a.eps} &
  \includegraphics[width=0.42\textwidth]{f8b.eps} \\
 \end{tabular}
 \end{center}
 \caption{Differential magnitude distribution (left panel; bin width $0.25$ mag) and cumulative
  distribution of the selected Hilda population of 6,094 asteroids known as of Jan~2024
  (Sec.~\ref{datag96}). Black line is the total population downloaded from the MPC database,
  blue are detections by CSS during the phase~I, red are detections by CSS during the phase~II
  operations. For sake of reference the upper abscissa translates the absolute magnitude value
  to size using a constant geometric albedo $p_V=0.055$, mean value in the Hilda population
  reported by \citet{grav2012}. This information is relevant to understand what part of the
  population is being tested in the long-term simulations in Sec.~\ref{stabi}.}
 \label{fig8}
\end{figure*}
%%%%%%%%%%%%%%%%%%%%%%%%%%%%%%%%%%%%%%%%%%%%%%%%%%%%%%%%%%%%%%%%%%%%%%%%%%%%%%%%%%%%%%%%%%%%%%%%%%%%%%%

\noindent{\it Population in families.-- }Collisional families were primarily identified and analyzed in the space of synthetic proper element space (Appendix~\ref{propel}). Their membership has been mapped to the space of proper parameters. Using these data, we constructed 1-D distributions $D^{\rm fam}_i(p_i)$ for $(a_{\rm p},e_{\rm p},\sin I_{\rm p})$. Finally, $D^{\rm fam}_i(p_i)$ were represented with suitable smooth functions, which will be used in Sec.~\ref{res} for bias-corrected population analysis. In the case of the families, $D^{\rm fam}_i(p_i)$ remain fixed, and optimization refers only to the magnitude distribution $D^{\rm fam}_H(H)$.

Color-coded lines in Fig.~\ref{fig7} show $D^{\rm fam}_i(p_i)$ for the three major families in the Hildas, named Schubart, Hilda and Potomac. Specific features are as follows:
\begin{itemize}
\item The semimajor axis $a_{\rm p}$ distributions for Hilda and Potomac may be well approximated with a simple Gaussian. However, the Schubart family indicates an interesting asymmetric feature whose origin may be twofold.  First, the Schubart family resides close to the exact resonance (stationary) solution.  This suggests that the initial fragments could have been launched, on both sides from this location in osculating semimajor axis, while the proper parameter $a_{\rm p}$ only maps it on one side \citep[see also Figs.~16-19 of][for a more detailed analysis]{bv2008}.  Second, results from our numerical simulation shown on Fig.~\ref{figyy} indicate that fragments initially launched to orbits with large $a_{\rm p}$ value are long-term unstable and might have been destabilized over the $\simeq 1.5-1.7$~Gyr age of this family \citep[e.g.,][]{bv2008,pisa2017};
\item The eccentricity $e_{\rm p}$ distributions of all of the presented families show an interesting structure, namely (i) the distribution of the Schubart family has two maxima, at all likelihood created by diffusion of smaller fragments due to the Yarkovsky effect%
\footnote{Indeed, the double-peak profile disappears when limiting to the large bodies in the Schubart family, say $H<14$ (see Fig.~\ref{fig1911} for more details)}
\citep[being reminiscent of the similar effect discovered for moderately old asteroid families in the main belt, e.g.,][but in the resonant configuration maps on the eccentricity, rather than semimajor axis; \citet{bv2008}]{vok2006}, (ii) the distribution of Hilda family is well represented using a broad Gaussian, but near $e_{\rm p}\simeq 0.2$ it is punctuated by a peculiar concentration of members (possibly an artifact related to simple proper parameters or a young subcluster in the family), and (iii) the distribution of the Potomac family is deficient in fragments at small $e_{\rm p}$ values, likely caused by the long-term instability shown in Fig.~\ref{figyy};
\item The inclination $\sin I_{\rm p}$ distribution is well represented using simple Gaussians except for the Schubart family, where a "leg" toward slightly larger values than the median is seen.  This feature is likely produced by a slow chaotic diffusion of Schubart members at large $a_{\rm p}$ values.
\end{itemize}
We did not include contributions from the small clusters Francette and Guinevere, which have about 150 (or less) known fragments (see Table~\ref{mb_fams_2023}).  The reason is that they would not be reliably represented in our solution.

\subsection{Magnitude distribution}\label{magdist}
Figure~\ref{fig8} shows the differential (left panel) and cumulative (right panel) distribution of the absolute magnitude values for the 6,094 selected Hilda members (black line), and those detected during the phase~I (blue line) and phase~II (red line) operations of CSS. While the effect of signal saturation for the brightest objects in the population (Fig.~\ref{fig3}) threatened their detection probability, all were resolved during both phases. Smaller Hildas were generally detected up to magnitude $\simeq 16$ during phase~II, but a small fraction was missed independently of the magnitude value during phase~I. This difference is clearly due to the CSS's fields-of-view confined to small ecliptic latitudes, such that objects on higher-inclination orbits might have simply escaped the CSS viewing geometry.

The simplest model, adopted for instance by \citet{wb2017}, is to use a single power-law
\begin{equation}
 D_H(H) = N\, 10^{\gamma\,(H-H_0)}\; ,\label{dh1}
\end{equation}
containing two adjustable parameters: (i) population normalization $N$ at reference magnitude $H_0$, which we always fix to magnitude $16$ (one can also set $N=1$ and adjust $H_0$ as a free parameter), and (ii) the slope $\gamma$. For the sake of making a comparison, we perform an initial simulation with the power-law model above.  With some of our data extending to magnitude $H\simeq 17$ (Fig.~\ref{fig8}), our follow-up approach can be more ambitious and potentially more accurate at the same time.

Therefore, as in \citet{vok2024}, we adopt a more flexible approach originally introduced to study the magnitude distribution of the near-Earth asteroid population by \citet{neomod1}. In particular, we represent the cumulative magnitude distribution ${\rm log}_{10} N(<H)$ using cubic splines in the interval $7.5<H<18$. This range is constrained by available data: (i) there are no Hildas in our sample with $H<7.5$ (the largest being (153)~Hilda with $H\simeq 7.78$), and (ii) there are too few G96 detections beyond the magnitude $\simeq 17.75$ (the faintest being 2021~QF34 with magnitude $\simeq 18$). Nevertheless, we expect reliable results to be obtained all the way down to magnitude $\simeq 17$. 

In practice, we divide this interval into a certain number of segments. In each, we consider a mean slope parameter to be a free parameter (for $n$ segments this implies $n$ independent slopes to be fitted). Furthermore, we add the $(n+1)$-th parameter representing the absolute normalization $N_{\rm back}(<H_0)$ of the population at a certain reference magnitude $H_0$; as mentioned above, we
use $H_0=16$. There are no additional parameters to be fitted since the solution for
${\rm log}_{10} N(<H)$ is then fixed by enforcing the continuity (including the first derivative) at the segment boundaries. 

After a brief tests of the model, we set up $5$ segments, which implies $6$ free parameters. Comparing the degrees-of-freedom in our approach to the traditional method (\ref{dh1}), we find that our work does not over-parameterize the problem, yet it does gain a significant benefit in generality.

We adopt the same approach to describe the magnitude distribution of the background population and collisional families. In the latter case, however, we use fewer magnitude segments, typically $2-4$ (depending on the size of the largest body in the family).

% -------------------------------------------------------------
% SEC ???
\section{Adjustment of the model parameters via calibration on the CSS observations} \label{method}
The remaining part of this paper concerns the parametric space $(\mathbf{p};H)$. The model $\mathcal{M}$ and the available observations have been defined or mapped to $(\mathbf{p};H)$ as discussed in the previous section. However, to connect them, we need one more piece of information, namely a bridge in the form of the detection probability ${\cal P}(\mathbf{p};H)$ of the survey used. Once known, we can map the ideal-world model prediction $dN_{\rm model}(\mathbf{p};H)$ in $(\mathbf{p}+d\mathbf{p};H+dH)$ cell to the imperfect-world reality $dN_{\rm pred}(\mathbf{p};H)$ represented by the observations. 

The essence of the imperfection is basically two-fold.  First, survey operations might have been insufficient to observe objects throughout the parametric space.  The missed portion must be quantified. Second, the instrument itself exhibits a plethora of imperfections.  For example, it has an apparent magnitude limit beyond which fainter objects are not detected.  Another example is a limit below which brighter objects are not detected as moving bodies that can be linked to the Hilda population.  Both of these difficulties also depend on the apparent motion of the image on the detector. 

Fortunately, these two issues have been carefully evaluated for CCS operations between 2013 and 2022 by \citet{nes2023} and successfully used in the sister study of Jupiter Trojans by \citet{vok2024}. For that reason, we refer to those papers for further details. Here we only briefly outline how they help us construct ${\cal P}(\mathbf{p};H)$; this is presented in Sec.~\ref{det_prob}. 

The predicted number of observed objects in the $(\mathbf{p}+d\mathbf{p}; H+dH)$ cell is
\begin{equation}
 dN_{\rm pred}(\mathbf{p};H) = {\cal P}(\mathbf{p};H)\;dN_{\rm model}(\mathbf{p};H) \label{biased}
\end{equation}
This value may be directly compared to the real number of observations. In the process of this comparison, we seek to adjust free coefficients of the model to achieve the best match. Our numerical method and tools for this optimization are presented in Sec.~\ref{opti}.
% Tab 2 %%%%%%%%%%%%%%%%%%%%%%%%%%%%%%%%%%%%%%%%%%%%%%%%%%%%%%%%%%%%%%%%%%%%%%%%%%%%%%%%%%%
\begin{deluxetable}{ccccc}[t]
\tablecaption{\label{bins}
 Orbital parameters and absolute magnitude space considered in our model}
\tablehead{
 \colhead{parameter} & \colhead{min} & \colhead{max} & \colhead{bin width} &
 \colhead{number of bins} }
\startdata
 $a_{\rm p}$ (au) & $3.95$ & $4.05$ & $0.0025$   &  40 \\
 $e_{\rm p}$      & $0$    & $0.32$ & $0.01$     &  32 \\
 $\sin I_{\rm p}$ & $0$    & $0.35$ & $0.007$    &  50 \\
 $\varpi_{\rm p}$ & $0^\circ$ & $360^\circ$ & $30^\circ$ & 12 \\
 $H$              & $7.5$  & $18$   & $0.10$     & 105 \\ 
\enddata
\tablecomments{The second and third columns specify minimum and maximum values of the parameter, the fourth column is the width of the bin, and the fifth column gives number of bins.}
\end{deluxetable}
%%%%%%%%%%%%%%%%%%%%%%%%%%%%%%%%%%%%%%%%%%%%%%%%%%%%%%%%%%%%%%%%%%%%%%%%%%%%%%%%%%%%%%%%%%%

\subsection{Detection probability}\label{det_prob}
With the approach described above, we now describe our procedure to determine ${\cal P}(\mathbf{p};H)$ as a sequence of steps.

In the first step, we consider the orbital part of the parametric space $\mathbf{p}$ divided into small bins $(\mathbf{p},\mathbf{p}+d\mathbf{p})$ as specified in Table~\ref{bins}. In each of them, we generate $N_{\rm orb}=2,000$ orbits that have their proper parameters uniformly distributed in the bin. Next, we use the algorithm described in the Appendix~\ref{map2} to assign heliocentric orbital elements to each of these synthetic orbits. The epoch of osculation is MJD 60,200.0, or Sep~13, 2023, close to the epoch of data in the MPC catalog used in Sec.~\ref{datag96} to identify the currently known population of Hildas. Each of these orbits is numerically propagated backwards in time for slightly more than a decade, namely to Jan~1, 2013, which predates the first available frame of the phase~I observations of CSS. 

In the second step, we consider information about CSS operations during phases~I and II (Sec.~\ref{datag96}). This includes a full list of frames with a specific field-of-view obtained at certain times: there are (i) $N_{\rm FoV}=61,585$ such frames in the phase~I, and (ii) $N_{\rm FoV}=162,280$ such frames in the phase~II operations of CSS (together making $223,865$ frames available). Next, we analyze the orbital history of the synthetic trajectories propagated from the orbital bins in $\mathbf{p}$ and consider whether they geometrically appear in the field-of-view of any of the CSS frames. To do this, we use the publicly available {\tt objectsInField} code (oIF) from the Asteroid Survey Simulator package \citep{naidu2017}. For frames not crossed by the particular orbit, we assign a detection probability $\epsilon=0$ of that orbit.

In the third step, we consider the opposite situation, namely when a given orbit appears to be in the field-of-view of a certain frame, we need to evaluate the true detection probability. We turn to the work of \citet{nes2023}, who conducted a detailed analysis of moving objects identified in each of these frames and determined their detection probability $\epsilon(m,w)$ as a function of the apparent magnitude $m$ and of the rate of motion $w$. The only modification of this initial analysis concerns dependence on $w$. While \citet{nes2023} used the CSS observations to characterize the near-Earth asteroid population, large values of $w$ presented a problem for a detection.  In the case of distant populations like Jupiter Trojans and Hildas, small values of $w$ lead to difficulties. This issue was analyzed by \citet{vok2024} (e.g., their Fig.~1) and we adopt their method for analyzing the Hildas.  Fortunately, this problem is not severe, as Hildas typically move slightly faster than the Trojans.

At this point in our analysis, we also add the magnitude component of parametric space (the previous geometric considerations do not depend on $H$). We test a sufficient range of absolute magnitudes $H$ between $7.5$ and $18$ using $0.1$ magnitude bins (Table~\ref{bins}) and compute the apparent magnitude $m$ on the frame using Pogson's relation:
 \begin{equation}
  m=H+5\log(R\Delta)-P(\alpha)\; . \label{pogson}
\end{equation}
 As expected, $m$ is directly related to the absolute magnitude $H$, which satisfies the dependence of ${\cal P}$ on $H$, along with several correction factors. The second term on the right-hand side of (\ref{pogson})  follows from the flux decreasing with heliocentric distance $R$ and observer distance $\Delta$ (these values are known from the simulation and epoch of the observation). 
 
 The last term is the phase function, which depends on the phase angle $\alpha$ (i.e., Sun-asteroid-observer). We adopt a simple $H-G$ magnitude phase function system and the slope parameter $G=0.15$ \citep[see, e.g.,][]{bow1989}. This is mainly because the absolute magnitude values $H$ used here are taken from the MPC database that uses this setup. Alternatively, we also tested a linear phase function $P(\alpha)\propto \alpha$ with $\simeq 0.05$~magnitude per degree slope.  This approximation matches the few detailed observations available for Hilda members from the Kharkiv group \citep[e.g.,][]{phase3,phase1,phase2}.  
 
Modeling additional uncertainty factors in $m$, such as the role of the rotation phase in the detection of small, irregularly shaped asteroids, represents a difficult problem, and we do not include it in our analysis \citep[see also discussion on page~16 in][]{vok2024}.  

As a result of our three steps, we have computed the corresponding detection probability $\epsilon_{j,k}$ for (i) each of the synthetic orbits $j=1,\ldots,N_{\rm orb}$, (ii) each of the frames $k=1,\ldots,N_{\rm FoV}$ and (iii) all the $H$ magnitudes within the needed range. Finally, the representative detection probability assigned to the $(\mathbf{p}+d\mathbf{p};H+dH)$ bin over the whole duration of the survey is expressed as
\begin{equation}
 {\cal P}\left(\mathbf{p};H\right) = \frac{1}{N_{\rm orb}}\sum_{j=1}^{N_{\rm orb}}\left\{1-\prod_{k=1}^{N_{\rm FoV}}
  \left[1-\epsilon_{j,k}\right]\right\}\; . \label{bias}
\end{equation}
Note that ${\cal P}$ is actually evaluated as a complementary value to the nondetection of the body, which on each frame reads $1-\epsilon_{j,k}$. Similarly, we define the representative rate of detection
\begin{equation}
 {\cal R}\left(\mathbf{p};H\right) = \frac{1}{N_{\rm orb}}\sum_{j=1}^{N_{\rm orb}} \sum_{k=1}^{N_{\rm FoV}}\epsilon_{j,k}\; . \label{rate}
\end{equation}
providing the mean number of frames in which the synthetic Hilda asteroid should be detected by the survey.
% FIG 1 %%%%%%%%%%%%%%%%%%%%%%%%%%%%%%%%%%%%%%%%%%%%%%%%%%%%%%%%%%%%%%%%%%%%%%%%%%%%%%%%%%%%%%%%%%%%%%%
\begin{figure*}[t!]
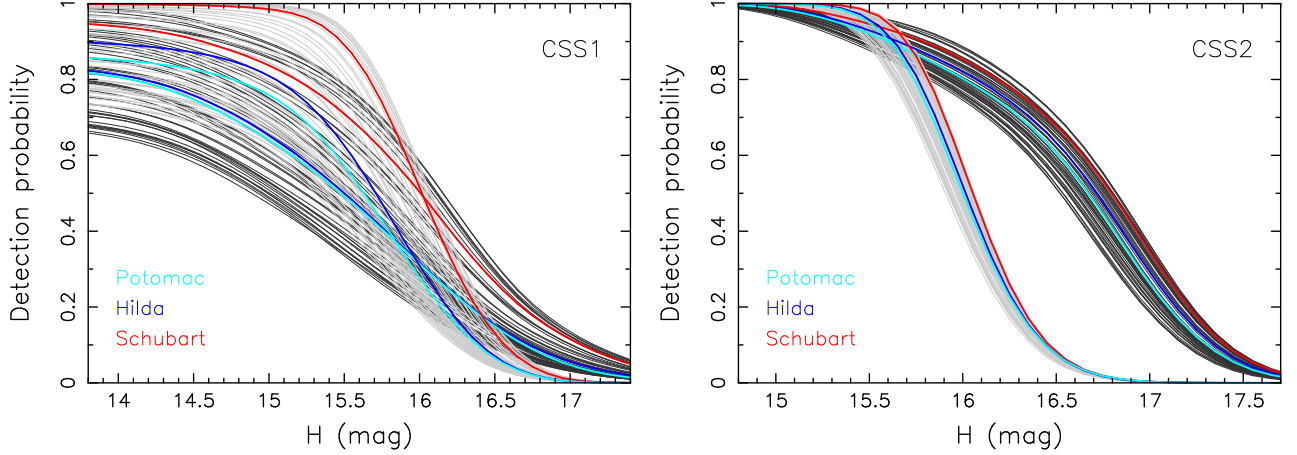

 \begin{center}
 \begin{tabular}{cc}
  \includegraphics[width=0.46\textwidth]{f9a.eps} &
  \includegraphics[width=0.46\textwidth]{f9b.eps} \\
 \end{tabular}
 \end{center}
 \caption{Detection probability ${\cal P}\left(\mathbf{p};H\right)$  for Hilda asteroid having orbital parameters $a_{\rm p}=3.99$~au, $e_{\rm p}=0.155$, all sampled values of $\sin I_{\rm p}$ (smaller values larger detection probability for a given $H$), and two values of $\varpi_{\rm p}=45^\circ$ (dark gray curves) and $225^\circ$ (light gray curves) as a function of the absolute magnitude $H$. Left panel for phase~I, right panel for phase~II of CSS operations. The color-coded curves are for the approximate inclinations of the major families: (i) red for Schubart, (ii) blue for Hilda, and (iii) cyan for Potomac families. The detection probability for $\varpi_{\rm p}=225^\circ$ are shifted by up to $\simeq 0.5$~magnitude in $H$ compared to $\varpi_{\rm p}=45^\circ$ values for faint objects in the phase~II (see also Fig.~\ref{fig5} are related discussion). The spread with orbital inclination is small for the phase~II data thanks to field-of-view sampling of high ecliptic latitudes (Fig.~\ref{fig1}). The phase~I operations concentrated more closely to the ecliptic zone, thus the detection probability falls more rapidly with increasing inclination value. Additionally, even the lowest-inclination probability of phase~I is somewhat smaller and the $\varpi_{\rm p}$ trend less evident due to its considerably shorter timespan.}
 \label{fig9}
\end{figure*}
%%%%%%%%%%%%%%%%%%%%%%%%%%%%%%%%%%%%%%%%%%%%%%%%%%%%%%%%%%%%%%%%%%%%%%%%%%%%%%%%%%%%%%%%%%%%%%%%%%%%%%%
% FIG 1 %%%%%%%%%%%%%%%%%%%%%%%%%%%%%%%%%%%%%%%%%%%%%%%%%%%%%%%%%%%%%%%%%%%%%%%%%%%%%%%%%%%%%%%%%%%%%%%
\begin{figure*}[t!]
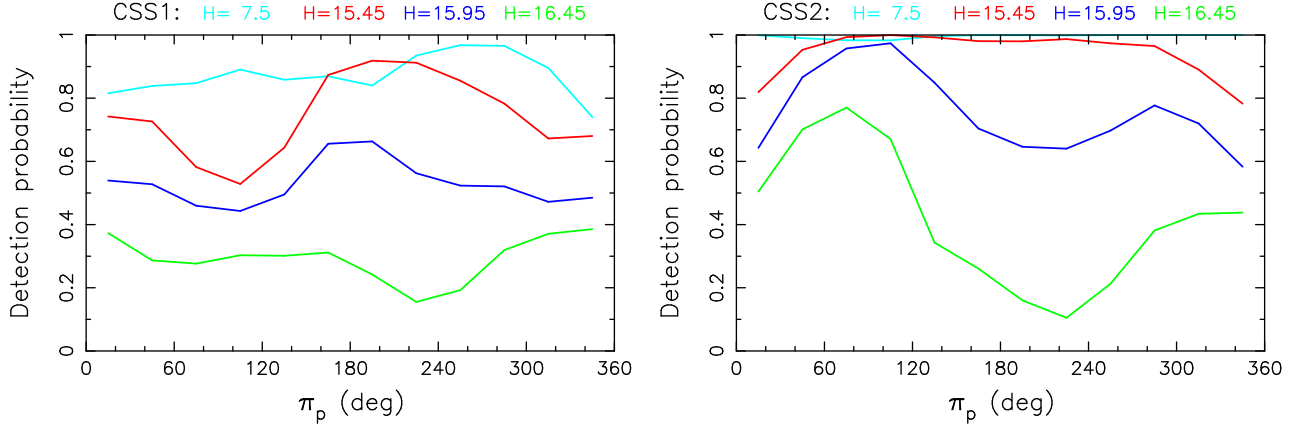

 \begin{center}
 \begin{tabular}{cc}
  \includegraphics[width=0.46\textwidth]{f10a.eps} &
  \includegraphics[width=0.46\textwidth]{f10b.eps} \\
 \end{tabular}
 \end{center}
 \caption{Detection probability ${\cal P}\left(\mathbf{p};H\right)$ for Hilda asteroid having orbital elements $a_{\rm p}=3.99$~au, $e_{\rm p}=0.155$, $\sin I_{\rm p}=0.05$ as a function of $\varpi_{\rm p}$ for four values of the absolute magnitude $H$: (i) $7.5$ (cyan curve), (ii) $15.45$ (red curve), (iii) $15.95$ (blue curve), and (iv) $16.45$ (green curve). In the CSS phase~II case (right panel), ${\cal P}\left(\mathbf{p};H\right)$ for the groups (i) and even (ii) ---brightest from the faint group-- does not depend on $\varpi_{\rm p}$ (or only weakly). Conversely, ${\cal P}\left(\mathbf{p};H\right)$ for fainter groups (iii) and (iv) becomes strongly dependent on $\varpi_{\rm p}$. Statistics of the observed Hildas (Fig.~\ref{fig5}) match the predicted profile with maximum at $\varpi_{\rm p}\simeq 45^\circ$ and minimum at $\varpi_{\rm p}\simeq 225^\circ$. The trends in the CSS phase~I are affected by field-of-view incomplete coverage of sky and its short duration.}
 \label{fig10}
\end{figure*}
%%%%%%%%%%%%%%%%%%%%%%%%%%%%%%%%%%%%%%%%%%%%%%%%%%%%%%%%%%%%%%%%%%%%%%%%%%%%%%%%%%%%%%%%%%%%%%%%%%%%%%%

\subsection{Optimization of model parameters}\label{opti}
Consider now a certain bin $(\mathbf{p}+d\mathbf{p};H+dH)$ indexed $j$ in which CSS detected $n_j (\geq 0)$ individual Hildas and our model $\mathcal{M}$ predicted $\lambda_j=dN_{\rm pred}(\mathbf{p};H)$ such objects (see Eq.~\ref{biased}). Assuming that the conditions of Poisson statistics are satisfied, drawing $n_j$ objects out of $\lambda_j$ expected obeys a probability distribution
\begin{equation}
 p_j(n_j) = \frac{\lambda_j^{n_j}\exp(-\lambda_j)}{n_j!}\; . \label{prob0}
\end{equation}
Combining the information from all bins in orbital-magnitude parameter space, and assuming no correlations among the bins, the joint probability of the model prediction versus data reads
\begin{equation}
 P = \prod_{j} \frac{\lambda_j^{n_j}\exp(-\lambda_j)}{n_j!}\; . \label{prob1}
\end{equation}
Finally, the target function (log-likelihood) is defined as
\begin{equation}
 {\cal L} = \ln P = -\sum_j \lambda_j + \sum_j n_j \ln \lambda_j\; , \label{prob2}
\end{equation}
where a constant term $-\sum_j \ln(n_j!)$ has been dropped representing just an absolute normalization of ${\cal L}$ that cannot affect the optimization of the model coefficients.  For reference, if the results of two independent surveys are to be combined, such as Hilda observations from the phase~I and phase~II operations of CSS, we simply sum their respective ${\cal L}$ values. The optimization procedure aims to maximize ${\cal L}$ in the space of free coefficients of $\mathcal{M}$. 

We use a powerful and well-tested {\tt MultiNest}%
\footnote{\url{https://github.com/JohannesBuchner/MultiNest}}
code to perform the optimization procedure, namely parameter estimation and error analysis \citep[e.g.,][]{feroz2008,feroz2009}. The popularity and power of {\tt MultiNest} stems from its ability to deal efficiently with complex parameter space that can be endowed with degeneracies in high dimensions. For the sake of brevity, we refer to the literature quoted above to learn more about this versatile package.
% FIG 1 %%%%%%%%%%%%%%%%%%%%%%%%%%%%%%%%%%%%%%%%%%%%%%%%%%%%%%%%%%%%%%%%%%%%%%%%%%%%%%%%%%%%%%%%%%%%%%%
\begin{figure*}[t!]
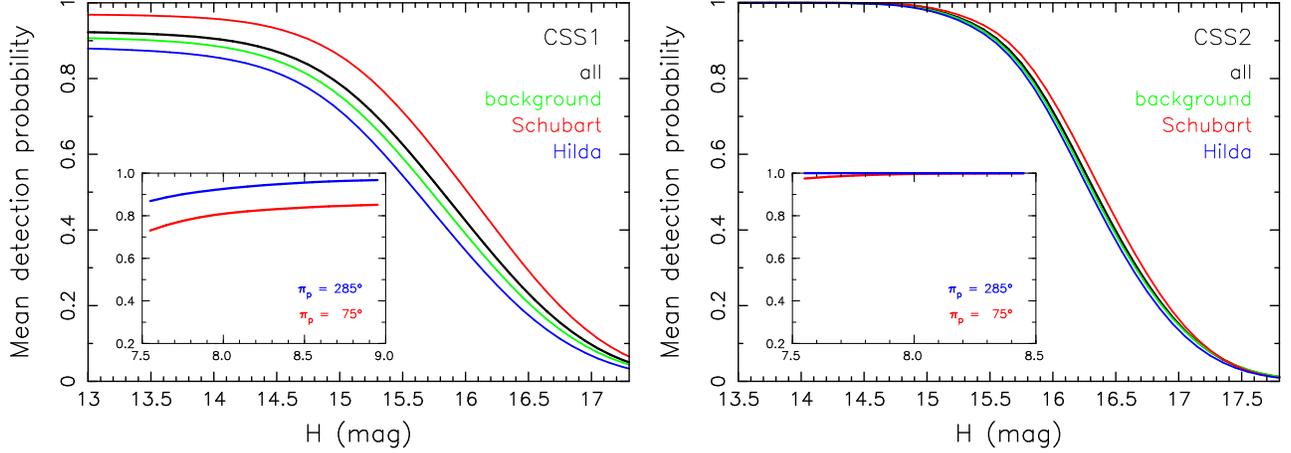

 \begin{center}
 \begin{tabular}{cc}
  \includegraphics[width=0.46\textwidth]{f11a.eps} &
  \includegraphics[width=0.46\textwidth]{f11b.eps} \\
 \end{tabular}
 \end{center}
 \caption{Population-wise weighted detection probability  $\overline{\mathcal{P}}(H)$ computed using Eq.~(\ref{ave1}) (left panel for phase~I, right panel for phase~II of the CSS operations). Different curves for specific choice of the weighting function $w(\mathbf{p})$: (i) background (green), (ii) Schubart family (red), and (iii) Hilda family (blue). The composite total is the black line.  The embedded plots show the behavior of $\overline{\mathcal{P}}(H)$ for the brightest Hildas and two specific values of $\varpi_{\rm p}=75^\circ$ (red) and $285^\circ$ (blue). Here the detection probability may again drop from unity due to signal saturation, especially when the asteroid is sensed near perihelion of its orbit (red curve). This problem, and the sky coverage incompleteness, manifest primarily in the CSS phase~I due to its short duration. In the twice longer phase~II, even the bright Hildas may eventually appear to be observable close to the aphelion of their orbit, increasing their apparent magnitude to allow detection.}
 \label{fig11}
\end{figure*}
%%%%%%%%%%%%%%%%%%%%%%%%%%%%%%%%%%%%%%%%%%%%%%%%%%%%%%%%%%%%%%%%%%%%%%%%%%%%%%%%%%%%%%%%%%%%%%%%%%%%%%%

% -------------------------------------------------------------
% SEC ???
\section{Results}\label{res}

\subsection{Detection probability of CSS observations}
Before presenting our main results, we briefly discuss the behavior of the detection probability ${\cal P}\left(\mathbf{p};H\right)$. Our goal is to help future researchers avoid mistakes in formulation and coding, as well as provide a check on the expected features hinted at by observations.
% FIG 1 %%%%%%%%%%%%%%%%%%%%%%%%%%%%%%%%%%%%%%%%%%%%%%%%%%%%%%%%%%%%%%%%%%%%%%%%%%%%%%%%%%%%%%%%%%%%%%%
\begin{figure*}[t!]
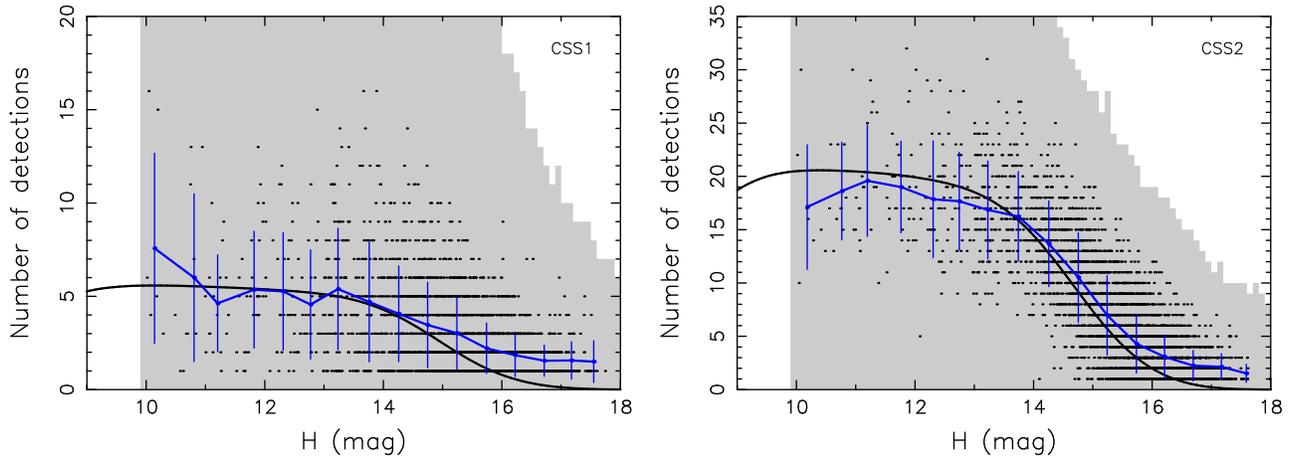

 \begin{center}
 \begin{tabular}{cc}
  \includegraphics[width=0.46\textwidth]{f12a.eps} &
  \includegraphics[width=0.46\textwidth]{f12b.eps} \\
 \end{tabular}
 \end{center}
 \caption{Number of detections for each unique Hilda in the CSS phase~I (left; $2,970$ bodies) and phase~II (right; $4,317$ bodies) plotted as a function of the absolute magnitude. Each black dot stands for one observed Hilda member. The blue curve is the mean number of detections in a $0.25$ wide magnitude bin, and the vertical bars are statistical dispersion within the bin. The black line is the population-wise weighted $\overline{\mathcal{R}}(H)$ prediction from our model. The grey area delimits our simple estimate of a range defining in $0.1$-wide bins the minimum to maximum number of detections as a function of $H$ (see the text for more details).}
 \label{fig12}
\end{figure*}
%%%%%%%%%%%%%%%%%%%%%%%%%%%%%%%%%%%%%%%%%%%%%%%%%%%%%%%%%%%%%%%%%%%%%%%%%%%%%%%%%%%%%%%%%%%%%%%%%%%%%%%

In Fig.~\ref{fig9} we have fixed semimajor axis and eccentricity to $a_{\rm p}=3.99$~au and $e_{\rm p}=0.155$. From there, we display ${\cal P}\left(\mathbf{p};H\right)$ for a variety of $\sin I_{\rm p}$ values and two specifically chosen values of $\varpi_{\rm p}=45^\circ$ and $225^\circ$. We expect and the results confirm two distinct trends, especially in the longer-lasting phase~II of CSS operations.

First, higher-inclination orbits have a lower detection probability. This correlation is stronger in the phase~I data due to their confinement to low-ecliptic latitudes (Fig.~\ref{fig1}).  Second, the $\varpi_{\rm p}=45^\circ$ branch data reach higher ${\cal P}\left(\mathbf{p};H\right)$ values compared to the $\varpi_{\rm p}=225^\circ$ branch data for faint Hildas ($H\geq 15.5$). This behavior, produced by Jupiter's systematic perturbations on the orbits of the Hilda asteroids, was hinted at by the observations (Fig.~\ref{fig5}).  We also note that ${\cal P}\left(\mathbf{p};H\right)$ never reaches unity in the phase~I data for inclination $\sin I_{\rm p}\geq 0.15$, even for bright Hildas. This outcome is a purely geometrical effect of sampling in the fields-of-view only low ecliptic latitudes. In effect, we are missing some asteroids that reach higher latitude values. Some others are also missed due to the 3~yr span of this early phase, which obscures trends in the $\varpi_{\rm p}$ dependency of ${\cal P}$.

Another look at the $\varpi_{\rm p}$ dependence in our analysis is given by Fig.~\ref{fig10}. Here we keep the same $a_{\rm p}$ and $e_{\rm p}$ values as before and additionally fix $\sin I_{\rm p}=0.05$, which corresponds to the Schubart family. We plot ${\cal P}\left(\mathbf{p};H\right)$ as a function of $\varpi_{\rm p}$ for four values of absolute magnitude $H$: (i) a very bright case with $H=7.5$, and (ii) a series of three $H$ values where faint detections are made. 

Again, the tendencies are best exemplified during the CSS phase~II. As hinted at by the data in Fig.~\ref{fig5}, the detection probability of Hilda asteroids only weakly depends on $\varpi_{\rm p}$ for bright objects. In contrast, the fainter Hildas gradually exhibit a dependence on $\varpi_{\rm p}$, with the non-detection interval located near $\varpi_{\rm p}\simeq 180^\circ-280^\circ$. These orbits have small values of the heliocentric osculating eccentricity, and in perihelia they are too far from the observer to overcome the telescope detection limit. The phase~I results are affected by two drawbacks: its short duration and incomplete sky coverage of the fields-of-view.

% --- global value
In order to simply communicate the probability with which CSS detects Hilda objects of a given absolute magnitude value $H$, we define the average detection probability $\overline{\mathcal{P}}(H)$ by integration over the orbital space
\begin{equation}
 \overline{\mathcal{P}}(H) = \frac{\int d^4\mathbf{p}\,w(\mathbf{p})\,{\cal P}(\mathbf{p};H)}{\int d^4\mathbf{p}\,w(\mathbf{p})}\; , \label{ave1}
\end{equation}
where the weight is given by the product of the 1-D distribution functions $D_i(p_i)$ of proper orbital parameters from Eqs.~(\ref{model2}) and (\ref{distp})
\begin{equation}
 w(\mathbf{p})=\prod_{i=1}^{3} \,D_i\left(p_i\right)  \label{ave2}
\end{equation}
and $d^4\mathbf{p}$ is the volume element of the parametric space (this time including also $\varpi_{\rm p}$). We can also choose the weighting specific to a particular subpopulation of Hildas, namely the background population of a certain collisional family. The formal integrals in (\ref{ave1}) are simplified by summation over the bin grid defined in Table~\ref{bins}. We also define the average detection rate $\overline{\mathcal{R}}(H)$ of a representative Hilda member of magnitude $H$ using (\ref{rate}) in the integrand of Eq.~(\ref{ave1}) numerator.

Figure~\ref{fig11} shows $\overline{\mathcal{P}}(H)$ for both phases~I and II, as well as for various choices of the weighting kernel $w(\mathbf{p})$ (from individual families to the whole population). The expected features include overall better CSS performance during phase~II, with minimum dependence on orbital inclination (represented by the low-inclination Schubart family and higher-inclination Hilda family). They also show formal completion at $H\simeq 15$. Another expected feature comes from the detection probability during phase~I.  It is smaller than unity, even for small $H$ values (bright Hildas), due to the CSS fields-of-view missing high ecliptic latitudes. 

The inset panels in Figure~\ref{fig11} show the behavior of $\overline{\mathcal{P}}(H)$ for the brightest Hildas, namely for $H\leq 7.5-9$. Here the probability is smaller than unity, even in phase~II due to signal saturation to a small degree. This is most prominent when $\varpi_{\rm p}\simeq 60^\circ-150^\circ$, which forces some of the objects to near-perihelion locations where they would undergo putative detection.

Figure~\ref{fig12} shows complementary information to the detection probability discussed above, namely the population averaged rate of Hilda asteroids detection $\overline{\mathcal{R}}(H)$ for both phases~I and II (shown by the black line).  They can be compared to the real CSS detections (black dots). The reasonable match between the lines provides an independent justification of our model. $\overline{\mathcal{R}}(H)$ drops for both bright and faint objects for the reasons discussed above, namelt that the bright Hilda asteroids tend to have signals that sometimes saturate on the CCD chip, while the faint ones drop below the detection limit of the telescope. 

This behavior is similar to Jupiter Trojans \citep[see Fig.~19 in][]{vok2024}, but Hildas exhibit an interesting difference, namely a long tail of detections beyond absolute magnitude $16$. This stems from the large orbital eccentricities of some Hildas, allowing smaller bodies to occasionally be detected near perihelion. 

The drawback of our detection rate definition in (\ref{rate}) is that small (but non-zero) probabilities, when added together, may eventually result in a number smaller than unity for faint Hildas ($H\geq 17$). True detections, however, increment in integer numbers. In order to account for this fact, we also ran our survey simulator, turning the field-of-view probabilities $\epsilon_{j,k}$ ($j$-th simulated orbit in $k$-th field-of-view) into detections (or non-detections) in a Poissonian manner. By estimating the population in each $0.1$-wide magnitude bin from Fig.~\ref{fig8} (left panel), we could consider, in each $\mathbf{p}$-space bin, the minimum and maximum possible integer detections from this sample. The corresponding range over all orbital space is shown by the gray area in Fig.~\ref{fig12}. Although this admittedly pushes the range of detection rates to extreme values, the envelope successfully captures the trend defined by true CSS detections.
% FIG 1 %%%%%%%%%%%%%%%%%%%%%%%%%%%%%%%%%%%%%%%%%%%%%%%%%%%%%%%%%%%%%%%%%%%%%%%%%%%%%%%%%%%%%%%%%%%%%%%
\begin{figure}[t!]
 \begin{center}
  \includegraphics[width=0.47\textwidth]{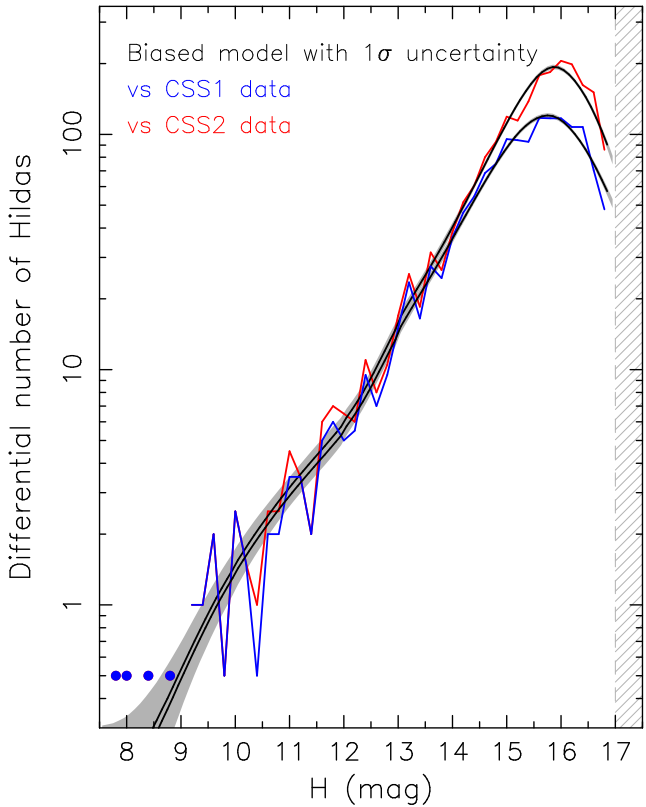} 
 \end{center}
 \caption{Differential magnitude distribution of the G96 observations (blue for phase I and red for phase II) compared with the best-fitting biased model (black line with a $\sigma$ interval, based on analysis of 10,000 posterior random samples of the model, depicted by the gray zone). The ordinate normalization corresponds to the magnitude $0.1$~mag bins used for the model. The data, originally allocated to $0.2$~mag bins, were smeared into the neighboring $0.1$~mag wide bins for direct comparison with the model (this also implies that the lowest occupancy is "$0.5$~body"). The largest detected objects with $H<9$~mag (four in both phases~I and II) are shown individually by symbols (the model predicts slightly less than one object in this range, but the fit is dominated by $H>9$~mag objects).}
 \label{fig_biased}
\end{figure}
%%%%%%%%%%%%%%%%%%%%%%%%%%%%%%%%%%%%%%%%%%%%%%%%%%%%%%%%%%%%%%%%%%%%%%%%%%%%%%%%%%%%%%%%%%%%%%%%%%%%%%%
% FIG 1 %%%%%%%%%%%%%%%%%%%%%%%%%%%%%%%%%%%%%%%%%%%%%%%%%%%%%%%%%%%%%%%%%%%%%%%%%%%%%%%%%%%%%%%%%%%%%%%
\begin{figure*}[t!]
 \begin{center}
  \includegraphics[width=0.95\textwidth]{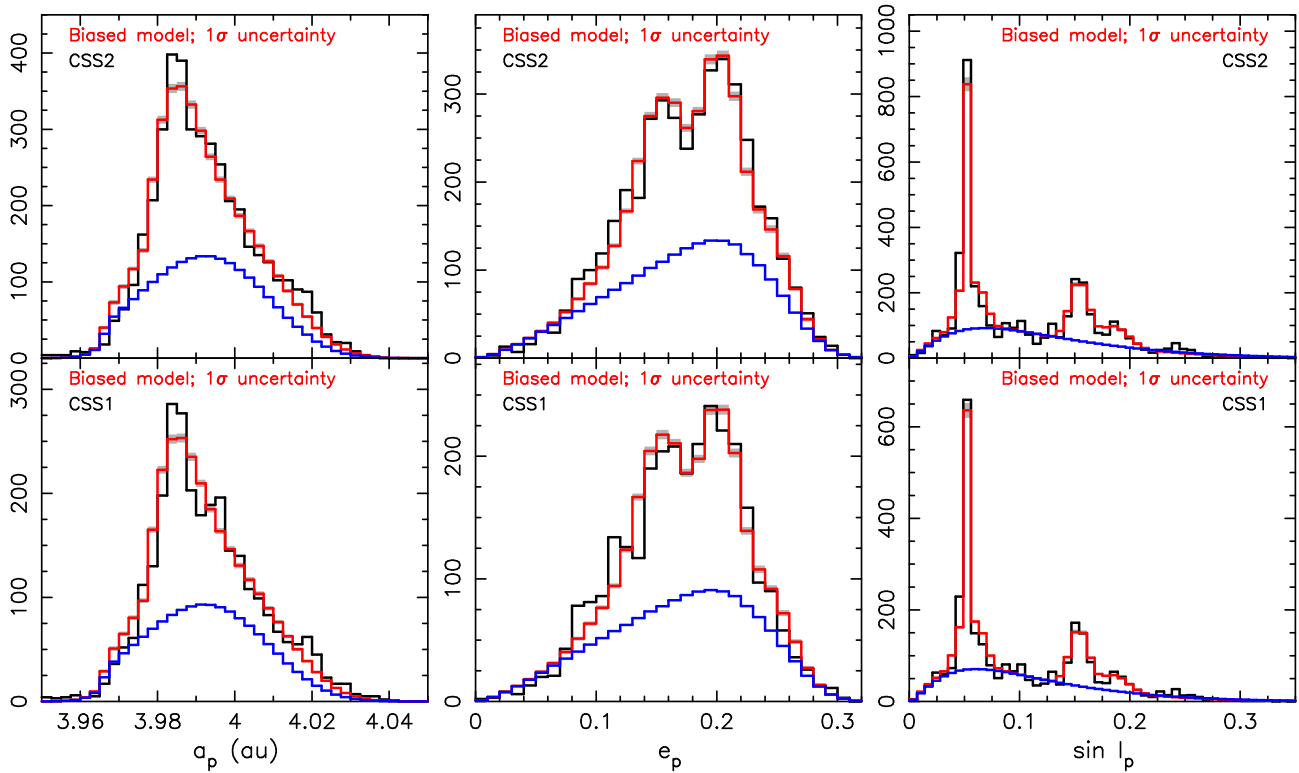} 
 \end{center}
 \caption{Projected distributions of the proper orbital parameters $a_{\rm p}$ (left), $e_{\rm p}$ (middle), and $\sin I_{\rm p}$ (right) of the biased Hilda population from our model (red lines) and the G96 observations (black histogram) using the phase I (bottom panels) and phase II (top panels) data. The shaded gray area delimits the $\sigma$ region of the solution (created by 10,000 posterior random samples of the model). The blue histogram is the best-fitting model of the background population. The residual mismatch is most likely expression of two effects: (i) a slight inaccuracy (or simplicity) in the template masks of the families accounted for and absence of smaller clusters, and (ii) possible correlations of the orbital distribution in the orbital elements.}
 \label{fig_orbs_biased}
\end{figure*}
%%%%%%%%%%%%%%%%%%%%%%%%%%%%%%%%%%%%%%%%%%%%%%%%%%%%%%%%%%%%%%%%%%%%%%%%%%%%%%%%%%%%%%%%%%%%%%%%%%%%%%%
% FIG 1 %%%%%%%%%%%%%%%%%%%%%%%%%%%%%%%%%%%%%%%%%%%%%%%%%%%%%%%%%%%%%%%%%%%%%%%%%%%%%%%%%%%%%%%%%%%%%%%
\begin{figure}[t!]
 \begin{center}
  \includegraphics[width=0.47\textwidth]{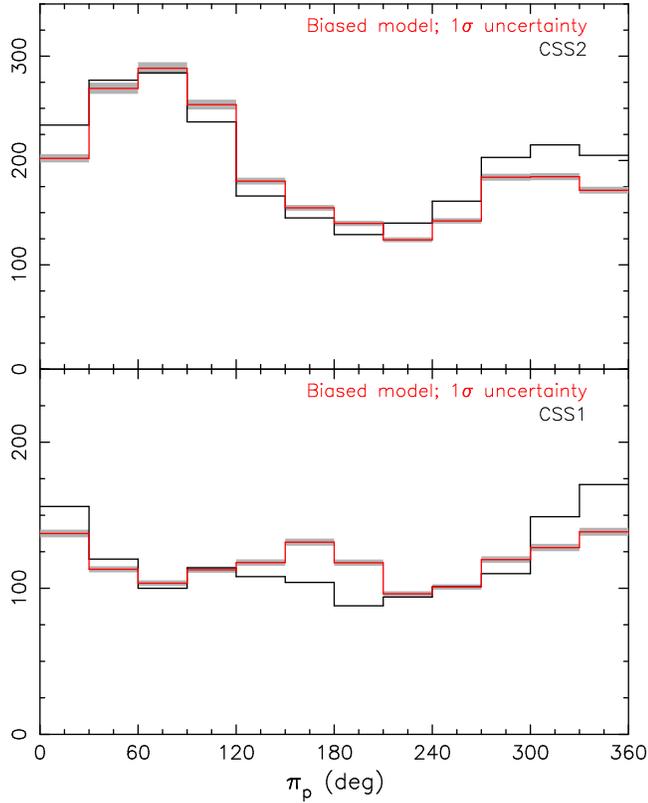} 
 \end{center}
 \caption{Projected distributions of the proper orbital longitude of perihelion $\varpi_{\rm p}$ of the biased Hilda population from our model (red line) and the G96 observations (black histogram) using the phase I (bottom panel) and phase II (top panel) data. We used here data for faint objects, $H$ value in between $15.5$ and $17$. The shaded gray area delimits the $\sigma$ region of the solution (created by 10,000 posterior random samples of the model). The quality of the phase~II (both in timespan and sky coverage) results in a rather good match between the observations and the model. Except for a mismatch in the $\varpi_{\rm p}$ range $120^\circ-180^\circ$, the model performs acceptably well even in the phase~I (bottom panel).}
 \label{fig_varpi_biased}
\end{figure}
%%%%%%%%%%%%%%%%%%%%%%%%%%%%%%%%%%%%%%%%%%%%%%%%%%%%%%%%%%%%%%%%%%%%%%%%%%%%%%%%%%%%%%%%%%%%%%%%%%%%%%%
% FIG 1 %%%%%%%%%%%%%%%%%%%%%%%%%%%%%%%%%%%%%%%%%%%%%%%%%%%%%%%%%%%%%%%%%%%%%%%%%%%%%%%%%%%%%%%%%%%%%%%
\begin{figure}[t!]
 \begin{center}
  \includegraphics[width=0.47\textwidth]{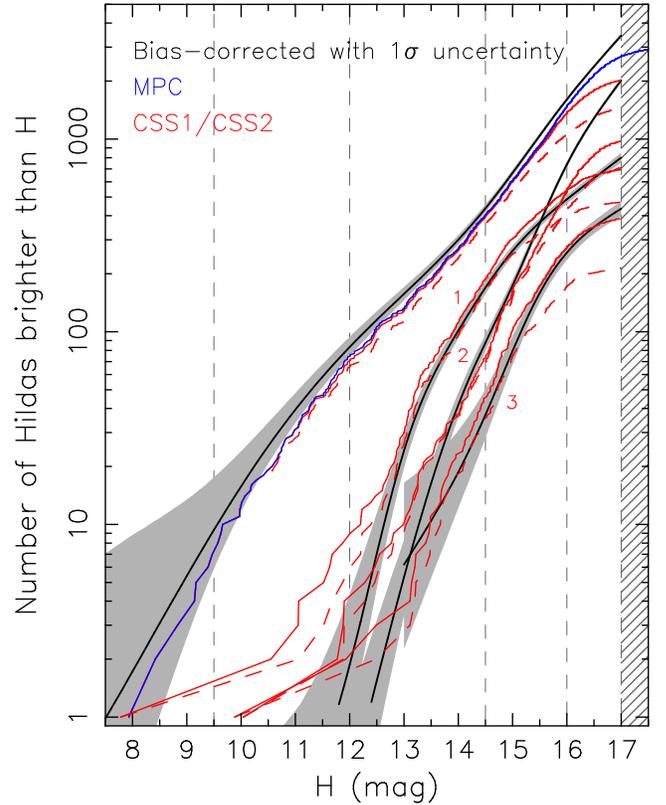} 
 \end{center}
 \caption{Bias-corrected cumulative magnitude distribution of the Hilda population based on G96 observations. The background population, shown by the upper curve, is separated from the major families (see Table~\ref{mb_fams_2023}): Hilda (label 1), Schubart (label 2), and Potomac (label 3). The red lines are the CSS observations (solid are the data in phase II, dashed are the data in phase I), and the black line is the bias-corrected model with $\sigma$ interval (gray zone). The blue line is the whole background population from the MPC catalog. The vertical dashed lines show the magnitude segments used for the representation of the background population (Table~\ref{model_back}).}
 \label{fig_h_unbiased}
\end{figure}
%%%%%%%%%%%%%%%%%%%%%%%%%%%%%%%%%%%%%%%%%%%%%%%%%%%%%%%%%%%%%%%%%%%%%%%%%%%%%%%%%%%%%%%%%%%%%%%%%%%%%%%
% FIG 1 %%%%%%%%%%%%%%%%%%%%%%%%%%%%%%%%%%%%%%%%%%%%%%%%%%%%%%%%%%%%%%%%%%%%%%%%%%%%%%%%%%%%%%%%%%%%%%%
\begin{figure*}[t!]
 \begin{center}
  \includegraphics[width=0.95\textwidth]{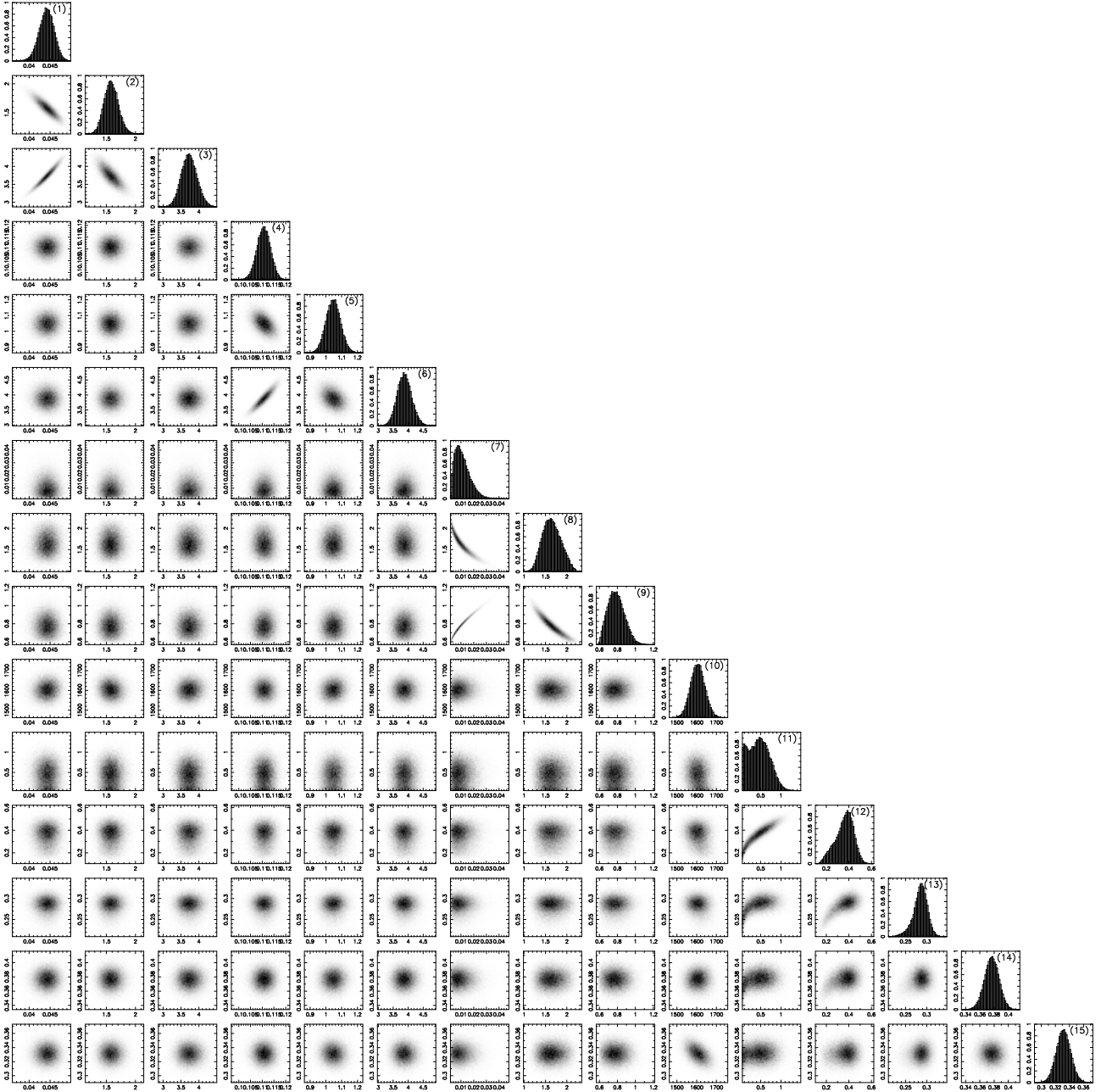} 
 \end{center}
 \caption{The posterior distribution of $15$ model parameters $(p_1,\ldots,p_{15})$ characterizing the background population of Hildas from our nominal fit to G96 observations (also called the “corner diagram”). The individual plots are labeled (1)–(15) following the model parameter sequence given in Table~\ref{model_back}; the first nine parameters determine the orbital distribution in proper parameters, and the last six parameters determine the absolute magnitude cumulative distribution (with the tenth parameter being the normalization at 16 mag, and the last five parameters the mean slope values in the chosen segments). The near-to-circular distribution of the parameter solution indicates uncorrelated parameters. In some cases though correlations exist and are shown by quasi-linear parameter dependence (e.g., $p_1$ vs $p_2$, $p_1$ vs $p_3$, etc.). These are well understood from the functional form of the distribution functions $D_i(p_i)$ given in (\ref{distp}).}
 \label{fig_params_corr}
\end{figure*}
%%%%%%%%%%%%%%%%%%%%%%%%%%%%%%%%%%%%%%%%%%%%%%%%%%%%%%%%%%%%%%%%%%%%%%%%%%%%%%%%%%%%%%%%%%%%%%%%%%%%%%%

\subsection{Fitting the observed Hilda population}\label{fit_nominal}
As part of our code testing, we first performed simulations using CSS observations acquired separately during phases I and II. Once we verified that these results were satisfactorily compatible, we performed a global simulation that took into account all CSS observations. In what follows, we only present and discuss the results obtained in this global fit.

The model parameters were optimized by seeking the maximum of the target function (\ref{prob2}) in the parameter space $(\mathbf{p};H)$ introduced above. Technically, $(\mathbf{p};H)$ was divided into more than 70 million bins specified in Table~\ref{bins}, where the parameter boundaries are also provided. The only exception was the maximum value of the absolute magnitude $H$, which was taken to be $17$ in our nominal simulation (this is because the number of available data beyond that limit is already too small; see Fig.~\ref{fig8}). 

The optimization algorithm is based on matching CSS observations projected to $(\mathbf{p};H)$ and the prediction of the biased model $dN_{\rm pred}(\mathbf{p};H)$ given in (\ref{biased}). The bias-corrected part of the model, $dN_{\rm model}(\mathbf{p};H)$, is made up of a direct product of 1-D distributions individual to each of the parameters. Additionally, the model linearly superposes contributions from the background population and the populations in collisional families in the orbital (4-D) sector (given by ${\mathbf{p}}$) and the absolute magnitude (1-D) sector (given by $H$). In our simulations, we only used three major families (Hilda, Schubart, and Potomac). The remaining three clusters listed in Table~\ref{mb_fams_2023} contain too few members at this time to be well resolved by the model.

The adjusted model coefficients define (i) the spline representation of the cumulative $H$-distribution of both the background and families' populations (Sec.~\ref{magdist}), and (ii) the orbital distribution of the background population (Sec.~\ref{orbdist}). The localization of the families in orbital space has been predefined by template functions matching their identification in the Appendix~\ref{propel}. However, note that the CSS observations themselves are not divided into background and family components. This separation is only performed by the {\tt MultiNest} code.

Figures~\ref{fig_biased} and \ref{fig_orbs_biased} show a comparison of the CSS data distributions, for both phases~I and II, and the distribution of the biased model $dN_{\rm pred}(\mathbf{p};H)$. Here we present 1-D distributions for the parameters by simply summing over all other parameters. On the model side, we show the best-fit solution and a $1\sigma$ surrounding zone obtained by analyzing 10,000 posteriori solutions with similar values of the log-likelihood function ${\cal L}$ provided by {\tt MultiNest}. 

The match between the model and the observations in these simple 1-D representations is arguably good. Some outlier cases in the orbital proper-parameter distributions in Fig.~\ref{fig_orbs_biased} may be due to missing collisional families or a slight mismatch in our template functions for the included families (see, for instance, the signal near $a_{\rm p}\simeq 4.02$~au, $e_{\rm p}\simeq 0.08-0.13$ or $\sin I_{\rm p}\simeq 0.24$).

A modest deficiency in the Schubart family population is a notable feature of the right panel of Fig.~\ref{fig_orbs_biased}. This feature is probably related to its correlation with other orbital proper parameters and a likely small inaccuracy of Schubart-family template functions. As an aside, in reality, the true localization of the Schubart family in $\mathbf{p}$ may not be well represented by a simple product of 1-D distributions. However, generalizations beyond this model are left for future work.

We assumed that each of the components in the Hilda population (background and families) has a uniform intrinsic distribution in $\varpi_{\rm p}$. However, the observed faint Hildas show a selection bias toward certain $\varpi_{\rm p}$ values that make it possible for them to be detected near perihelion, if they have a sufficiently large eccentricity (Fig.~\ref{fig5}). Therefore, an additional test of the bias function ${\cal P}(\mathbf{p};H)$ is provided by comparing the $\varpi_{\rm p}$ distributions determined by observations and the model (Fig.~\ref{fig_varpi_biased}). The match is very good for data of the phase~II (expressing the principal variation with a maximum for $\varpi_{\rm p}\simeq \varpi^\prime$ due to Jupiter's forced term; see (\ref{eprop2}) and related discussion in Appendix~\ref{map1}), but not-accounted for correlations with the distribution in other orbital parameters may contribute to differences in several bins that are slightly above statistical noise. As expected, the agreement is slightly worse in the phase~I data.
% Tab 2 %%%%%%%%%%%%%%%%%%%%%%%%%%%%%%%%%%%%%%%%%%%%%%%%%%%%%%%%%%%%%%%%%%%%%%%%%%%%%%%%%%%
\begin{deluxetable}{l|ccc}[t] 
 \tablecaption{\label{model_back}
  Median and uncertainties of Hilda background population parameters in the nominal model.}
 \tablehead{
 \colhead{} &\colhead{Fixed} &\colhead{Median} & \colhead{$\pm\sigma$} }
%\decimals
\startdata
 \rule{0pt}{3ex}
 & \multicolumn{3}{c}{{\it -- Semimajor axis: $D_1(a_{\rm p})$ --}} \\ [2pt]
  $c_a$        & $3.95$ & --    & -- \\
  $\alpha$     & --     & $1.58$ & $0.12$ \\
  ${\bar c}_a$ & $3.95$ & --    & -- \\
  $\sigma_a$   & --     & $0.044$ & $0.002$ \\
  $\beta$      & --     & $3.73$ & $0.22$ \\
 & \multicolumn{3}{c}{{\it -- Eccentricity: $D_2(e_{\rm p})$ --}} \\ [2pt]
  $c_e$       & $0$    & --    & -- \\
  $\alpha$     & --     & $1.05$ & $0.05$ \\
  ${\bar c}_e$ & $0.12$ & --    & -- \\
  $\sigma_e$   & --     & $0.11$ & $0.03$ \\
  $\beta$      & --     & $3.87$ & $0.24$ \\ 
 & \multicolumn{3}{c}{{\it -- Inclination: $D_3(\sin I_{\rm p})$ --}} \\ [2pt]
  $c_{\sin I} $ & $0$    & --    & -- \\
  $\alpha$     & --     & $1.65$ & $0.21$ \\
  ${\bar c}_{\sin I}$ & $0$ & --    & -- \\
  $\sigma_{\sin I}$  & --  & $0.011$ & $0.006$ \\
  $\beta$      & --     & $0.79$ & $0.09$ \\ [3pt] 
 & \multicolumn{3}{c}{{\it -- Magnitude distribution parameters --}} \\ [2pt]
 $N_{\rm back}$ & -- & $1605$ & $36$  \\ 
 $\gamma_1$ & -- & $0.47$ & $0.24$ \\
 $\gamma_2$ & -- & $0.36$ & $0.08$ \\
 $\gamma_3$ & -- & $0.29$ & $0.02$ \\
 $\gamma_4$ & -- & $0.37$ & $0.01$ \\
 $\gamma_5$ & -- & $0.33$ & $0.01$ \\ [2pt]
\enddata

\tablecomments{The six parameters at the bottom part of the Table specify the cumulative magnitude distribution: (i) $\gamma_i$ are mean slopes on five magnitude segments defined by intervals ($7.5-9.5$, $9.5-12$, $12-14.5$, $14.5-16$, $16-17$), and (ii) $N_{\rm back}=N_{\rm back}(<H_0)$ is the complete population up to $H_0=16$ magnitude. The nine parameters in the upper part of the table specify the coefficients of the orbital distribution $D_i(p_i)$ functions given by Eq.~(\ref{distp}) for $\mathbf{p}=(a_{\rm p},e_{\rm p},\sin I_{\rm p})$.}
\end{deluxetable}
%%%%%%%%%%%%%%%%%%%%%%%%%%%%%%%%%%%%%%%%%%%%%%%%%%%%%%%%%%%%%%%%%%%%%%%%%%%%%%%%%%%%%%%%%%%
% FIG 1 %%%%%%%%%%%%%%%%%%%%%%%%%%%%%%%%%%%%%%%%%%%%%%%%%%%%%%%%%%%%%%%%%%%%%%%%%%%%%%%%%%%%%%%%%%%%%%%
\begin{figure}[t!]
 \begin{center}
  \includegraphics[width=0.47\textwidth]{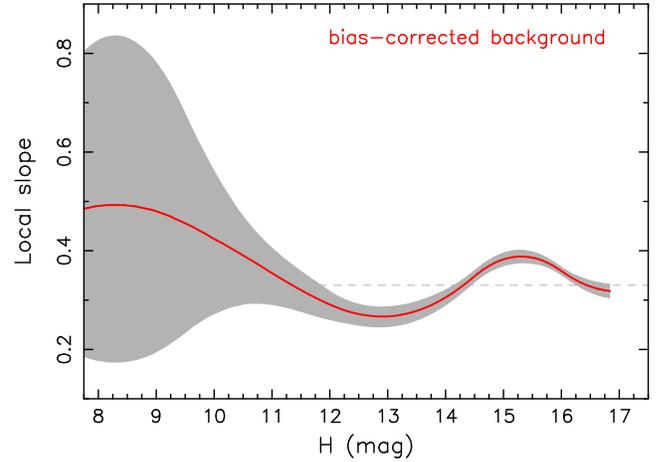} 
 \end{center}
 \caption{The local slope of the cumulative absolute magnitude distribution for the bias-corrected background population of Hildas. The gray zone is the $\sigma$ interval of the solution. The horizontal dashed line at $0.34$ slope is the  value found by \citet{wb2017} using the simple power-law fit.}
 \label{fig_gamma}
\end{figure}
%%%%%%%%%%%%%%%%%%%%%%%%%%%%%%%%%%%%%%%%%%%%%%%%%%%%%%%%%%%%%%%%%%%%%%%%%%%%%%%%%%%%%%%%%%%%%%%%%%%%%%%

\subsection{Bias-corrected Hilda population}\label{bc_popul}
Removing the multiplicative factor ${\cal P}(\mathbf{p};H)$ from the $dN_{\rm pred}(\mathbf{p};H)$ distribution is a direct way to obtain the intrinsic (bias-corrected) model distribution $dN_{\rm model}(\mathbf{p};H)$ (Eq.~\ref{biased}).

Figure~\ref{fig_h_unbiased} shows the intrinsic cumulative magnitude distribution of the background population and the three families included in the simulation (black lines). For comparison, we also show the cumulative distributions of the biased observations. The observed asteroids in the background population are represented by the CSS data (red lines) and the MPC data (blue line). Note that the latter contains significantly more faint objects, with contributions included from large-aperture surveys such as Pan-STARRS or dedicated efforts using big telescopes such as Subaru or Keck. Still, the MPC observed population remains at all magnitudes smaller than the intrinsic population predicted by our model, thereby providing a good sanity check on our work. In fact, it may appear surprising that the model prediction for the background population stays higher than the MPC set for bright objects like those with magnitudes $\simeq 12-13$. At these brightnesses/sizes, the Hilda population is complete. The observed difference stems from our ability to separate the background population from members in collisional families. Indeed, when we compare the model prediction of the complete bias-corrected population (background together with families), it closely matches the MPC sample to magnitude $H\simeq 16.4$.

The solution for the three main families, Schubart, Hilda, Potomac, is shown by the bottom curves identified with the labels 1, 2, and 3, respectively. Here, the result is modestly less satisfactory, especially in the case of the Hilda family in which the model undershoots the population identified by the clustering method in the proper element space (Appendix~\ref{propel}). 

We suspect that this difference is related to an inaccuracy in the mask function of the family within the $\mathbf{p}$ space.  We use the simple product of three 1-D distributions in $(a_{\rm p},e_{\rm p},\sin I_{\rm p})$ for both background and family representation, but in reality the exact family structure in this space is more complicated, especially for extended and diffuse families like Hilda and Potomac. As a consequence, if we compare the intrinsic model solution for the Hilda and Potomac families, we find that both fall short of the population from the hierarchical clustering method shown in Fig.~\ref{figww}. Conversely, some of the Hilda and Potomac members that leaked into the background population in our model are the probable source of the overabundance discussed above.

Our solution for the more compact Schubart family is better. In the first magnitude segments used in this case, namely $12-14$, $14-15$ and $15-16$, we found mean slope values of $1.00\pm 0.18$, $0.62\pm 0.04$ and $0.63\pm 0.02$, respectively, and a population of $738\pm 22$ with $H\leq 16$. At this magnitude limit, it represents $46$\% of the background population. In the magnitude range $16-17$, the slope becomes shallower and goes to $0.44\pm 0.02$. Here again, however, the family template function that we used in orbital space was probably too simple, such that some of Schubart's small and more dispersed members were incorrectly associated with the background population. A more complex analysis of this issue is left for future work.

The three collisional families included in our fit contain $1451\pm 65$ members with $H\leq 16$. Together with the asteroid clusters not accounted for, and the probable leakage of some family members to the background population, the families include approximately $50$\% of the total Hilda population at this magnitude limit.

Table~\ref{model_back} and Fig.~\ref{fig_params_corr} provide information on the numerical coefficients (including formal uncertainty and mutual correlations) that define the background population. While those of the orbital segment are less straightforward, those characterizing the magnitude distribution are more easily understandable. We find that the background population of Hildas has $1605\pm 36$ objects with $H\leq 16$. The cumulative distribution is steeper for large Hildas, and becomes shallower for objects having magnitudes greater than $H\simeq 10$. This change in slope is reminiscent of the Jupiter Trojan magnitude distribution, but on both sides of the dividing magnitude $H \simeq 10$, Hildas have shallower slope exponents. 

For example, in the first segment ($7.5\leq H\leq 9.5$), we find $\gamma_1=0.46\pm 0.24$, with a large formal uncertainty due to its small population. It bends in the second segment ($9.5\leq H\leq 12$) to $\gamma_2=0.36\pm 0.08$. In the comparable magnitude interval, Jupiter Trojans have slopes of $\simeq 0.65$ and $\simeq 0.43$ \citep[see][]{vok2024}. For $H > 12$ bodies, Jupiter Trojans maintain an average slope of $\simeq 0.44$, while Hildas have a mean value of only ${\bar \gamma}=0.32\pm 0.04$ between magnitudes $12$ and $17$. 

Analysis of the cumulative magnitude distribution defined by cubic splines allows us to determine a locally defined slope $\gamma$ for the entire range of $H$ values described by the model. In particular, at each of the $dH=0.1$~magnitude interval used by the model (see Table~\ref{bins}), we determine an increment $dN$ of the intrinsic population and define $\gamma=\log(dN)/dH$. The result is shown in Fig.~\ref{fig_gamma}. The limited Hilda population for $H\leq 11$ makes the respective value of $\gamma$ uncertain, but for smaller Hildas $\gamma$ is reasonably well determined. Although the mean value is $\simeq 0.32$, the formal uncertainty of the locally defined $\gamma$ is small enough to resolve the possible dependence on $H$, the principal characteristic being a maximum at $H\simeq 15$. We discuss this behavior of $\gamma(H)$ further in Sec.~\ref{concl}.

\subsection{Variant simulations}
In Secs.~\ref{fit_nominal} and \ref{bc_popul}, we presented the results of our nominal model. Although the fit to the data provided each model parameter with their reported uncertainty (e.g., Table~\ref{model_back}), these values must be considered formal. The reason is because the model itself represents a certain choice. Only when a variety of such models with different parameter choices are used can more realistic uncertainties be evaluated. In this work, we do not make extensive efforts in this direction, but instead restrict ourselves to a few examples.
\smallskip

\noindent{\it Restricted range of magnitude distribution.-- } Our nominal model includes the largest Hildas, but there is only a small number of them scattered irregularly throughout the Hilda stability zone. Given that our approach couples orbital and magnitude distributions, the small number of the largest objects may affect the results in an inconvenient manner. Therefore, in the first variant of the nominal model, we omitted asteroids of magnitude $H< 10$. In practice, we dropped the first magnitude segment and shifted the lower boundary of the second segment (now becoming the first) to $H=10$. Otherwise, the model remained identical to that presented in Secs.~\ref{fit_nominal} and \ref{bc_popul}.

The results are essentially identical to the nominal model, showing that the weight given to the largest Hildas is minimal. For example, the background population with $H\leq 16$ now has $1606\pm 38$ objects, and the counts of members in the three families are also well within the uncertainty interval of the nominal solution.
\smallskip

\noindent{\it Simplified magnitude representations: broken and single power-law models.-- }Next, we opted to keep the orbital distribution and collisional family modeling as before, but we simplified the way we express the cumulative magnitude distribution of the background population. In this variant, instead of a five-sector model with a normalization parameter $N_{\rm back}$, we now use a simpler broken power law representation between the $H=7.5$ and $17$ magnitudes. This means two sectors with mean slopes $\gamma_1$ and $\gamma_2$ divided at a breakpoint $H_{\rm break}$ and a normalization constant $N_{\rm back}$ at $H_0=16$. Accordingly, rather than using a six-parameter magnitude model, our new approach has four adjustable parameters $(\gamma_1,\gamma_2,H_{\rm break},N_{\rm back})$. 

We focus on results related to the magnitude sector (subject to the principal change). Upon convergence, we obtained $N_{\rm back}=1785\pm 103$, $H_{\rm break}=10.1\pm 1.0$ and $\gamma_2=0.30\pm 0.02$ (with $\gamma_1=0.2\pm 0.1$ only poorly constrained). We compared Bayesian-based evidence factors $\ln P$ (Eq.~\ref{prob2}) of the best-fit solutions provided by
{\tt MultiNest}. The relative preference of one over the other models is directly
determined by the value of $\exp \left[\Delta \left(\ln P\right)\right]$. Since $\Delta \left(\ln P\right)=-6.5$ in this case, the likelihood of the simple broken power-law model is $\sim 1.5\times 10^{-3}$. This value indicades this model variant is clearly an inferior solution. Indeed, the differential data in Fig.~\ref{fig_biased} exhibit a curvature near $H\simeq 13$ that calls for more than a single exponent $\gamma$ of the model.

We also considered an even simpler case, in which we discarded all Hildas with $H<10$ and fit the background population using a single power law model in between magnitudes $10$ and $17$. We obtained the slope $\gamma=0.34\pm 0.06$ \citep[very close to the results in][]{wb2017} and a population of $1570\pm 35$ for $H\leq 16$. As before, the simple power law magnitude models have problems dealing with the curvature in the distribution over a large range of values.
\smallskip

\noindent{\it Extension of the orbital modeling.-- }In the next variant case, we  represent the Hilda magnitude distribution in the same way as in the nominal model, but we slightly extend the orbital model. In particular, the eccentricity distribution function $D_2(e_{\rm p})=e_{\rm p}^\alpha\,\exp[-0.5 ((e_{\rm p}-{\bar c}_e)/\sigma_e)^\beta]$ had a fixed value ${\bar c}_e=0.12$. We have now made this parameter adjustable, thus extending the set of solved for parameters by one more.

This allowed us to obtain ${\bar c}_e = 0.15\pm 0.04$, a slightly higher value than in our nominal model presented above. As expected, with one more free parameter in the model, the fit to the eccentricity distribution (middle panels in Fig.~\ref{fig_orbs_biased}) improved. The Bayesian evidence score is superior by $\Delta \left(\ln P\right)=-9.3$ over the nominal model, which implies that the likelihood of the nominal model is only $\sim 9\times 10^{-5}$ compared to this extended variant. 

Despite this impressive improvement, the overall population did not change (basically all parameters of the magnitude distribution remain as in the Table~\ref{model_back}). Our takeaway message is that there are avenues to improve the orbital model, but the background population statistics is reasonably robust. This may change if the distribution functions for collisional families were reevaluated, but this complex task awaits future work.

\section{Discussion and conclusions} \label{concl}
The main results of our work can be summarized as follows:

(1) \citet{vok2024} developed a novel formalism to simultaneously determine the orbital and magnitude distribution of a small-body population.  It was applied to the case of the Jupiter Trojans. Here we used the same approach to characterize the orbital architecture and magnitude distribution of the Hilda population, which resides in the stable region of the 3/2 mean motion resonance with Jupiter. 

Although the magnitude distribution description is the same for each study, with both using a cubic-spline representation of the cumulative distribution, the orbital model is different. First, the simplest semi-analytical model for Hilda dynamics is more complex than those used for Jupiter Trojans. Second, Hilda orbits have their largest osculating eccentricity correlated with their longitude of perihelion, and both are restricted to a limited interval of values. In our analysis, this effect required us to extend the 3-D orbital space of semimajor axis, eccentricity, and inclination by longitude of perihelion. The population distribution functions, smooth in the model formulation, were represented by their discrete values in bins that parsed 4-D orbital and 1-D magnitude space (we use more than 750,000 bins for orbits and 100 bins in magnitudes). Similarly to the Trojan study, we characterized the background (dispersed) population separated from the populations within the collisional families.

(2) We provide a new catalog of proper orbital elements for the Hilda population (Appendix~\ref{propel}). This allowed us to identify statistically significant clusters using hierarchical clustering analysis, which we interpret as collisionally-born asteroid families. We found six cases, of which five were previously reported in the literature. The increased number of asteroids in these populations led to a more accurate description of their nature, with the most outstanding case being that of the Schubart cluster.  In general, asteroids in families represent more than $60$\% of the Hilda population, a fraction that is exceptionally large among other small body populations.

(3) We applied our model to the observations of the CSS station G96 taken between January~2013 and June~2022. More than 220,000 documented fields of view were well characterized in terms of detection probability, which is a function of each body's apparent magnitude and rate of motion across the sky. The G96 camera upgrades in May~2016 made us split the survey period into phases I and II, which we consider as two independent surveys. During phase I, G96 detected 2970 individual Hildas (with 8686 detection in total). These numbers increased during phase II to 28,930 detections of 4317 individual Hildas. In total, a significant fraction (more than 70\%) of the entire Hilda population known today was thus detected by CSS. Following \citet{vok2024}, we determined the detection probability defined over the orbital elements and the absolute magnitude range used by our model. The detection probability is a convolution of the aforementioned detection probabilities within a given field of view and the probabilities that occur in any of the survey fields of view.

(4) We applied the {\tt MultiNest} code to modify the free parameters of our model to reach a maximum likelihood match to CSS observations. This outcome leads to a suitably robust algorithm to tackle the multi-parametric task with existing correlations. Conveniently, {\tt MultiNest} also provides posterior distributions of the solved-for parameters, enabling us to evaluate the statistical significance of the model and its components.

(5) Although the quantitative solution of the orbital part of our model is interesting on its own, the principal finding concerns the absolute magnitude distribution. By separating out collisional family members from the smooth background, we found that the mean power-law exponent of the background population in the range $(11,17)$ is only $0.32\pm 0.04$. This is significantly shallower than found over a similar range for Jupiter Trojans \citep[see][and references therein]{vok2024}. Alternately, we can also express the same result in absolute population numbers. \citet{vok2024} found the following background biased corrected populations of Jupiter Trojans with $H\leq 14.5$: (i) $3951\pm 44$ in the L4 swarm and (ii) $2664\pm 36$ in the L5 swarm. In contrast, our nominal model of the Hilda population only has $437\pm 21$ asteroids in the biased-corrected background population. This low value is a surprise considering that models indicate that the number of bodies captured from the primordial Kuiper belt for the Trojans and Hildas should be coparable to one another \cite {vok2016}. Although the difference between the L4 and L5 populations represents a problem of its own, the substantially smaller Hilda population over this magnitude range is an additional puzzle that needs to be understood.

In what follows, we briefly comment on some of our results presented above.  We also set the stage for further studies, building on results of the present paper.
\smallskip
% FIG 1 %%%%%%%%%%%%%%%%%%%%%%%%%%%%%%%%%%%%%%%%%%%%%%%%%%%%%%%%%%%%%%%%%%%%%%%%%%%%%%%%%%%%%%%%%%%%%%%
\begin{figure*}[t!]
 \begin{center}
  \includegraphics[width=0.9\textwidth]{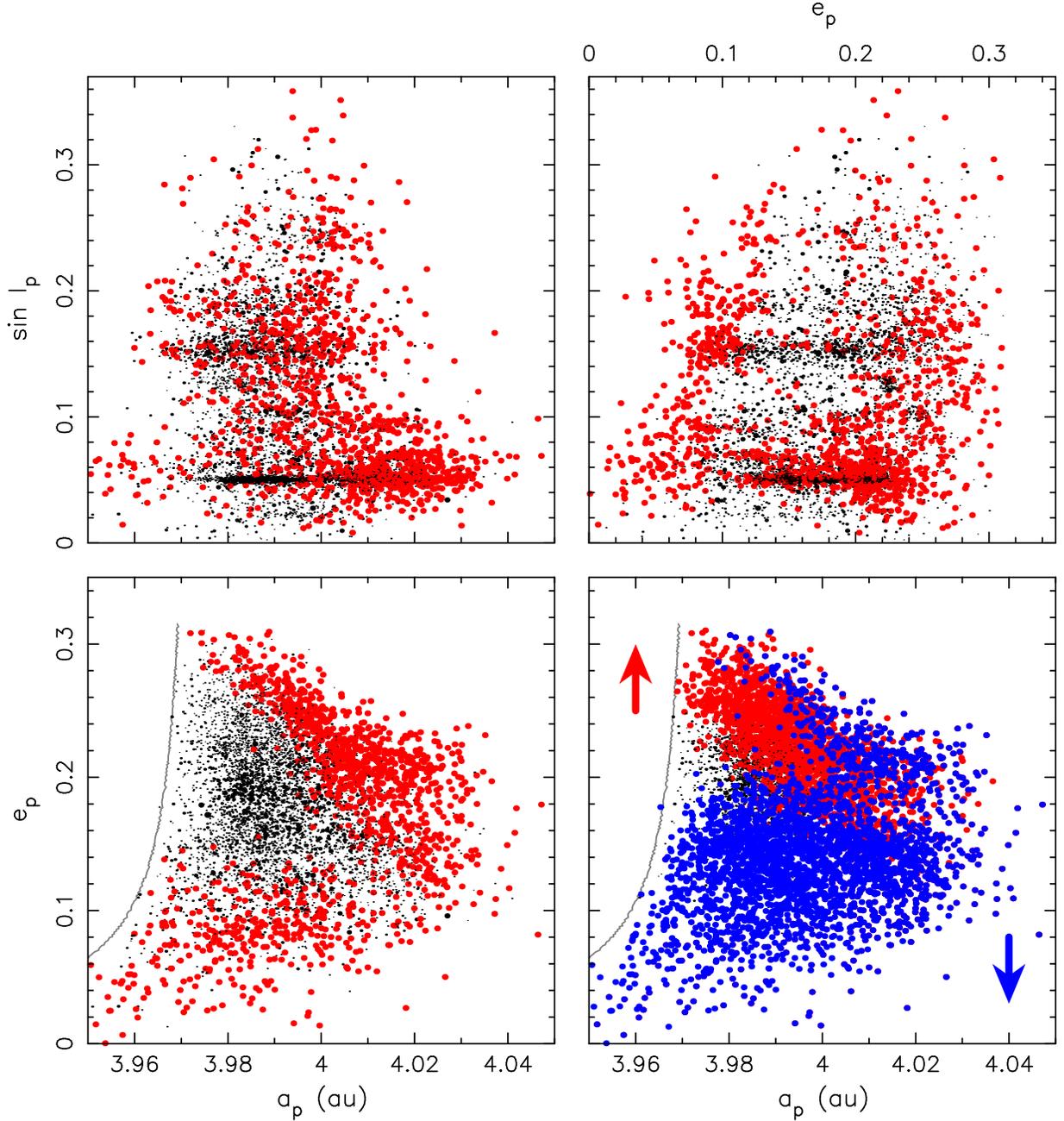} 
 \end{center}
 \caption{Top panels and bottom left panel: The same as in Fig.~\ref{fig6}, but now the long-term unstable Hilda asteroids were highlighted with red symbol. These objects were eliminated from the population in a numerical simulation spanning $4$~Gyr and accounting for gravitational perturbations from all planets. The currently observed population of 6,094 Hilda asteroids represented the initial data of the simulations. The eliminated Hildas were located near the border of the stability zone (see, in particular, the bottom left panel and compare with a similar results in Fig.~6 of \cite{bv2008}). An excessive fraction of the Schubart family members at high-eccentricity orbits would be eliminated confirming its age younger than $4$~Gyr \citep[see][]{bv2008}. Bottom right panel: $(a_{\rm p},e_{\rm p})$ projection of the
 known Hilda population as on the bottom left panel, but now the color-coded asteroids are those, which were eliminated in $4$~Gyr lasting simulation that included the effect of the thermal accelerations. In this case all bodies were given $\simeq 4$~km size and maximum estimated value of the thermal accelerations: (i) red symbols had positive along track acceleration, and (ii) blue symbols had negative along track acceleration. The usual secular drift in semimajor axis transfers to a drift in eccentricity due to the resonance lock \citep[e.g.,][]{bv2008}; the sense of the drift is indicated by the arrows.}
 \label{figyy}
\end{figure*}
%%%%%%%%%%%%%%%%%%%%%%%%%%%%%%%%%%%%%%%%%%%%%%%%%%%%%%%%%%%%%%%%%%%%%%%%%%%%%%%%%%%%%%%%%%%%%%%%%%%%%%%

% slope increase of the background population

\subsection{Slope variation of the Hilda magnitude distribution}
We first consider the exponent of the local slope of the cumulative absolute magnitude $\gamma(H)$ distribution. We find that the local slope of the background population $\gamma$ increases for $H > 14.5$.  This behavior is interesting and requires an explanation (Fig.~\ref{fig_gamma}).  

One speculative possibility is that there is no change in the power-law slope of the size frequency distribution, but instead the mean albedo increase for smaller objects near this transition.  For example, if we assume (i) $p_V=0.06$ for $H\leq 14$ and (ii) $p_V\simeq 0.06 +0.015\, (H-14)$ for $H\geq 14$, we would obtain exactly the observed increase in $\gamma(H)$ between magnitudes $14$ and $15.5$.  Although such behavior cannot be ruled out at this time, we find it unlikely.  

A second possibility, one that we consider more plausible, is that this is where we are seeing the contribution of collisional families either not accounted for or not treated well enough in our simulations. For example, the medium and small families listed in Table~\ref{mb_fams_2023}, namely Francette and Guinevere, were not included in our fit. Furthermore, some families may be so dispersed that the hierarchical clustering technique used in the Appendix~\ref{propel} may not be able to identify them.  A missing family might explain the local peak in the proper inclination distribution near $\sin I_{\rm p}\simeq 0.24$ (see the rightmost panels in Figs.~\ref{fig7} and \ref{fig_orbs_biased}). 

Finally, in Sec.~\ref{bc_popul} we noted that the intrinsic population of the major families included in our simulation is underestimated at magnitudes $H\geq 14.5$, a problem likely related to their complicated structure in proper element space.  In effect, it is impossible for them to be represented by a simple product of 1-D distribution functions in each of the proper elements alone. As a consequence, some family members missing in our solution for the dispersed Hilda and Potomac clusters could have leaked to the background population, therefore increasing the background slope $\gamma$.

% Long-term stability
\subsection{Long-term dynamical stability of the Hilda population}\label{stabi}
As a preliminary step toward understanding the shallow power law slope of the background population, and possibly that of old and dispersed families like Hilda, we conducted the following set of numerical experiments. 

In the first simulation, we considered 6,094 Hilda objects identified in Sec.~\ref{datag96} and propagated their orbits forward in time for $4$~Gyr. We used the {\tt swift} integrator and a similar setup as in Sec.~\ref{datag96}.  All planets were included as massive perturbers, with initial data from the JPL ephemerides DE420.  A short timestep of $3$~days was used. Hildas were eliminated from the simulation when their heliocentric distance was smaller than $1$~au or greater than $10$~au, or when they hit Jupiter or Saturn. To speed up this computation, we also distributed the simulation across 50 CPUs, each of which propagated $\sim  120$ Hildas. 

Figure~\ref{figyy} shows the situation at the end of the simulation, namely the bodies projected onto the 2D planes of proper parameters $(a_{\rm p},e_{\rm p},\sin I_{\rm p})$ as in Fig.~\ref{fig6}, but now with the long-term unstable orbits (eliminated during the simulation) highlighted by red symbols. Clearly, the red unstable objects are not randomly distributed. Instead, they are concentrated towards the border of the populated region in the $(a_{\rm p},e_{\rm p})$ projection and somewhat less in the $(e_{\rm p},\sin I_{\rm p})$ projection. This confirms that the dynamically stable zone of the J3/2 resonance is fully filled. Of particular interest is the large-$a_{\rm p}$ and large-$e_{\rm p}$ end of the prominent Schubart family,  which has been significantly depleted over $4$~Gyr. The existing population of this family on such orbits points towards its younger age \citep[see][who argue this family is $1.7\pm 0.7$~Gyr old based on a similar simulation]{bv2008}. 

Another issue clarified by the results shown in Fig.~\ref{figyy} relates to the data in Fig.~\ref{fig4}, where we have shown that the orbits of the largest Hilda objects ($H\leq 12$, say) only reach moderate eccentricity values. Although less evident, this remains true when the osculating eccentricity at the abscissa of Fig.~\ref{fig4} is replaced with the proper parameter $e_{\rm p}$.  If we assume that the orbits of the brightest Hildas survived $>4$~Gyr, such that they are essentially primordial \citep[e.g.,][]{betal2011}, they must be located in the long-term stable portion of the J3/2 resonance. In fact, we verified that these large Hilda objects are located in the zone truncated by the red-flagged orbits in Fig.~\ref{figyy}.  That zone is filled in a uniform fashion, as expected with their origin by capture.
% FIG 1 %%%%%%%%%%%%%%%%%%%%%%%%%%%%%%%%%%%%%%%%%%%%%%%%%%%%%%%%%%%%%%%%%%%%%%%%%%%%%%%%%%%%%%%%%%%%%%%
\begin{figure}[t!]
 \begin{center}
  \includegraphics[width=0.47\textwidth]{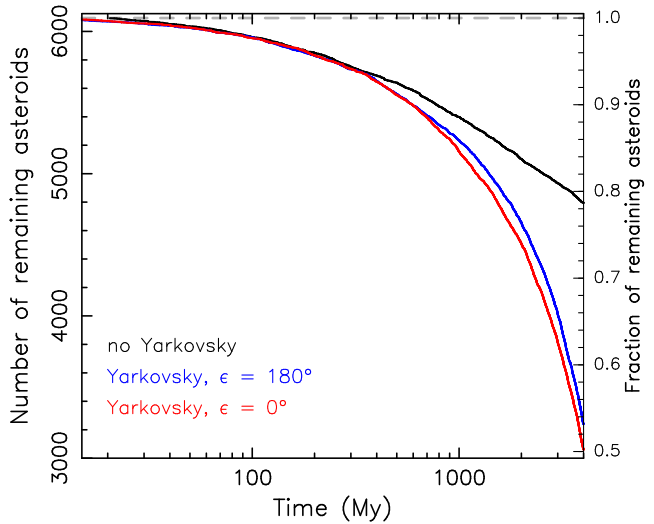} 
 \end{center}
 \caption{Depletion of the observed Hilda population with time (6,094 initial bodies identified in Sec.~\ref{datag96} initially, horizontal dashed line). Solid lines show results from three simulations: (i) only gravitational perturbations from planets (black line), (ii) thermal accelerations as if all bodies had $H\simeq 16$ and an estimated maximum outward drift (red line), and (iii) thermal accelerations as if all bodies had $H\simeq 16$ and an estimated maximum inward drift (blue line). The runs with the thermal accelerations thus correspond to bodies with sizes modestly smaller than $4$~km for $p_V\simeq 0.055$ \citep[e.g.,][]{grav2012} and assume the objects have bulk densities of $1.5$ g~cm$^{-3}$.}
 \label{figzz}
\end{figure}
%%%%%%%%%%%%%%%%%%%%%%%%%%%%%%%%%%%%%%%%%%%%%%%%%%%%%%%%%%%%%%%%%%%%%%%%%%%%%%%%%%%%%%%%%%%%%%%%%%%%%%%

Next, we repeated the long-term integration simulation of the 6,094 Hildas using an idealized model in which, in addition to gravitational perturbations from the planets, we also accounted for Yarkovsky thermal accelerations \citep[i.e., the Yarkovsky effect,][]{vetal15}. To see how the bodies react to a maximum possible Yarkovsky drift, we assign the propagated Hildas drift rates of $da/dt=\pm 5.4\times 10^{-5}$~au~My$^{-1}$.  We also used the simplified model presented by \citet{far2013}. Given the characteristic scaling of the Yarkovsky effect with size $D$, heliocentric semimajor axis $a$ and bulk density, namely $da/dt \propto 1/(D a^2\rho)$, this corresponds to about $D\simeq 4$~km Hildas or $H\simeq 16$ for $p_V=0.055$. 

The possibility of positive and negative signs in $da/dt$ means that we had to perform two simulations, one for objects that would drift outward and the other for those that would drift inward. As discussed in the Appendix of \citet{bv2008}, the resonant lock of the semimajor axis transfers the drift to the proper eccentricity such that: (i) orbits which are subject to the positive Yarkovsky drift $da/dt$ secularly increase their $e_{\rm p}$ value and (ii) vice versa (see Fig.~\ref{figyy}). 

We speculate that smaller Hildas, subject to larger thermal accelerations, are pushed to higher eccentricity orbits, thereby explaining the trend seen in Fig.~\ref{fig4}.  This would work in conjunction with the YORP effect, which helps to explain why lost objects are found at the most extreme distances within the Schubart family (Fig.~\ref{fig1911}). 

Figure~\ref{figzz} shows the number of Hildas remaining in each of our three long-term simulations as a function of time. In the base run, where only gravitational perturbations have been taken into account, the population is depleted by $\simeq 23$\% after $4$~Gyr. This value is comparable to the depletion of Jupiter Trojan clouds studied by \citet{holt2020}. 

In contrast, the depletion in the simulations containing thermal accelerations is considerably larger. In both runs, setting $da/dt$ positive or negative leads to the loss of $> 50$\% of the objects on $4$~Gyr. Obviously, larger Hildas would get depleted to a lesser degree, such that the decay curve would eventually approach that of the simulation containing just the gravitational perturbations. 

Overall, the strong depletion of smaller Hildas should help make the primordial size/magnitude distribution shallower with time.  Although promising, this effect would need to be included in a model of collisional evolution to see if it can explain the slope differences between the Hildas and Jupiter Trojans \citep[note that the Yarkovsky effect produces significant depletion of Jupiter Trojan clouds only on much smaller sizes, e.g.,][]{hell2019}.

% Next
\subsection {Implications}
While both the present study and that of \citet{vok2024} focus on describing the existing populations of Hildas and Jupiter Trojans, their results pose interesting science questions which need to be followed up in future studies. The puzzles arise from a mutual comparison of the L4- and L5-swarm Trojan populations discussed in \citet{vok2024}.  Here we highlight the Hilda to Trojan differences. 

We consider the paradigm in which the initial populations of Hildas and Trojans were captured in their respective resonances during the giant planet instability period \citep[e.g.,][]{lev2009,vok2016,n2018}. These models suggest that the initial captured populations had comparable sizes.  Several aspects of the currently observed populations support this point of view.  For example, the numbers of the largest Hildas and Trojans are similar to each other, and their populations fill the entire dynamically stable region at their resonance locations.  The physical characteristics of the bodies of the two populations are also similar to each other. 

The main disagreement between the two populations today consists of different magnitude (and size) distributions for objects dimmer (smaller) than $H\simeq 10$. While Jupiter Trojans have a mean magnitude slope $\gamma\simeq 0.44$ in the $10$ to $15$ magnitude range, Hildas have $\gamma\simeq 0.34$ in the overlapping $11$ to $17$ magnitude range. 

Here we have identified the dynamical leakage of small Hildas due to the Yarkovsky effect as a culprit to explain this difference. However, we suspect that a significant role could also be played by differences in collisional evolution between the two populations over the past $4-4.5$~Gyr \citep[see already][]{lev2009}. \citet{bot2023} developed recently a detailed model of a collisional model for Jupiter Trojans as part of a larger study whose other focus was on a collisional evolution of the scattering disk population. This allowed them to explain the size frequency distribution of the currently observed Trojan population. Their work did not consider the Hilda population, so this task remains an interesting avenue for future work. In this case, however, the collisional evolution needs to be coupled with a dynamical depletion.

Finally, while continuous operations of CSS will provide valuable data for Jupiter Trojans and Hildas modeling (profiting from a long temporal extend of a well-characterized source), our approach is readily applicable to other future surveys. In this respect we may mention the Vera C. Rubin observatory that promises to further revolutionize understanding of the inventory of Solar system small-body populations \citep[e.g.,][]{schwa2023}.

\acknowledgments
The Catalina Sky Survey is funded by NASA’s Near-Earth Object Observations (NEOO) Program under grant 80NSSC24K1187. The simulations were performed on the NASA Pleiades Supercomputer. The authors thank the NASA NAS computing division for continued support and the referee for valuable comments on the submitted manuscript. The work of DV and MB was supported by the Czech Science Foundation (grant 25-16507S). DN’s work was supported by NASA's Solar System Workings program. WB’s work in this paper was supported by NASA's SSERVI Program through cooperative agreement 80NSSC23M0176 and NASA's Lucy mission through contract NNM16AA08C.

%%%%%%%%%%%%%%%%%%%%%%%% REFERENCES %%%%%%%%%%%%%%%%%%%%%%%%%%
%\newpage

\bibliographystyle{aasjournal}

% \bibliography{lit}{}
%\bibliographystyle{aasjournal}

\appendix

\section{A brief review of Hilda population}\label{hildas}

\noindent{\it Dynamics, population and its orbital architecture.-- } Studies of Hilda-type motion, only somewhat later expanded to the Hilda population, have a long history. Soon after Johann Palisa discovered the asteroid (153) Hilda in November 1875 \citep{p1875}, Franz K\"uhnert collected the available single-opposition observations to determine its orbit \citep{k1876}. His interest in this object was piqued
because it seemed to approach within $\simeq 0.6$~au of Jupiter, the closest distance of all known asteroids to date. Even though his orbital description did not take into account this object's resonant-protection mechanism (and its actual closest approaches to Jupiter only occur at $\simeq 1.9$~au), the dynamics of Hilda-type objects remained in the mind of many astronomers. It was hoped that the strong perturbing effects of Jupiter on these worlds could help improve what was known about Jupiter's mass \citep[e.g.,][]{laves1904}. 

From there much of the first half of the 20th century was characterized by contradictory opinions about the nature of the few known asteroids having orbits similar to (153) Hilda. The theorists disputed whether the Hildas librate about the strictly periodic resonant J3/2 state or whether they are ultimately unstable, such that they only temporarily visit Jupiter's vicinity. An encyclopedic overview of these efforts can be found in \citet{hagi1972} (Sec.~9 of Volume~2, in particular).

As often occurs in science, progress in our understanding of the Hilda population resulted from a synergy of research in the mid 1960s: (i) the number of known asteroids clustered about the J3/2 commensurability with Jupiter increased to 17, constituting the core of an actual population, and (ii) electronic computers allowed numerical techniques to overcome existing problems characterized by poor convergence of traditional series representation of the perturbing function, and even principal non-convergence for orbits formally swapping heliocentric distance with Jupiter \citep[for both, see,][]{schub1964}. Just as important, Joachim Schubart focused his attention on Hildas and
wrote a series of papers that defined our early knowledge of Hilda-type dynamics and population characteristics \citep[see][to mention just the most important ones]{schub1964,schu1968,schu1982a,schu1982b,schu1991}. 

As computing power increased in the 1980's and 1990's, it became possible to attempt more ambitious studies of Hilda dynamics. These new works took advantage of novel analytical methods, such as symplectic maps, to produce more accurate descriptions of planetary dynamics. Several stability studies of the phase space associated with Jupiter's mean motion resonances are found in \citep[e.g.,][]{m1986,wis1987,mfm1995,mfm1996,nfm1997}. 

The phase space of the J3/2 resonance was found to harbor a large, long-term stable region bounded by several secular resonances that triggered borderline chaos and instability \cite[e.g.,][]{mm1993}. Unlike the J2/1 resonance, it was shown that the Hilda region had the ability to hold a significant population of asteroids.  However, the question of whether nature actually filled the Hilda region with objects was left to observers.   

New discoveries slowly increased the Hilda population, but as of 1997 it still contained only 79 members, all deep inside the stable region of the Hilda phase space \citep{nfm1997}.  The game changer came from powerful all-sky near-Earth object (NEO) surveys that became operational at the beginning of the 2000s.  While finding NEOs, the surveys also contributed immensely to our knowledge of numerous small body populations across the solar system. As an example, \citet{grav2012} were able to determine the size frequency distribution and albedos for more than 1,000 Hilda objects. The present-day number of known Hildas has increased to more than 6,000 bodies. 

\citet{bv2008} used numerical methods borrowed from the Trojan \citep{m1993} and Hecuba \citep{retal2002} populations to determine the synthetic proper elements for Hilda objects that were more accurate than the traditional proper parameters introduced by \citet{schu1982a,schu1982b,schu1991}. Their sample contained 1,197 resonant objects, enough to provide two novel findings: (i) they showed that the Hilda population fills the stable zone of phase space up to large eccentricity and inclination values, and (ii) the proper element distribution revealed evidence for two major collisional families \citep[a strongly concentrated cluster about (1911)~Schubart and a more diffuse cluster about (153)~Hilda itself, see also][]{betal2011}. For the former, a possible clustering of orbits about asteroid (1911)~Schubart had been previously suggested by \citet{schu1982a,schu1991}, but definitive evidence only came from modern datasets, with the cluster containing more than a thousand bodies. The substantial fraction of Hilda objects in the two families was further confirmed by \citet{vin2015}, despite using somewhat less accurate proper orbital parameters \citep[basically returning to the old concepts set by][]{schu1982a,schu1982b,schu1991}, and by \citet{pisa2017}. 

The work of \citet{vino2020} using the increase in population count revealed further smaller clusters in the Hilda population. Although their investigation used somewhat less accurate orbital parameters and only considered a 2D proper parameter space of eccentricity and inclination, our new work below has confirmed their results.  We have even identified two additional families.  All of this underlines our primary take-away message, namely that collisional clusters play a prominent role in the Hilda population (see more details in Sec.~\ref{concl}).
\smallskip

\noindent{\it Physical studies of Hilda population.-- }The rapidly increasing population of Hilda objects prompted studies of their physical parameters, with comparisons made to their neighboring resonant population of Jupiter Trojans. The initial effort to accumulate physical data on the Hilda population was carried out by the Uppsala group of Mats Dahlgren. \citet{da1995} and \citet{da1997} conducted the first systematic visible spectroscopic surveys of the Hildas. They found that most bodies had D- or P-type taxonomies, with a slight majority for the D-types.  A very small component of objects with flat spectra was also found to be compatible with C-type taxonomy.

Further taxonomic information has been collected by \citet{ghb2008}, \citet{wb2017} and \citet{wbe2017}. \citet{wb2017} used the Sloan Digital Sky Survey broad-band photometric observations to infer two color indexes defined in \citet{ro2008}. They again found a bi-modality in the population, sorting them into two broad less-red and red groups (aka the C-/P-type taxonomies and the D-type taxonomy).  Intriguingly, the Schubart and Hilda collisional families were dominated by less-red objects. \citet{wbe2017} used the IRTF/SpeX spectrograph to extend the known spectral information to the near infrared band for 25 large Hildas. They showed featureless spectra that (i) confirmed a less-red vs. red bimodality, and (ii) indicated a close similarity to the Jupiter Trojans data.

The absolute magnitude $H$ distribution of Hilda population was also tracked by \citet{wb2017}. Aware of the collisional families in the population (whose identification was downloaded from the Planetary Data System system as of 2015), they analyzed the families and the background populations separate from one another.  Given that they did not have a procedure to correct for observational biases, they conservatively restricted their study to bright Hildas with $H\le 14$. This range of $H$ values, when plotted as a differential absolute magnitude distribution, was compared to a power law $\propto 10^{\gamma H}$ with a single, solved-for exponent $\gamma$.  The exponents of the two families were  $\gamma_1=0.40^{+0.04}_{-0.03}$ for Hilda family and $\gamma_2=0.43^{+0.07}_{-0.03}$ for the Schubart family.%
\footnote{In Sec.~\ref{res} we find even steeper exponents, confirming an earlier result of \citet{bv2008}.}
These values were found to be steeper than the exponent of the background population of $\gamma=0.34^{+0.02}_{-0.01}$. More importantly though, the magnitude distribution of the Hilda population was found to be considerably shallower than the magnitude distribution of Jupiter Trojans in the same range \citep[$\gamma=0.46\pm 0.01$, see][]{wetal2014,vok2024}. Additionally, the Trojan less-red population was found to have a steeper magnitude distribution than the red population. Interestingly, no such difference was found for the different color sub-populations of Hildas.

As for the fainter end of the Hilda magnitude distribution, \citet{ty2018}  analyzed single night observations taken by the Hyper SuprimeCam instrument mounted on the $8.2$~m Subaru Telescope. They identified $130$ moving objects having orbits roughly compatible with Hildas, but 5\% were deemed to be outer main belt interlopers. The final sample they analysed had $91$ objects. Assuming that the objects have $\simeq 0.05$ albedo values, their sizes would range from $1$ km to $\sim 14$~km. Fitting a single power law to the differential absolute magnitude distribution, these authors infer $\gamma=0.38\pm 0.02$. Given the uncertainty about real membership of all these objects to the Hilda population and that the background and families' data were not separated, the result is reasonably close to the bright end slope. Curiously, it is also close to what was found for faint Jupiter Trojans in this size range \citep[see also analysis in][]{yo2019}.

Finally, \citet{grav2012} used the non-cryogenic phase of the WISE mission to characterize the albedo and size properties of 1,023 Hildas. They found a mean albedo of $p_V=0.055\pm 0.018$, with only a slight variation with size and family membership. While not performing a full-fledged debiasing effort, they suggested that a single-power law fits the observations down to about $5$~km diameter objects, with the slope of the cumulative distribution being $-1.7\pm 0.3$. Data from the two shortest-wavelength WISE bands was used to define the taxonomic class of the bodies, with a split between D-type objects (about 2/3) and C- and P-type objects (about 1/3, in which the flatter spectra suggested that the C-type taxonomy only represented a minor component).

The first dedicated survey of lightcurve observations of Hildas was completed by \citet{da1999}, who obtained data for 47 objects. When compared to the main belt asteroids, these authors noted (i) a markedly non-Maxwellian distribution of the rotation periods (with a significant tail to slow-rotators) and (ii) systematically larger mean lightcurve amplitudes. These findings were recently confirmed and strengthened by \cite{sza2020}, who used photometric data of the K2 extension of the Kepler mission to determine the rotation parameters of 125 Hilda members.
The latter survey also found (i) a lack of fast rotators combined with a substantial fraction of slow rotators ($P>100$~hr periods made up  $\sim 18$\% of the population), (ii) no difference between the lightcurve parameters of the less-red and red objects, and (iii) a number of double-period lightcurves that they interpreted as binaries; if so, Hildas hold a large fraction of binary systems. In summary, they concluded that the rotation parameters of Hildas are similar to Jupiter Trojans. 

Interesting data from a sample of 17 small ($D< 3$~km) Hildas were obtained by \citet{chang2022}. They discovered several fast rotators in this category that were close to the spin barrier at $\simeq 3$~hr (i.e., the spin period where gravitational aggregates would begin to shed mass by centrifugal forces). They suggested that this may be evidence of a larger proportion of C-type asteroids among smaller Hildas. 

Finally, \citet{gaia2023} analyzed the sparse photometric data from the Gaia DR3 catalog to determine the spin state of 61 Hilda asteroids (see their Fig.~4). The affinity of asteroid obliquity values toward the $0^\circ$ and $180^\circ$ values is a characteristic fingerprint of the thermal YORP torques commonly seen in the main-belt population. This observation suggests thermal accelerations \citep[known as the Yarkovsky effect, e.g.,][]{bv2008,vetal15} should affect the orbital architecture of small Hildas (Sec.~\ref{concl}).

Taken together, the observations of physical parameters available to-date reveal a large degree of similarity between the Hilda and Jupiter Trojan populations. A notable exception is the significantly shallower magnitude distribution of Hildas (see also Sec.~\ref{concl}). 
\smallskip

\noindent{\it Origin of Hilda population.-- }Studies of small bodies populations have provided compelling evidence that the orbits of giant planets reconfigured themselves sometime within the first $100$~Myr after their formation \citep[e.g.,][]{n2018}.  This so-called "giant planet instability" influenced the Hilda population in two ways: (i) any putative population of Hildas that existed prior to the instability were destabilized by that event \citep[e.g.,][]{betal2011,rn2015}, and (ii) a new Hilda population was captured within the stable zone within the J3/2 resonance \citep[e.g.,][]{lev2009,vok2016}. The captured objects originated from a massive trans-Neptunian planetesimal disk, or primordial Kuiper belt, that was dispersed by an outward migrating Neptune \citep[a minor, perhaps few percent level, contribution could have also originated in the outer main belt region, e.g.,][]{rn2015}. The broad distribution of eccentricities and inclinations, delimited only by the stability zone of the resonance, is a natural component of this capture scenario and in fact provides a strong supporting argument in its favor. Interestingly, \citet{vok2016} found approximately the same capture probability for Jupiter Trojans and Hildas, implying that their populations should be comparable both in number and in physical parameters. The available observations, briefly reviewed above, mostly match this prediction. The primary observation that differs comes from the shallower magnitude distribution of the Hildas for $H\leq 17$ objects.

For the sake of completeness, we also mention a few origin scenarios for Hildas that avoid the giant planetary instability. For example, Hildas might have been captured if Jupiter's experienced substantial inward migration within a gas disk, with the planetesimals scooped up by the  J3/2 resonance. This possibility was investigated by \citet{fra2004} and \citet{pir2019}. The former concluded that the eccentricity and libration angle distributions of the Hilda and Thule populations, and the large disparity in population between the J3/2 and J4/3 resonances, respectively, require Jupiter's slow migration by at least $0.45$~au. However, these authors did not investigate how their models would explain the most challenging constraint, namely the dispersed inclination distribution of the Hildas.

\section{Proper orbital parameters for Hilda-type orbits}\label{propar}
Here we summarize the variables, the proper orbital parameters $\mathbf{p}$ in the terminology of this paper, used in the main text to describe the orbital architecture of the Hilda population. The main difficulty in choice of $\mathbf{p}$ consists of conflicting requirements: ideally, we need some sort of heliocentric elements (i) that are long-term stable to provide accurate information about structure of Hilda population, and (ii) in the same time admitting the simplest possible correspondence to the osculating orbital elements $\mathbf{e}$. The latter facilitates mapping the synthetic population of Hilda asteroids to the pipeline of detection probability evaluation. Furthermore, we aim to keep the dimensionality of $\mathbf{p}$ as low as possible, certainly lower than that of $\mathbf{e}$ (which is dim$(\mathbf{e})=6$). In accordance with various systems of proper elements for non-resonant and resonant asteroid populations, dim$(\mathbf{p})=3$ seems optimum. Furthermore, $\mathbf{p}=(a_{\rm p},e_{\rm p},\sin I_{\rm p})$ with
a semimajor axis parameter $a_{\rm p}$, eccentricity parameter $e_{\rm p}$, and sine of inclination parameter $\sin I_{\rm p}$ would make these parameters easily understandable. In accordance with the discussion in Sec.~\ref{datag96}, we extend this set of proper parameters by a proper value of the longitude of perihelion $\varpi_{\rm p}$. Although of different flavor, this parameter is required to accurately describe the detection probability for orbits reaching high osculating eccentricity. Our final proper parameter space is therefore four-dimensional, dim$(\mathbf{p})=4$, and is made up of $\mathbf{p}=(a_{\rm p},e_{\rm p},\sin I_{\rm p},\varpi_{\rm p})$ vectors.

The analytical methods to describe fine details of motion in the first-order mean motion resonances with Jupiter have been found to be notoriously difficult. Therefore, we pay a sacrifice from both sides, accuracy in (i) and simplicity in (ii). Our final orbital element toolkit is approximated in a number of respects. While we encourage readers to develop more involved methods in future work, we find our approach justified enough to provide reasonable results.

In the next section, we briefly recall the Hamiltonian approach, and related conjugated variables, suitable for description of asteroids in the 3/2 mean motion resonance with Jupiter. Equipped with this knowledge, we then address the two fundamental matters of their use:
\begin{description}
\item[$\mathbf{e}\rightarrow \mathbf{p}$ mapping] needed to represent 
 the observed Hilda population in the $\mathbf{p}$-space (Appendix~\ref{map1}), and
\item[$\mathbf{p}\rightarrow \mathbf{e}$ mapping] needed to transform 
 a sample of synthetic orbits of Hilda population to their heliocentric osculating counterpart, in order to assign detection probability to various cells in the $\mathbf{p}$-space using the CSS detection pipeline (Appendix~\ref{map2}).
\end{description}

\subsection{Canonical variables and averaging over the fastest variable}\label{cano}
The use of canonical variables for the description of asteroid motion in first-order mean-motion resonances with Jupiter has a long tradition \citep[see already][and references
therein]{schub1964}, which has reached a masterful level especially in the literature of the 1980s to 1990s on the topic. We thus
adopt the Hamiltonian formalism as the backbone of our approach.
Although Delaunay variables $(L,G,H;\ell,g,h)$ are traditionally the starting choice
of the orbital elements \citep[e.g.,][]{md1999}, it is also useful to recall their
correspondence to the widespread Keplerian elements: semimajor axis $a$ using
$L=\sqrt{(1-\mu)a}$, eccentricity $e$ using $G=L\sqrt{1-e^2}$, inclination $I$
using $H=G\cos I$, longitude of node using $\Omega=h$, longitude of perihelion using $\varpi=g+h$, and longitude in orbit $\lambda$ using $\lambda=\varpi+\ell$. These
variables, or their combination, are going to represent the $\mathbf{e}$-parameters introduced above. The choice
of units assumes Jupiter's semimajor axis and heliocentric mean motion equal to one,
the mass of the Sun $1-\mu$ and the mass of Jupiter equal to $\mu$, and the gravitational constant unity. Description of asteroid dynamics in the 3/2 mean motion resonance with Jupiter further requires the following angles: (i) $\sigma=3\lambda'-2\lambda-\varpi$, the principal resonant angle, (ii) $\nu = -3\lambda'+2\lambda+\varpi'$, and (iii) $\sigma_z = 3\lambda'-2\lambda-\Omega$, non-resonant but slowly changing angles. Jupiter's heliocentric orbital elements are denoted by a prime. The model developed in this section assumes a framework of a restricted elliptic problem (the Sun and Jupiter massive primaries, and the asteroid a mass-less particle). Therefore, the orbital elements of Jupiter are constant, obviously except for the mean longitude in orbit $\lambda'$ (equal to time in our system of units, as both $a'=1$ and the mean motion $n'=1$).

The Hamiltonian $\mathcal{H}$ describing heliocentric motion of the asteroid reads
\citep[e.g.,][]{lh1988,mm1993}
\begin{equation}
 \mathcal{H} = L' - \frac{(1-\mu)^2}{2L^2}-\mathcal{R}\; , \label{ham1}
\end{equation}
where
\begin{equation}
 \mathcal{R} = \mu \left(\frac{1}{|{\bf r}-{\bf r}'|} - \frac{{\bf r}\cdot {\bf r}'}{r'^3}\right) \label{pert1}
\end{equation}
is the Jupiter perturbation, ${\bf r}$ and ${\bf r}'$ are the respective heliocentric position vectors of the asteroid and Jupiter (both of which should be evaluated using Delaunay elements). Traditional methods of analytical celestial mechanics proceed by developing $\mathcal{R}$ in either powers of eccentricities and inclinations (Kaula-type approach), or powers of ratio between asteroid and planet semimajor axes (Laplace-type approach). However, since the early studies these classical approaches have been found inadequate for Hilda-type motion for two reasons \citep[e.g.,][]{laves1904,ha1937,m1986,fm1988,mm1993}: (i) both eccentricity and inclinations of Hilda asteroids may be quite large, and (ii) the orbits reach a heliocentric distance $r$ which is temporarily larger than that of Jupiter $r'$. In this situation, if precision is desired, $\mathcal{R}$ must be evaluated numerically.

The specifics of the motion in the 3/2 mean motion resonance motivate us to introduce a new set of canonical variables $(Z,N,S_z;z,\nu,\sigma_z)$. The two angles $\nu$ and $\sigma_z$ were already given above, while their conjugated momenta are
\begin{eqnarray}
 N & = & \frac{3}{2}\,L-H -b \sqrt{2\left(L-G\right)}\cos\left(\varpi'-\varpi\right)
  +\frac{b^2}{2}\; , \\
 S_z & = & G-H \; .
\end{eqnarray}
The remaining two variables $Z$ and $z$ defined using a powerful reducing transformation \citep[e.g.,][]{h1986,wis1986,sb1988}
\begin{eqnarray}
 \sqrt{2 Z}\,\cos z & = & \sqrt{2\left(L-G\right)}\,\cos\sigma-b \,\cos\nu\; , \label{red1} \\
 \sqrt{2 Z}\,\sin z & = &  \sqrt{2\left(L-G\right)}\,\sin\sigma+b \,\sin\nu    \label{red2} 
\end{eqnarray}
\citep[see also][who considered a planar variant of the Hilda-type motion]{lh1988}.
Here, $b$ is an arbitrary constant. When $b=0$, $Z=L-G$ and $z=\sigma$, otherwise $z$ constitutes a new resonant angle with better behavior than $\sigma$ for the elliptic orbit of Jupiter. For that reason $b\propto e'$, with the exact relation being discussed below. 

Next, we consider a canonical transformation to a final set of variables
$(\Phi,\Psi,\Psi_z;\phi,\psi,\psi_z)$ defined by
\begin{equation}
 \begin{array}{llll}
  \Phi & =  N-Z-S_z\; , &  \phi & = -z \; , \\
  \Psi & =  N \; ,      &  \psi & = z+\nu \; , \label{final_var} \\
  \Psi_z & =  S_z \; ,  &  \psi_z & = \sigma_z-z \; .
 \end{array}
\end{equation}
A notable simplicity, originating in properties of the reducing transformation, is the relation of the momentum $\Phi$ to the orbital semimajor axis $a$, namely
$\Phi=\frac{1}{2}\sqrt{(1-\mu)a}$. Furthermore, $N\simeq \frac{3}{2}L-H + \mathcal{O}(eb,b^2)$. Observing further that the orbital inclination $I$ values for Hilda-type orbits are not too large and $H\simeq G + \mathcal{O}(\sin^2 I/2)$, the inclination and node dynamics are to a large degree decoupled
from the other elements. In particular, the conjugated pair of
variables $(\Phi,\phi)$ describe the fundamental resonant dynamics, and the conjugated pair of variables $(\Psi,\psi)$ make it coupled to the dynamics of the eccentricity and longitude of perihelion. 

So far we focused on orbital variables of the Hilda-type asteroid, but
the Hamiltonian $\mathcal{H}$ in (\ref{ham1}) depends also on parameters
describing Jupiter's motion. In our approximation of the restricted elliptic problem,
$(\Lambda',\lambda')$ constitute the active conjugated pair of Jupiter-related variables with $\Lambda'=L'+\frac{3}{2}L$. Further reduction of degrees-of-freedom of the problem follows from observation of hierarchy among the timescales
on which the coordinates change. Jupiter's mean longitude in orbit $\lambda'$ is the fastest of all, followed by $\phi$ whose characteristic timescale is at least an order of magnitude longer than Jupiter's orbital period. As the short-term orbital effects are not relevant for our application, we average the Hamiltonian over $\lambda'$. Since the longitude in orbit $\lambda$ of the asteroid is resonantly coupled to $\lambda'$, the averaging involves both angles, preserving though the slower variables such as $\phi$ (or related $\sigma$ and $\nu$ from
Eqs.~\ref{red1} and \ref{red2}). The average value of the perturbing potential $\mathcal{R}$
is evaluated numerically \citep[we use a simple Romberg method with tightly controlled accuracy, see, e.g.,][]{nr2007} using
\begin{equation}
 {\overline \mathcal{R}} = \frac{1}{6\pi} \int_0^{6\pi} d\ell \,\mathcal{R}
  \left(\ell,\ell'\right)\; ,
\end{equation}
with $\ell'(\ell)= \frac{2}{3}\left(\ell-\sigma\right)-\nu$ \citep[e.g.,][but see already \citet{schub1964}]{moons1994}.
Dropping the constant $\Lambda'$, the average Hamiltonian then reads
\begin{equation}
 {\overline \mathcal{H}} = \mathcal{H}_0 -{\overline \mathcal{R}} =
 -3\Phi -\frac{(1-\mu)^2}{8\Phi^2}-{\overline \mathcal{R}}\; .
\end{equation}
Since ${\overline \mathcal{R}}$ depends on all variables listed in Eq.~(\ref{final_var}), the problem still
represents a three-degrees of freedom task (therefore the ${\overline \mathcal{H}}=C$ integrals of motion are rather complex hyper-surfaces in a six-dimensional space). Nevertheless, the unperturbed part $\mathcal{H}_0$ depends only on $\Phi$, which at least allows to define a zero-order resonance location $\Phi_\star$
from Hamilton's equation $\partial \mathcal{H}_0 /\partial \Phi =0$, with a straightforward solution $\Phi_\star = \left[\left(1-\mu\right)^2/12\right]^{1/3}$. Alternately, we may also use the $\Phi=\Phi(a)$ correspondence to express the approximate location of the resonance using $a_\star=(2/3)^{2/3} \left(1-\mu \right)^{1/3}\simeq 0.7629$ \citep[in units of Jupiter's semimajor axis $a'=1$; e.g.,][]{m1986}. Near $\Phi_\star$, in particular within the $\delta \Phi\propto \sqrt{\mu}$ vicinity, the resonant angle $\phi$ changes slowly, and the proper resonant regime is distinguished by libration of $\phi$ about the stable stationary point at $\phi=0$ for reasonably small values of orbital eccentricity \citep[e.g.,][]{mm1993}. 

% FIG 1 %%%%%%%%%%%%%%%%%%%%%%%%%%%%%%%%%%%%%%%%%%%%%%%%%%%%%%%%%%%%%%%%%%%%%%%%%%%%%%%%%%%%%%%%%%%%%%%
\begin{figure*}[t!]
 \begin{center}
  \includegraphics[width=0.9\textwidth]{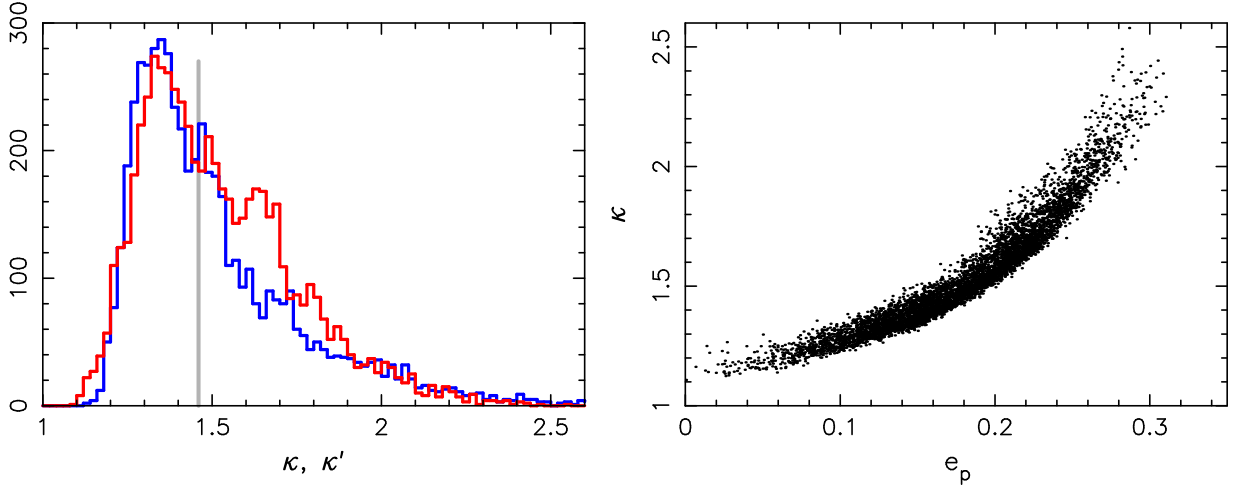} 
 \end{center}
 \caption{Left panel: Statistical distribution of the eccentricity 
  factor $\kappa$ (Eq.~\ref{eprop2}, red) and the inclination factor $\kappa'$ (Eq.~\ref{pi}, blue) numerically determined for $6,094$ orbits of the observed Hilda population. Their median values $\simeq 1.47$ are close to each other and highlighted by the gray vertical line. Right panel: Correlation of the eccentricity factor $\kappa$ (ordinate) and the proper eccentricity parameter $e_{\rm p}$ (abscissa) numerically determined from the sample of $6,094$ observed Hildas.}
 \label{figk}
\end{figure*}
%%%%%%%%%%%%%%%%%%%%%%%%%%%%%%%%%%%%%%%%%%%%%%%%%%%%%%%%%%%%%%%%%%%%%%%%%%%%%%%%%%%%%%%%%%%%%%%%%%%%%%%

\subsection{Osculating orbits to proper parameters: $\mathbf{e}$ to $\mathbf{p}$}\label{map1}
In order to achieve further progress, at least in an approximate way, we continue to follow the timescale hierarchy. The period of $\phi$ (typically $250$ to $350$~yr) is now the shortest, followed by the intermediate period of $\psi$, and the longest period of $\psi_z$. As a rule of thumb, there is always about an order-of-magnitude separating each of the consecutive timescales. Since the core of resonant dynamics consists in $(\Phi,\phi)$ variable, we now take the opposite standpoint to the procedure of averaging over $\lambda'$.
Namely, we consider $(\Psi,\Psi_z;\psi,\psi_z)$ approximately constant over libration timescale of $\phi$. Freezing two degrees of freedom, ${\overline \mathcal{H}}={\overline \mathcal{H}}(\Phi,\phi)$ is integrable, though not in analytical terms. However, ${\overline \mathcal{H}}=C$ first integral allows us to numerically map arbitrary initial conditions to a
reference point on the $C$-isoline which will be taken as a first proper-parameter. In particular, we consider the crossing of ${\overline \mathcal{H}}=C$ with $\phi=0$ at maximum $\Phi=\Phi_{\rm p}$ value. Because of direct correspondence between $\Phi$ and the semimajor axis, we find more instructive to interpret $\Phi_{\rm p}$ in terms of proper semimajor axis
\begin{equation}
 a_{\rm p}=4\Phi_{\rm p}^2/(1-\mu)\; . \label{aprop}    
\end{equation}
Next, we turn to Eqs.~(\ref{red1}) and (\ref{red2}), which we express at the reference point $z=\phi=0$ and $\Phi_{\rm p}$. We set $b=\sqrt{2\Phi_\star}\,\kappa\, e'$ (introducing thus a new arbitrary, but constant factor $\kappa$ instead of $b$), $e_{\rm p}=\sqrt{(\Psi-\Phi_{\rm p}-\Psi_z)/\Phi_\star}$ and $-\varpi_{\rm p}=-3\lambda'+2\lambda$, rewriting thus Eqs.~(\ref{red1}) and (\ref{red2}) to a more familiar form dependent on the heliocentric orbital elements
\begin{equation}
 \sqrt{2\,\frac{\Phi_{\rm p}}{\Phi_\star}\left(1-\eta\right)}\,\exp(\imath \varpi) = \kappa\, e'\,\exp(\imath \varpi') + e_{\rm p}\,\exp(\imath \varpi_{\rm p})  \label{eprop1}
\end{equation}
($\eta=\sqrt{1-e^2}$). For small-eccentricity orbits we may further approximate $\sqrt{2(1-\eta)}\simeq e$, and with $\sqrt{\Phi_{\rm p}/\Phi_\star}\simeq 1$, we obtain
\begin{equation}
 e\,\exp(\imath \varpi) = \kappa\, e'\,\exp(\imath \varpi') + e_{\rm p}\, \exp(\imath \varpi_{\rm p}) \; . \label{eprop2}
\end{equation}
This relation, suggested already by \citet{schu1968} who analysed results of short-term numerical simulations of few Hilda-type orbits known at that time \citep[see also][]{schu1982a,schu1991}, resembles secular dynamics of
non-resonant asteroids in the main belt. The two terms on the right hand side represent (i) the forced eccentricity term due to Jupiter perturbation, and (ii) the free (proper) eccentricity term. Since the forced eccentricity $e_{\rm f}$ is not equal to Jupiter's osculating eccentricity $e'$ at the current epoch, we need to consider the factor $\kappa$ such that $e_{\rm f}=\kappa\,e'$;
$\kappa$ may either be estimated using analytical or semi-analytical methods \citep[e.g.,][]{fm1988,sb1988}, or could be determined purely numerically \citep[e.g.,][]{schu1968,schu1982a,schu1991}. For sake of simplicity, we took the second approach and estimated $\kappa$ numerically, by performing Fourier analysis of the Cartesian folding of the eccentricity and perihelion using $e\,\exp(\imath \varpi)$ on the left hand side of Eq.~(\ref{eprop2}) for $6,094$ orbits of the observed population of Hildas (Sec.~\ref{datag96}). To that end we performed  a $1$~Myr lasting propagation using the {\tt swift} package.
We used $5$~yr sampling and discrete Fourier transform algorithm by \citet{fm1981}. We identified terms with planetary frequencies, dominated by $g_5$ and $g_6$, and the proper frequency $g$ (not only $g$ is at least an order of magnitude larger, but it is also negative, corresponding to the retrograde precession of $\varpi_{\rm p}$). The amplitude of the $g$ terms provides an estimate of $e_{\rm p}$, and the composite amplitude of the forced terms defines $e_{\rm f}$. Given $e'=0.0483$, we can easily determine $\kappa$ factor of each orbit in the population. Their statistical distribution is shown on the left panel of Fig.~\ref{figk} (red histogram). There is a rather large spread in the individual $\kappa$ values spanning an interval of values $1$ to
$\sim 2.5$. We note $\kappa$ is correlated with the proper eccentricity parameter $e_{\rm p}$ (right panel on Fig.~\ref{figk}),
with smaller $\kappa$ values more typical for small $e_{\rm p}$ orbits and vice versa. In our method, however, we have to choose a single value of $\kappa$ (or $b$; Sec.~\ref{cano}). We thus determine the median value ${\overline \kappa}\simeq 1.47$, and use it in Eqs.~(\ref{eprop1}) or (\ref{eprop2}). While there is no dynamical reason for $\varpi_{\rm p}$ to not be uniform (and, indeed, the observed bright Hildas confirm this conclusion; left panel on Fig.~\ref{fig5}), detection probability of faint Hildas depends on this parameter. With that conclusion we thus extended the set of proper orbital parameters by $\varpi_{\rm p}$. Both $e_{\rm p}$ and $\varpi_{\rm p}$ are determined from either of the Eqs.~(\ref{eprop1}) or (\ref{eprop2}).

% FIG 1 %%%%%%%%%%%%%%%%%%%%%%%%%%%%%%%%%%%%%%%%%%%%%%%%%%%%%%%%%%%%%%%%%%%%%%%%%%%%%%%%%%%%%%%%%%%%%%%
\begin{figure}[t!]
 \begin{center}
  \includegraphics[width=0.47\textwidth]{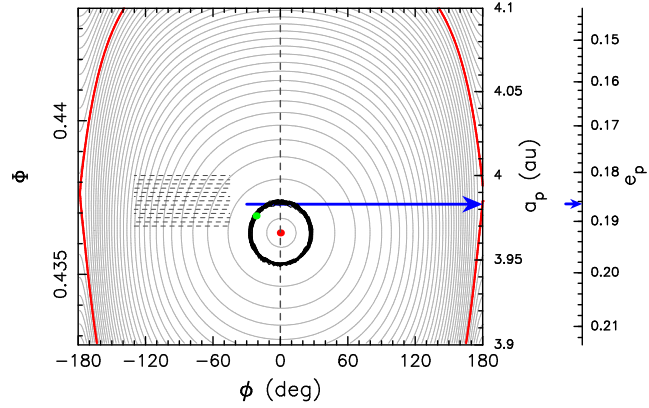} 
 \end{center}
 \caption{Resonant variables $\Phi$ vs $\phi$ for (1911) Schubart. The gray contours are isolines of constant Hamiltonian ${\overline \mathcal{H}}=C$ for different values of $C$. Critical curves are highlighted in red: (i) the stable stationary center of the resonance at $\phi=0$ corresponds to a maximum value of $C\simeq -1.965258$, and (ii) the separatrix of the resonant regime connects the unstable stationary centers at $\phi=\pm 180^\circ$ and has $C\simeq -1.966212$. The black curve is a result of a numerical integration of the real orbit of (1911) Schubart with perturbations from all planets taken into account (the green dot specifies the initial conditions of the simulation). The simulation span $5$~kyr time interval. In order to eliminate a slight blur by short-period perturbations, terms with periods smaller than $20$~yr have been digitally filtered out (equivalent to averaging the Hamiltonian over $\lambda_1$). Remaining behaviour is a simple libration about the stationary point (red dot) with a period of P$_\phi=262.5$~yr. The grid in $a_{\rm p}$, schematically shown by the dashed lines, corresponds to the bins used in our populations analysis (see Table~\ref{bins}).}
 \label{fig_1911_1}
\end{figure}
%%%%%%%%%%%%%%%%%%%%%%%%%%%%%%%%%%%%%%%%%%%%%%%%%%%%%%%%%%%%%%%%%%%%%%%%%%%%%%%%%%%%%%%%%%%%%%%%%%%%%%%
% FIG 1 %%%%%%%%%%%%%%%%%%%%%%%%%%%%%%%%%%%%%%%%%%%%%%%%%%%%%%%%%%%%%%%%%%%%%%%%%%%%%%%%%%%%%%%%%%%%%%%
\begin{figure*}[t!]
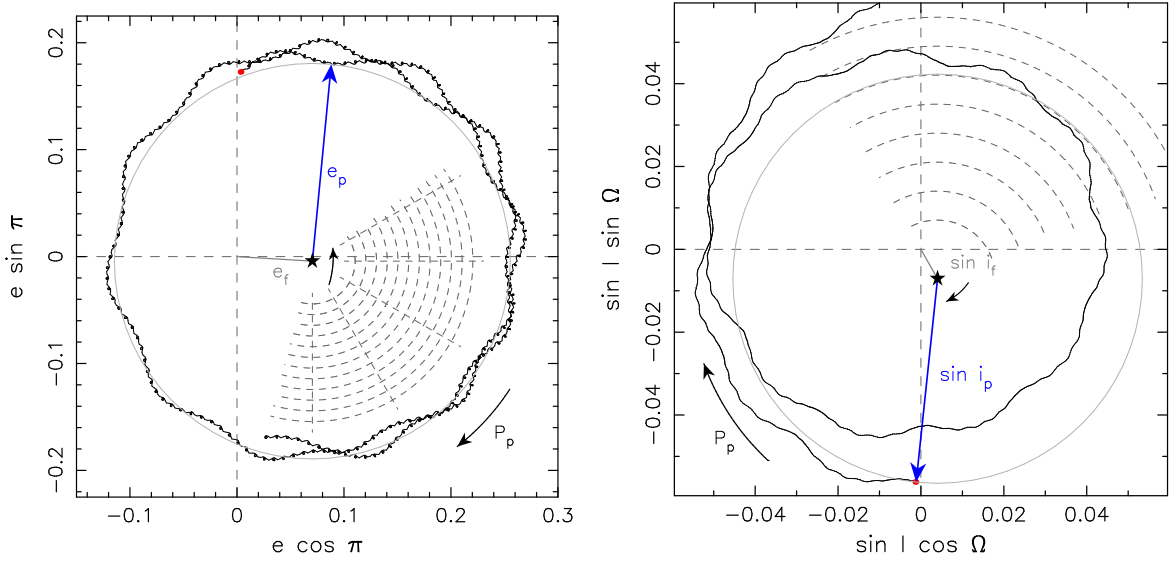

 \begin{center}
 \begin{tabular}{cc}
  \includegraphics[width=0.42\textwidth]{f23a.eps} &
  \includegraphics[width=0.42\textwidth]{f23b.eps} \\
 \end{tabular}
 \end{center}
 \caption{The orbit of (1911) Schubart numerically propagated and
  projected onto the Cartesian-like eccentricity-perihelion plane of $e\,\exp(\imath\,\varpi)$ (left panel) and inclination-node plane of $\sin I\,\exp\left(\imath\, \Omega\right)$ (right panel). The simulation spanned $5$~kyr in the first case and $30$~kyr in the second case and the output is shown by the black line (the initial epoch J2000.0 of the propagation is shown by red dot). These results confirm decomposition onto the forced part (shown by the start symbol) and the proper term (highlighted by the gray circle) from Eqs.~(\ref{eprop2}) and (\ref{pi}). The proper eccentricity parameter $e_{\rm p}$ and inclination parameter $\sin I_{\rm p}$ are the respective radii of the gray circle and shown by blue vector. The arrow with P$_{\rm p}$ label indicate the sense of time evolution driven mainly by proper precession frequency, $g$ and $s$ respectively. For (1911) Schubart they read P$_{\rm p}=3059.8$~yr for eccentricity and P$_{\rm p}=20879.5$~yr for inclination.}
 \label{fig_1911_2}
\end{figure*}
%%%%%%%%%%%%%%%%%%%%%%%%%%%%%%%%%%%%%%%%%%%%%%%%%%%%%%%%%%%%%%%%%%%%%%%%%%%%%%%%%%%%%%%%%%%%%%%%%%%%%%%

The orbital inclination $I$ and longitude of node $\Omega$ of Hildas are only weakly coupled to the J3/2 resonant dynamics. Cartesian-type combination of the inclination and node in $\sin I\,\exp\imath \Omega$ exhibits, with a good accuracy, decomposition to the forced and free (proper) terms typical to main-belt asteroid orbits
\citep[e.g.,][]{schu1982a,schu1982b}; \citet{schu1982b} showed that the largest resonance-coupling terms at approximately $2g$ frequency have amplitude of only $\simeq 0.1^\circ-0.15^\circ$, which is less than the width of bins of proper inclination we use (see Table~\ref{bins} and Fig.~\ref{fig_1911_2}) and may be thus safely neglected. Similarly to the eccentricity case (Eq.~\ref{eprop2}), we may thus represent \citep[see also][]{vin2015}
\begin{equation}
 \sin I\,\exp\left(\imath\, \Omega\right) = \kappa'\,\sin I'\,
  \exp\left(\imath\, \Omega'\right) + \sin I_{\rm p}
 \exp \left(\imath\,\Omega_{\rm p}\right) \; . \label{pi} 
\end{equation}
The principal forced effect, approximated here by the first term, has the $s_6$ frequency. The typical range of proper frequency values $s$ of the second term in (\ref{pi}) is twice to five times higher than $s_6$. Therefore, the characteristic periods of the forced and proper inclination terms are less separated than in the eccentricity case. However, the amplitude of the forced term $\sin I_{\rm f}= \kappa'\,\sin I'\simeq 0.008-0.01$ is quite small (corresponding to
$I_{\rm f}\simeq 0.5^\circ$). Except for the tail of very low-$I_{\rm p}$ orbits, not much populated, the right hand side of Eq.~(\ref{pi}) is nearly always dominated by the proper term.
Finally, we used the numerical propagation mentioned above of the known Hilda population to determine individual $\kappa'$ values for each orbit. The left panel of Fig.~\ref{figk} (blue histogram) shows that the distributions of the factors $\kappa$ and $\kappa'$ are similar (though not identical). However, their median values are close enough to each other, such that we use Eq.~(\ref{pi}) with $\kappa'={\overline \kappa'}=1.47$ to determine the proper inclination parameter $\sin I_{\rm p}$ for the observed population of Hildas.
\smallskip

\noindent{\it Exemplary case: (1911) Schubart.--} In order to illustrate the $\mathbf{e}$ to $\mathbf{p}$ transformation, namely mapping the observed Hilda population to the space of proper parameters, we consider the case of (1911) Schubart. The time evolution of various orbital parameters is taken from the numerical simulation of the Hilda population mentioned above.

Figure~\ref{fig_1911_1} shows the orbit of (1911) Schubart projected onto a 2-D plane of the resonant variables $(\Phi,\phi)$ (black curve). For better comparison with the theory outlined above, we digitally filtered terms having periods less than $20$~yr, a procedure equivalent to the averaging over $\lambda_1$. The left ordinate is $\Phi$, the right ordinate maps its value to $a$. The isolines of constant average Hamiltonian ${\overline \mathcal{H}}=C$ are shown in gray for selected values of $C$. For the maximum allowed $C$, they
degenerate to a single point at $\phi=0^\circ$, namely the stationary solution (``exact resonance'') shown by the red dot. The characteristic of resonant dynamics is the libration of the orbital $\phi$ about zero within a limited interval of values. The exact trajectory of (1911) Schubart follows very closely a specific $C$ isoline fixed by the initial conditions shown by the green dot. These initial conditions occur in a certain phase of the libration in $\phi$, but following the $C$-isoline, they can be numerically mapped to the point at $\phi=0^\circ$ where $\Phi$ achieves a maximum. The corresponding value $a=a_{\rm p}$, shown by the blue arrow, determines the proper semimajor axis parameter. Repeating this procedure for all known Hildas, we project them into the bins $a_{\rm p}$ specified in Table~\ref{bins} and shown here schematically by the dashed horizontal grid.

The left panel of Fig.~\ref{fig_1911_2} shows the orbit of (1911) Schubart projected onto a Cartesian-type space defined by the orbital eccentricity and perihelion, namely $e\,\exp(\imath\,\varpi)$. The black curve shows $5$~kyr long time interval of the orbit evolution with the present-day initial conditions shown by the red dot. At the zero approximation, the orbit evolution over this timescale is well represented by the ``epicyclic law'' in (\ref{eprop2})
in which (i) the forced part is fixed and shown by the black star (evolving in a prograde sense in the direction of the inner arrow with a period of $\simeq 300$~kyr of the $g_5$ planetary frequency), and (ii) the free (proper) part evolves in a retrograde sense with a proper period of P$_{\rm p}=3059.8$~yr. In principle, the radius of this free-epicycle is the desired eccentricity parameter. However, a more accurate Eq.~(\ref{eprop1}), or Eqs.~(\ref{red1}) and
(\ref{red2}), depend also in the resonant $\Phi$ variable producing the smallest-amplitude epicyclic wiggle with a period of P$_\phi=262.5$~yr. While small, this effect is significant compared to the width of the bins in the proper eccentricity parameters shown by the dashed polar grid about the forced center. For that reason, we specify $e_{\rm p}$ as the distance from the forced center when $\Phi=\Phi_{\rm p}$ (Eq.~\ref{eprop1}). The analysis in Sec.~\ref{datag96} also indicated that large eccentricities of Hildas require us to take into account the phase $\varpi_{\rm p}$ of the proper term into account. The $30^\circ$ bins in this parameter are also shown by the excerpt of dashed polar grid.

Finally, the right panel of Fig.~\ref{fig_1911_2} shows the orbit of (1911) Schubart projected onto the Cartesian-type space defined by the orbital inclination and node, namely $\sin I\,\exp(\imath\,\Omega)$. Here, the black curve shows the numerically propagated orbit over a longer interval of $30$~kyr. Eq.~(\ref{pi}) again represents a very good zero-order description of the evolution with now less separation between the circulation of (i) the forced center (black star) with the fundamental period of $\simeq 49200$~yr ($s_6$ frequency), and (ii) the proper term with the fundamental period of P$_{\rm p}=20879.5$~yr (both precess in a retrograde sense as indicated by the arrows). For that reason, there is a notable shift in the forced center in the P$_{\rm p}$ period. The exact trajectory also exhibits a very small-amplitude term at the $2g$ frequency due to a very weak resonant coupling \citep[e.g.,][]{schu1982b}, but we neglect it in our analysis since it is smaller than the characteristic width of the proper inclination bin highlighted by the dashed polar circles centered at the forced term (see also Table~\ref{bins}).

\subsection{Proper parameters to osculating orbits: $\mathbf{p}$ to $\mathbf{e}$}\label{map2}
We now turn to the inverse problem, namely assigning a heliocentric orbit, with orbital elements $\mathbf{e}$, to a chosen $\mathbf{p}$-space orbit. This task is required in Sec.~\ref{det_prob} in the process of determining the probability of detection to a specific bin in $\mathbf{p}$. Because the dimensionality of the $\mathbf{p}$-space is lower than the $\mathbf{e}$-space (4 versus 6), the method also requires the addition of two parameters.

Assume that we select a certain $\mathbf{p}$-space bin. The bin size is small enough to justify a uniform distribution of trial (synthetic) orbits. Thus, we start with a vector of $(a_{\rm p},e_{\rm p},\sin I_{\rm p},\varpi_{\rm p})$ values. The algorithm for obtaining the heliocentric elements $\mathbf{e}$ is as follows:
\begin{description}
\item[eccentricity $e$ and longitude of perihelion $\varpi$] 
 these two elements are easily obtained by either of the Eqs.~(\ref{eprop1}) or (\ref{eprop2});
\item[inclination $I$ and longitude of node $\Omega$] here the 
 Eq.~(\ref{pi}) is used with $\Omega_{\rm p}$ drawn with uniform distribution from the interval $(0^\circ,360^\circ)$;
\item[semimajor axis $a$ and longitude in orbit $\lambda$] the proper configuration of the orbit corresponds to the extremal semimajor axis $a=a_{\rm p}$ at which $\phi=0$; however, there is a plethora of other orbits along the same averaged Hamiltonian ${\overline \mathcal{H}}=C$ curve that map to the same proper $a_{\rm p}$ (see, e.g., Fig.~\ref{fig_1911_1}); we assume a uniform distribution of orbit occurrence along the $C$-integral in polar angle measured from the resonance stationary point and pick $(\Phi,\phi)$ for which ${\overline \mathcal{H}}(\Phi,\phi)=C$, keeping the previously mentioned parameters constant; finally, the new set of orbital parameters, now with new $(\Phi,\phi)$ is transformed to the semimajor axis $a$ and longitude in orbit $\lambda$ values according to $a=4\Phi^2/(1-\mu)$ and Eqs.~(\ref{red1}) and (\ref{red2}); when solving $\lambda$ from the condition $\sigma=3\lambda'-2\lambda-\varpi_{\rm p}$ one should not hastily adopt just $\lambda=(3\lambda'-\varpi_{\rm p}-\sigma)/2$, but allow also for an equally possible solution $\lambda=(3\lambda'-\varpi_{\rm p}-\sigma)/2+\pi$. 
\end{description}
Heliocentric orbital elements $\mathbf{e}$ are primarily given in the invariable frame of the solar system throughout this paper. However, the available software needed for detection probability computation requires one to know the orbit in the ecliptic frame as well. For that reason, we specify the transformation between these two systems, as used here, in Appendix~\ref{trans}.
% FIG 1 %%%%%%%%%%%%%%%%%%%%%%%%%%%%%%%%%%%%%%%%%%%%%%%%%%%%%%%%%%%%%%%%%%%%%%%%%%%%%%%%%%%%%%%%%%%%%%%
\begin{figure*}[t!]
 \begin{center}
 \begin{tabular}{cc}
  \includegraphics[width=0.48\textwidth]{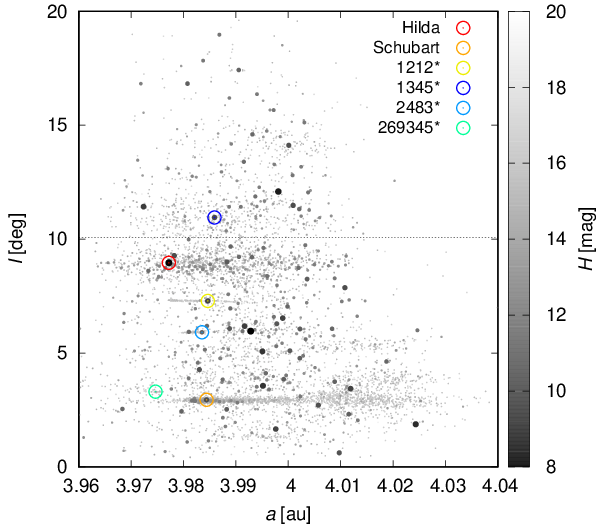} &
  \includegraphics[width=0.48\textwidth]{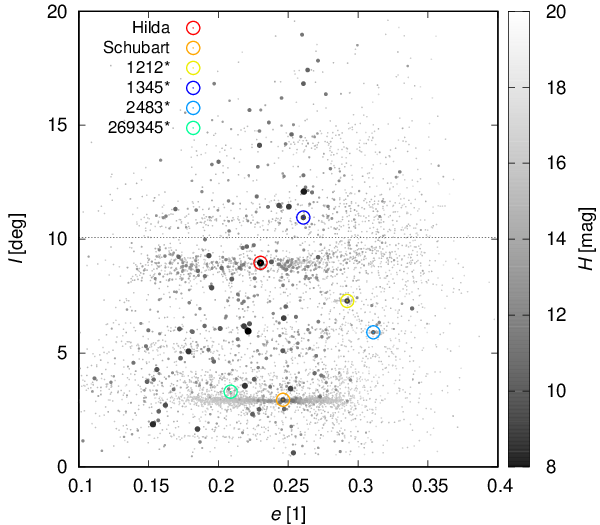} \\
 \end{tabular}
 \end{center}
 \caption{Synthetic proper elements for all Hildas shown in 2-D projections: $(a_{\rm p},I_{\rm p})$ (left panel), and $(e_{\rm p},I_{\rm p})$ (right panel) (width and grayscale of the symbol roughly proportional to the absolute magnitude; see the bar). Largest members of the collisional families are indicated by circles, color-coded and identified by the label. The horizontal dashed line at $10^\circ$ proper inclination is the auxiliary boundary used to separate the dispersed Hilda and Potomac families.}
 \label{fig_synprop}
\end{figure*}
%%%%%%%%%%%%%%%%%%%%%%%%%%%%%%%%%%%%%%%%%%%%%%%%%%%%%%%%%%%%%%%%%%%%%%%%%%%%%%%%%%%%%%%%%%%%%%%%%%%%%%%
% FIG 1 %%%%%%%%%%%%%%%%%%%%%%%%%%%%%%%%%%%%%%%%%%%%%%%%%%%%%%%%%%%%%%%%%%%%%%%%%%%%%%%%%%%%%%%%%%%%%%%
\begin{figure*}[t!]
 \begin{center}
  \begin{tabular}{c}
   \includegraphics[width=1.\textwidth]{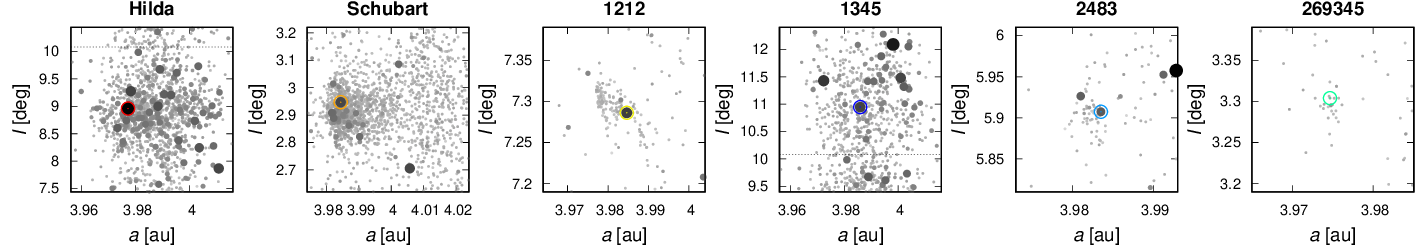} \\
   \includegraphics[width=1.\textwidth]{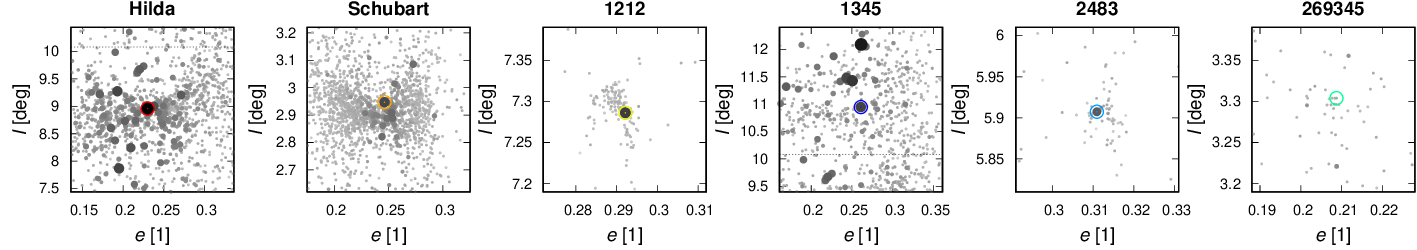} \\
  \end{tabular}
 \end{center}
 \caption{Synthetic proper elements of members in Hilda-population collisional families identified by the label at the top of the respective panel.}
 \label{fig_syn-prop_fams}
\end{figure*}
%%%%%%%%%%%%%%%%%%%%%%%%%%%%%%%%%%%%%%%%%%%%%%%%%%%%%%%%%%%%%%%%%%%%%%%%%%%%%%%%%%%%%%%%%%%%%%%%%%%%%%%

% ------------------------------------------------------------------
\section{Synthetic proper elements} \label{propel}
In this section, we determine synthetic proper elements for Hilda population using a fully numerical method described in \citet{bv2008} \citep[see also][where the same approach was used for the Hecuba population in the J2/1 resonance]{retal2002}. These elements are more accurate and stable than the proper parameters discussed in the previous section but require more intense numerical computation, and their relation to the osculating elements at a given epoch is less straightforward. For those reasons, they are not suitable for the main task of this paper, namely the principal variables of the parametric space of the Hilda-population model. However, their higher precision will allow us to identify the collisional families more accurately than the proper parameters. Since this effort was independent from the remaining parts of the paper, there are few differences in the sample statistics (not affecting though our main goal). Most importantly, the baseline set of candidate osculating orbits was downloaded from the {\tt AstOrb} catalog in mid July 2024, namely half a year later than the project started (Sec.~\ref{datag96}). We identified orbits for which the resonant angle $\sigma = 3\lambda' - 2\lambda - \varpi$ either librates, or slowly circulates. These were processed further by method briefly outlined below. The resulting catalog of proper elements for $6,393$ objects in the Hilda population, as well as membership in the families, is available at \url{https://sirrah.troja.mff.cuni.cz/~mira/tmp/hildas/} and on \url{https://doi.org/10.5281/zenodo.14959239} repository.

We used well tested software package {\tt swift} \citep{swift1994}, specifically the MVS2 integrator \citep{lr2001} that we extended by tools specific to the proper element computation. Our dynamical model included gravity of the Sun and four giant planets (Jupiter to Neptune). In order to preserve the orbital and secular frequencies of the planets (due to missing perturbations produced by terrestrial planets), we applied the barycentric correction and rotation to the reference system defined by the invariant plane.

In the first step, we applied an on-line digital filter to suppress short-period oscillations from the orbital elements, thereby constructing the mean orbital elements. In particular, the osculating elements were sampled with $1$~yr step and we applied the filter A described in \citet{q1991} with a decimation factor 10. Therefore, the intermediate sampling of the mean elements was $10$~yr. We could not use much longer values, since the shortest resonant oscillations that interest us have periods of $\simeq 150$~yr.

In the second step, we constructed resonant orbital elements by observing the behavior of the critical angle $\sigma = 3\lambda' - 2\lambda - \varpi$, where $\lambda'$ and $\lambda$ are the mean longitude in orbit of Jupiter and the asteroid, and $\varpi$ is the longitude of pericenter of the asteroid. Objects in the J3/2 mean motion resonance are distinct by libration of $\sigma$, although as reminded in Sec.~\ref{propar}, the libration center still exhibits secular oscillations about $0^\circ$ value due to Jupiter's perturbations. The relevant timescales range from hundreds to thousands of years. The oscillations of $\sigma$ are related with the correlated resonant evolution of the semimajor axis $a$ and eccentricity $e$, and to a lesser degree inclinations $I$. Following methods developed in \cite{retal2002}, we determined the resonant orbital elements, namely semimajor axis, eccentricity and inclination by recording the mean elements when the following set of conditions were satisfied:
\begin{equation}
 |\sigma| < 10^\circ \land \dot\sigma > 0 \land |\varpi - \varpi'| < 10^\circ \; .  \label{surfsec}
\end{equation}
Resonant elements were assigned for each $10$~kyr long interval in our integration. For practical reasons, we took a tolerance of $10^\circ$ in (\ref{surfsec}) to make sure the condition was satisfied in the chosen steps of $10$~kyr.

In the last step, we collected the resonant elements throughout the duration of the $10$~Myr integration and computed their average values. This eliminates the remaining secular periods induced by perturbations by Jupiter, and indirectly also other giant planets. Except for chaotic orbits, located close to the resonant separatrix or interacting with several tiny secular resonances, the resulting average values are approximate integrals of motion. Their values represent our synthetic proper semimajor axis $a_{\rm p}$, eccentricity $e_{\rm p}$ and inclination $I_{\rm p}$. In order to estimate their uncertainty, we extended the integration to $50$~Myr and repeated the procedure on every consecutive $10$~Myr long interval. Except for a small set of chaotic orbits, the typical median value of proper elements variations were $\Delta a_{\rm p} \simeq 7\times 10^{-5}\,{\rm au}$, $\Delta e_{\rm p} \simeq 9\times 10^{-5}$, and $\Delta I_{\rm p} \simeq 0.004^\circ$. We found this level of accuracy adequate to the purpose of the present work.

Figure~\ref{fig_synprop} shows the projection of the Hilda population in the $(a_{\rm p},I_{\rm p})$ and $(e_{\rm p},I_{\rm p})$ planes of the proper elements. The size and grayscale of the symbols correspond to the absolute magnitude of the object. Similarly to other small-body populations (such as the main asteroid belt or Jupiter Trojan clouds), the arrangement into distinct clusters is the principal feature of the Hildas distribution in the proper element space. As elsewhere, these groups are interpreted to be collections of collisionally-born fragments and according to the tradition we call them families. Using objective methods developed for analysis of similar groups found in the main asteroid belt \citep[e.g.,][]{zap1990,zap1995}, we identify collisional families in the Hilda population in the next Section. They also represent an important component in our effort to describe the bias-corrected population of Hildas in the main text of this paper.
% Tab 2 %%%%%%%%%%%%%%%%%%%%%%%%%%%%%%%%%%%%%%%%%%%%%%%%%%%%%%%%%%%%%%%%%%%%%%%%%%%%%%%%%%%
\begin{deluxetable}{rlrrr}[t] 
 \tablecaption{\label{mb_fams_2023}
  Statistically significant families in the Hilda population determined using the HCM and our new catalog of synthetic proper elements. Newly identified families are denoted with $\star$.}
 \tablehead{
  \colhead{number} & \colhead{designation} & \colhead{$v_{\rm cutoff}$} &
  \colhead{$N_{\rm mem}$} & Note \\ [-2pt]
  \colhead{} & \colhead{} & \colhead{(m s$^{-1}$)} & \colhead{} 
 }
% \decimalcolnumbers
\startdata
   153  & Hilda         & 90 & 1066 &  C-type \\
  1212  & Francette\tablenotemark{a} & 30 & 151 & --  \\
  1345  & Potomac$^\star$ & 140 &  506 & dispersed \\   
  1911  & Schubart      &  60 & 1882 & C-type \\
  2483  & Guinevere\tablenotemark{a}  & 40 & 54 & --  \\
269345  & 2008 TG106$^\star$\tablenotemark{b} & 54 & 17 & --  \\
\enddata
\tablenotetext{a}{Identified by \cite{vin2019,vino2020}.}
\tablenotetext{b}{Very compact cluster near Schubart family.}
\tablecomments{The number and designation correspond to the central body, $v_{\rm cutoff}$ is the velocity cutoff used in the HCM identification, $N_{\rm mem}$ is the number of associated members, and the last column provides additional brief information (such as a taxonomic class of the largest members).}
\end{deluxetable}
%%%%%%%%%%%%%%%%%%%%%%%%%%%%%%%%%%%%%%%%%%%%%%%%%%%%%%%%%%%%%%%%%%%%%%%%%%%%%%%%%%%%%%%%%%%
% FIG 1 %%%%%%%%%%%%%%%%%%%%%%%%%%%%%%%%%%%%%%%%%%%%%%%%%%%%%%%%%%%%%%%%%%%%%%%%%%%%%%%%%%%%%%%%%%%%%%%
\begin{figure}[t!]
 \begin{center}
 \includegraphics[width=0.47\textwidth]{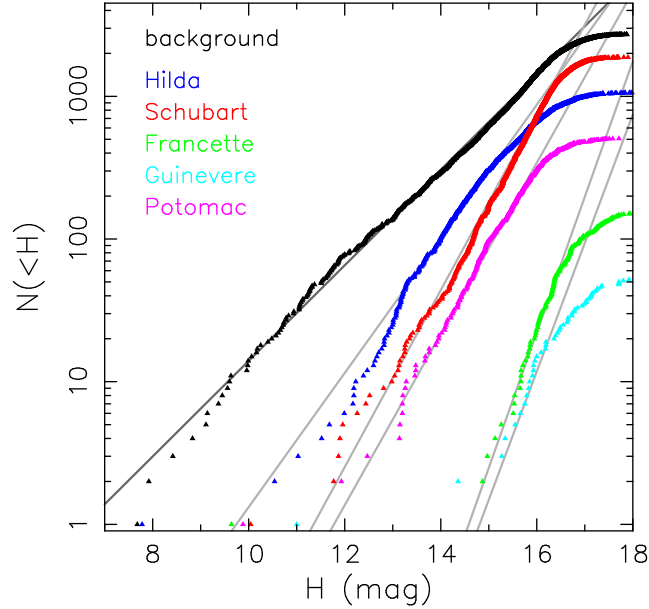} 
 \end{center} 
 \caption{Cumulative absolute magnitude distribution of the background and  the collisional families in the Hilda population. Black symbols is the background population, color-coded are the principal families: Hilda (blue), Schubart (red), Potomac (pink), Francette (green), and Guinevere (cyan). The dispersed nature of the Hilda family may be the reason for paucity of its members beyond $H\simeq 15$. The slanted lines are power-law approximations $N(<H)\propto 10^{\gamma H}$, with the following exponents: background $\gamma=0.34$, Hilda $\gamma=0.47$, Schubart $\gamma=0.6$, Potomac $\gamma=0.6$, Francette $\gamma=0.95$, and Guinevere $\gamma=0.9$. At small sizes  population in families becomes comparable (or even larger) to the background.}
 \label{figww}
\end{figure}
%%%%%%%%%%%%%%%%%%%%%%%%%%%%%%%%%%%%%%%%%%%%%%%%%%%%%%%%%%%%%%%%%%%%%%%%%%%%%%%%%%%%%%%%%%%%%%%%%%%%%%%
% FIG 1 %%%%%%%%%%%%%%%%%%%%%%%%%%%%%%%%%%%%%%%%%%%%%%%%%%%%%%%%%%%%%%%%%%%%%%%%%%%%%%%%%%%%%%%%%%%%%%%
\begin{figure}[t!]
% \plottwo{g96_nobs_L4.eps}{g96_nobs_L5.eps}
 \begin{center}
 \includegraphics[width=0.47\textwidth]{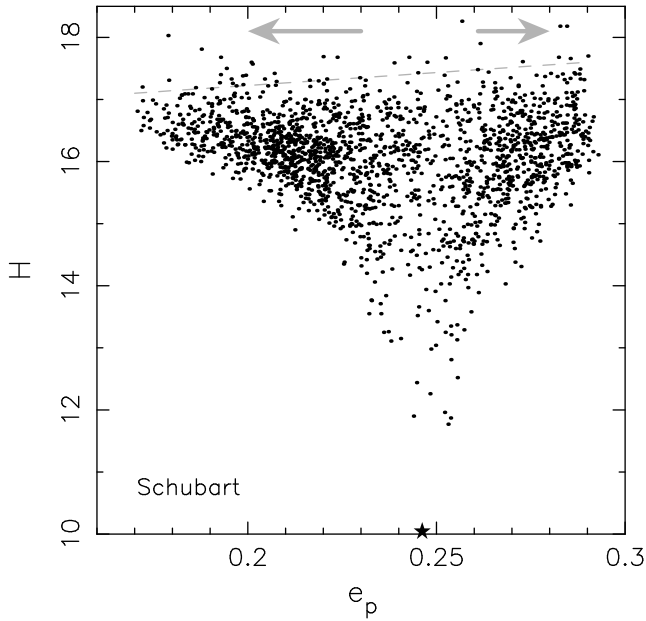} 
 \end{center} 
 \caption{Members of the Schubart family projected onto the plane defined by the proper eccentricity (abscissa) and absolute magnitude (ordinate); the star is (1911)~Schubart itself. The dashed line highlights higher detection efficiency at larger eccentricities (also discussed in Figs.~\ref{fig4} and \ref{fig5}). The gray arrows indicate suspected past flow of small Schubart members towards smaller/larger $e_{\rm p}$ values by synergy of the Yarkovsky and YORP effects \citep[see the Appendix of][]{bv2008}.}
 \label{fig1911}
\end{figure}
%%%%%%%%%%%%%%%%%%%%%%%%%%%%%%%%%%%%%%%%%%%%%%%%%%%%%%%%%%%%%%%%%%%%%%%%%%%%%%%%%%%%%%%%%%%%%%%%%%%%%%%

\subsection{Family identification}
We used the hierarchical clustering method (HCM; \citealt{zap1990,zap1995}) to identify the families. The cutoff velocity $v_{\rm cutoff}$ was individually adjusted for each of the families. The value reflected the statistically required contrast between the family and the background population of asteroids \citep[see discussion in][]{netal2015}. The resulting families were further analyzed and suspected interlopers were removed using: (i) taxonomic information based on broad-band photometry indexes, available spectra or albedo values \citep[data obtained from][]{parker2008,usui2011,nugent2015}, and (ii) projection of the family onto the plane of proper eccentricity $e_{\rm p}$ and absolute magnitude $H$. Asteroids too distant from the largest fragment in $e_{\rm p}$, for their given $H$, were removed from the family list. In particular, there exists a certain $C$-value for each of the families such that the valid members have $H\geq 5\log_{10}\left(|e_{\rm p}-e^\star_{\rm p}|/ C\right)$, with $e^\star_{\rm p}$ corresponding to the largest fragment in the family. This is reminiscent of the similar property of the families in the main belt, where, however, the proper eccentricity is replaced with the proper semimajor axis \citep[e.g.,][]{vok2006}. This is because the fragment distribution in the respective projections reflect either an initial velocity field (smaller members ejected with a larger velocity) or --more often-- the ability of small asteroids to drift farther from the family center by the Yarkovsky effect. This process affects principally the semimajor axis, but the resonant situation $a_{\rm p}$ is locked, and the thermal forces make the orbit evolve in $e_{\rm p}$ \citep[e.g.,][]{bv2008}. Finally, the Hilda and Potomac families were separated ''manually'' separated at $I_{\rm p}=10^\circ$. This is because both are rather dispersed clusters which require the largest values of $v_{\rm cutoff}$. But formally, at the required values, they would merge together. Yet, the distinct minimum in asteroid density near the chosen limit argues that they must be unrelated.

The final list of collisional families in the Hilda population is provided in Table~\ref{mb_fams_2023}, and their morphological appearance in the proper element projections $(a_{\rm p},I_{\rm p})$ and $(e_{\rm p},I_{\rm p})$ is shown in Fig.~\ref{fig_syn-prop_fams}.
Of the six families, two have been discovered in this work: the broad and dispersed Potomac family and a very compact 2008~TG106 family located very near the Schubart family. The dispersed nature of Hilda and Potomac families suggests that some of the assigned members may still be interlopers from the background populations. All other families are very compact with likely small interloper contribution. Taken at a face value, there is $3,676$ Hildas in the identified families of the $6,393$ population in total, which represents nearly $60$\% of the share. This fraction is very large, and it implies that we must take the families into account in the debiasing effort presented in the main part of this part. We include the three most populous families, namely Hilda, Schubart, and Potomac. Furthermore, families may become an even more important part of the population in smaller sizes due to their steeper magnitude distribution (Fig.~\ref{figww}).

Figure~\ref{fig1911} illustrates the structure of the most prominent Schubart family in the $(e_{\rm p},H)$ plane. The correlation between the smallest asteroids and proper eccentricity, highlighted by the dashed gray line, confirms trends also seen in Figs.~\ref{fig4} and \ref{fig5}. The concentration of smaller Schubart members ($H\geq 15$, say) towards the extreme $e_{\rm p}$ values, and paucity in the center, is another very important feature seen in Fig.~\ref{fig1911}. This is a counterpart of a similar effect detected in many main-belt families but in the proper semimajor axis values. \citet{vok2006} found that such concentrations of small family members result from a combined perturbing effect of the Yarkovsky and YORP effects (and also proved that they can help determine the age of the family). The fact that we see this effect in the Schubart family proves that both the Yarkovsky and YORP effects had enough time to affect the family structure. Future analysis may improve earlier estimates of the Schubart family age based on simpler modeling \citep[e.g.][]{bv2008,pisa2017}.

% ------------------------------------------------------------------
\section{Coordinate transformation between ecliptic and invariable systems}\label{trans}
In this paper, we use two heliocentric reference systems: (i) the ecliptic system of the J2000.0 epoch (e.g., the MPC catalog used for the initial search of known Hilda population, or the {\tt oIF} software used as a part of detection probability by CSS), and (ii) the invariable reference system, in which we represent orbits of the Hilda population in the remaining parts of our work. Since there are several definitions of the invariable system used in the literature, here we briefly specify our definition.

The $z-$axis of the invariable reference frame of the solar system is
defined by the total orbital angular momentum $\mathbf{L}$ of the
Sun and the planets. The orientation of the $x-$axis is most often set to be close to the vernal equinox of the ecliptic frame. This is also our choice.
In order to transform vectors between the two reference frames we use
a rotation matrix $\mathbb{R}$. In particular, any vector $\mathbf{V}$
in the ecliptic frame is expressed as $\mathbb{R}\cdot \mathbf{v}$,
where $\mathbf{v}$ is the same vector in the invariable frame
(the rotation matrix of the inverse transformation is just the transposed
form of $\mathbb{R}$). Denoting 
\begin{equation}
 \mathbf{L} = L\,\left( \begin{array}{c}
      \sin\theta \cos\phi \\
      \sin\theta \sin\phi \\
      \cos\theta 
      \end{array} \right) \; , \label{orbmom}
\end{equation}
the total orbital angular momentum of planets in ecliptic J2000.0 frame,
we find $\theta=1.5773756^\circ$ and $\phi=17.47808416^\circ$. Then
\begin{equation}
  \mathbb{R}=\mathbb{S}+\mathbb{A}\; , \label{rotm}   
\end{equation}
with a symmetric $\mathbb{S}=\mathbb{E}
+ 2 \sin^2\theta/2\,\mathbb{T}^2$ and an anti-symmetric $\mathbb{A}=\sin\theta\, \mathbb{T}$ parts. Here, $\mathbb{E}$ is the unit matrix, $\mathbb{T}^2=\mathbb{T}\cdot\mathbb{T}$ and
\begin{equation}
 \mathbb{T} = \left( \begin{array}{ccc}
      0 & 0 & \cos\phi \\
      0 & 0 & \sin\phi \\
      -\cos\phi & -\sin\phi & 0
      \end{array} \right) \; . \label{rotinv}
\end{equation}

%\listofchanges
\end{document}